\documentclass[%
 reprint,
superscriptaddress,
nofootinbib,
 amsmath,amssymb,
 aps,
 prd,
floatfix
]{revtex4-2}

\usepackage{hyperref}
\usepackage{orcidlink}
\usepackage{xcolor}
\usepackage[all]{hypcap}
\usepackage{ mathrsfs }
\usepackage{graphicx}
\usepackage{dcolumn}
\usepackage{bm}
\usepackage{multirow} 
\usepackage{array} 
\usepackage{cprotect}
\usepackage{dsfont}

\usepackage[deletedmarkup=xout,commentmarkup=uwave,commandnameprefix=ifneeded]{changes}

\definechangesauthor[name=Max, color=teal]{MI}
\definechangesauthor[name=Will, color=orange]{WF}
\definechangesauthor[name=Harrison, color=violet]{HS}

\begin{document}

\title{GW231123 ringdown: interpretation as multimodal Kerr signal}

\newcommand{\piaffil}{\affiliation{Perimeter Institute for Theoretical Physics, 31 Caroline St N, Waterloo, ON N2L 2Y5, CA}}
\newcommand{\ccaaffil}{\affiliation{Center for Computational Astrophysics, Flatiron Institute, New York NY 10010, USA}}
\newcommand{\cuaffil}{\affiliation{
 Department of Physics, Columbia University,
704 Pupin Hall, 538 West 120th Street, New York, New York 10027, USA
}}
\newcommand{\sbuaffil}{\affiliation{Department of Physics and Astronomy, Stony Brook University, Stony Brook NY 11794, USA}}

\author{Harrison Siegel \orcidlink{0000-0001-5161-4617}}
\email{hs3152@columbia.edu}
\piaffil
\cuaffil
\ccaaffil

\author{Nicole M. Khusid \orcidlink{0000-0001-9304-7075}}
\sbuaffil
\ccaaffil

\author{Maximiliano Isi \orcidlink{0000-0001-8830-8672}}
\affiliation{Department of Astronomy and Columbia Astrophysics Laboratory, Columbia University, Pupin Hall, New York, NY 10027, USA}
\ccaaffil

\author{Will M. Farr \orcidlink{0000-0003-1540-8562}}
\sbuaffil
\ccaaffil

\begin{abstract}
GW231123 is a short-duration, low-frequency gravitational wave signal consistent with a binary black hole coalescence and dominated by the merger-ringdown regime due to the high mass of the source. We demonstrate that fits of this ringdown signal using two quasinormal modes are statistically preferred over single-mode fits, for a broad range of fit start times. We also find that two-mode fits give remnant mass and spin measurements consistent with those of the inspiral-merger-ringdown model NRSur7dq4, whereas one-mode fits struggle to do so. Agreement of our fits with those of NRSur7dq4 is achieved by labeling the two quasinormal modes as the ${(\ell,m)=(2,2)}$ and ${(2,0)}$ Kerr prograde fundamental modes. However, we find some indications that fits with the ${(2,1)}$ quasinormal mode instead of the ${(2,0)}$ mode may describe the data better, hinting at possible NRSur7dq4 error or other systematics. When fitting at early times near the estimated peak strain, we find that the inclusion of a third mode, an ${(\ell,m,n)=(2,2,1)}$ prograde overtone, improves consistency with fits at later times. Finally, we perform a test of general relativity by searching for deviations from the Kerr frequency spectrum. Setting issues of systematics aside, we validate the Kerr frequency and damping rate spectrum to within $\pm10\%$ at the 90$\%$ credible level using a fundamental mode fit, and we also report  $\pm8\%$ constraints using a model with fundamental modes and an overtone fit at times near the peak strain. Understanding the systematic errors that may be affecting the most accurate analyses of GW231123 is crucial in the context of population and binary formation studies -- our ${(2,1)}$ mode fits return a significantly higher remnant mass and spin than all available inspiral-merger-ringdown models including NRSur7dq4, and this difference in parameter estimates may have astrophysical implications.
\end{abstract}

\maketitle


\section{\label{sec:intro} Introduction}

GW231123\_135430, henceforth GW231123,~\cite{GW231123_paper} is a short-duration, low-frequency gravitational wave signal measured by the LIGO-Virgo-KAGRA (LVK)~\cite{PhysRevD.111.062002_LIGOdetector, Virgo:2022fxr_virgopaper, 10.1093/ptep/ptaa125_KAGRApaper, LIGOScientific:2014pky} collaboration and included in Gravitational-Wave Transient Catalog 4 (GWTC-4)~\cite{LIGOScientific:2025slb}. When the signal is modeled as being sourced by a quasicircular precessing binary black hole (BBH) coalescence using the NRSur7dq4~\cite{Varma:2019csw_NRsurpaper} inspiral-merger-ringdown (IMR) model, the final source-frame black hole remnant mass is found to be ${M_f^s = 227^{+18}_{-27} \; M_{\odot}}$, the primary and secondary binary components have dimensionless spins of ${\chi_1 = 0.89^{+0.11
}_{-0.20}}$ and ${\chi_2 = 0.91^{+0.09}
_{-0.19}}$ respectively, and the matched filter signal-to-noise ratio (SNR) of the signal is ${\text{SNR}_\text{mf} = 22.6^{+0.3}
_{-0.3}}$ (reporting median and $90 \%$ highest posterior density values, see Table 3 of Ref.~\cite{GW231123_paper}).\footnote{The IMR posteriors we use in this work come from a version of Ref.~\cite{GW231123_paper} with an incorrect likelihood as described in Ref.~\cite{Talbot:2025vth}. The corrected parameter estimates are not sufficiently different to change the main conclusions of our work. See the journal-published version of Ref.~\cite{GW231123_paper} for the corrected estimates.} The signal is dominated by its merger-ringdown, due to the high detector-frame mass placing the inspiral regime in the less sensitive lower-frequency band of the LIGO detectors: we estimate the SNR of the signal from the peak of the strain onwards to be ${\text{SNR}_\text{post-peak} = 18.0^{+2.0}
_{-1.5}}~$\footnote{The post-peak SNR is computed in the time domain, as the distribution of optimal SNRs of 0.2 s duration template segments from their respective peaks onwards generated from 1000 random NRSur7dq4 posterior samples and whitened with the noise auto-covariance from our ringdown analysis~\cite{Isi:2021iql_analyzingbhringdowns, Siegel:2024jqd}. Our ${\text{SNR}_\text{post-peak}}$ uses a different noise model from the full-signal ${\text{SNR}_\text{mf}}$ reported in Ref.~\cite{GW231123_paper} (see Sec.~\ref{sec:conditioning}), so direct comparison should not be made, however ${\text{SNR}_\text{post-peak}}$ is the more relevant quantity for our specific ringdown analysis.} using NRSur7dq4, making the merger-ringdown of GW231123 the loudest of any signals until GW250114~\cite{GW250114_paper1, GW250114_paper2}.

In the theory of general relativity (GR), the final stage of a BBH coalescence can be described at sufficiently late times by perturbation theory via the Teukolsky equation~\cite{Teukolsky:1973ha}: this stage is the so-called ringdown. In this perturbative regime, the signal is quickly dominated by quasinormal modes (QNMs), which eventually decay and become subdominant to tails. The QNMs are observed as damped sinusoids possessing frequencies and damping rates determined solely by the mass and spin of the remnant black hole, and having amplitudes and phases related to the binary black hole configuration. Clear observation of more than one QNM in the spectrum emitted by a merger remnant allows for the inference of progenitor properties and validation of the Kerr metric \cite{testgr_ringdown_2,testgr_ringdownbayesian,testgr_overview, Zhu:2023fnf, Kamaretsos:2012bs,  Berti:2025hly}, and has recently been the subject of intense data analysis efforts \cite{Carullo:2019flw, IsiNoHair_GW150914,Isi_BHArea, LIGOScientific:2021sio_TGRGwtc-3, Isi_revisitGW150914,FinchMoore_GW150914, Cotesta_ringdownGW150914, Isi:2023nif, Siegel:2023lxl, Capano:2021etf, Capano:2022zqm, GW250114_paper1, GW250114_paper2, Brito:2018rfr, Pompili:2025cdc, Ma:2022wpv, Lu:2025mwp} following the dawn of gravitational-wave astronomy \cite{GW150914_detection,GWAstro_Review, Flanagan:2005yc}. This program of QNM-based inference is often referred to as black hole spectroscopy~\cite{Dreyer:2003bv}.

Ref.~\cite{GW231123_paper} reports in its text on indications that the post-peak signal of GW231123 is better fit by at least two long-lived QNMs as opposed to one. Here we verify these claims with a more complete analysis of the ringdown, using the \textsc{ringdown} code~\cite{Isi:2021iql_analyzingbhringdowns, Siegel:2024jqd, ringdowncode} to fit sums of damped sinusoids. Our results are reported with reference to the strain peak time and other parameter estimates given by the NRSur7dq4 IMR model. We show that multi-mode fits are preferred over single-mode fits by statistical goodness of fit metrics over a range of fit start times as late as 20.4 ms after the peak strain (or equivalently 14~$t_M$ in units of ${t_M = G M_f^d/c^3}$, corresponding to the detector-frame remnant mass $M_f^d$; see Sec.~\ref{sec:notation_conventions}).

We find that two long-lived QNMs are required to get remnant mass and spin measurements in agreement with NRSur7dq4 over almost all times in the signal where QNM fits recover non-zero amplitudes: this mass and spin agreement is achieved by labeling the two long-lived QNMs as the ${(\ell,m)=(2,2)}$ and ${(2,0)}$ prograde fundamental modes. The amplitudes of the two long-lived modes are comparable. These comparable amplitudes could be attributable to spin-orbit misalignment~\cite{Zhu:2023fnf, OShaughnessy:2012iol, Hamilton:2023znn, Nobili:2025ydt} and a preferential viewing angle to the source: but such a configuration would be at least somewhat finely-tuned. Due in part to the sparseness of highly spinning and precessing numerical relativity (NR) BBH simulations, it is hard at present to conclusively determine how astrophysically unlikely it is for the ${(2,0)}$ amplitude to be so large.

It is also possible that systematic waveform errors alter the mass and spin found by analysis with the NRSur7dq4 waveform, and therefore the QNM labeling required to match these parameters does not accurately represent the physical modes of the system associated with GW231123. Systematics are more of a concern in GW231123 than in most other gravitational wave signals: in Ref.~\cite{GW231123_paper}, five different IMR waveforms~\cite{Estelles:2021gvs, Thompson:2023ase, Ramos-Buades:2023ehm, Varma:2019csw_NRsurpaper, Pratten:2020ceb, Colleoni:2024knd} were found to all have significant systematic differences in parameter estimation for GW231123. High precession and heavy masses are known to be sources of systematic error in all current IMR waveforms~\cite{MacUilliam:2024oif, Dhani:2024jja, Varma:2019csw_NRsurpaper, GW231123_paper}, but we cannot definitively rule out nonstationary noise features or unmodeled physics contributing to the systematic differences between waveforms. Based on the mismatches to NR simulations shown in Refs.~\cite{GW231123_paper, Varma:2019csw_NRsurpaper}, NRSur7dq4 most likely outperforms all other available IMR models when fitting GW231123, and NRSur7dq4 is also not itself guaranteed to have systematic errors dominate over its statistical errors in this part of parameter space. For this reason we only explicitly show and make use of NRSur7dq4 parameter estimates, although the findings of this paper broadly hold regardless of which IMR waveform is considered.

To add to concerns of systematic errors, we find some evidence to suggest that GW231123 may actually be better fit by the pattern of the Kerr ${(\ell,m)=(2,2)}$ and ${(2,1)}$ prograde fundamental QNMs, although this model is in tension with the remnant mass and spin estimates of NRSur7dq4 and all other currently available IMR models. The parameter estimates of SEOBNRv5PHM and IMRPhenomTPHM are the closest to these QNM fits, but are still in significant tension. If a case can be more confidently made for the presence of systematic error in even the best-fitting IMR model, this could change the astrophysical interpretation of GW231123, with implications for population and binary formation channel studies using this signal~\cite{Mandel:2025qnh, Kiroglu:2025vqy, Tong:2025wpz, Paiella:2025qld, Li:2025pyo, Baumgarte:2025syh, Popa:2025dpz, Gottlieb:2025ugy, Delfavero:2025lup, Croon:2025gol, Bartos:2025pkv}. However, more work is still required to make definitive claims regarding NRSur7dq4 systematics for this signal. It is worth noting that GW190521~\cite{LIGOScientific:2020iuh, LIGOScientific:2020ufj, KAGRA:2021vkt}, another signal from a heavy and possibly highly precessing BBH, may also show preference for the ${(\ell,m)=(2,2)}$ and ${(2,1)}$ QNMs as the dominant modes~\cite{Siegel:2023lxl}. And it may be the case that precessing systems are more capable of exciting the ${(\ell,m)=(2,2)}$ and ${(2,1)}$ modes as the dominant QNMs as opposed to the ${(\ell,m)=(2,2)}$ and ${(2,0)}$~\cite{Zhu:2023fnf}, meaning the QNM models which are inconsistent with the IMR models for GW231123 may also be more physically plausible.

In addition to measurements of the two fundamental modes, we also find that the inclusion of an ${(\ell,m,n)=(2,2,1)}$ prograde overtone alongside both fundamental modes may help improve fits by making their parameter estimates and agreement with the Kerr spectrum more consistent over a wider range of fitting times.

In principle, the clear preference for multimodal fits to GW231123 enables tests of the Kerr metric. In practice, the possibility of IMR model systematic error or data quality issues makes interpretation of these tests more difficult. Nonetheless, taking the QNM model which best agrees with the Kerr metric and comparing it at $6~t_M$ to Kerr constraints reported at an equivalent time for GW250114~\cite{GW250114_paper1, GW250114_paper2}, the best-measured signal to date, we find significantly better constraints: at this fitting time we validate the Kerr frequency and damping rate spectrum of the $\left( \ell, m \right) = (2,1)$ and $(2,2)$ fundamental modes to within $\pm10\%$ at the $90\%$ credible level with GW231123, as opposed to the $\pm30\%$ constraint reported for GW250114. When fitting at earlier times, we find even tighter validations of the Kerr metric which are sub-10\% for a model with fundamental and overtone modes.

A key advantage of QNM fits is that their systematics are different from those of the IMR models. The IMR models are built to have rigidly prescribed relationships between inspiral and ringdown~\cite{Estelles:2021gvs, Thompson:2023ase, Ramos-Buades:2023ehm, Varma:2019csw_NRsurpaper, Pratten:2020ceb}, as well as built-in prescriptions of the possible ringdown mode excitations which are generally phenomenological or derived from fits to the available bank of NR simulations, which may not necessarily span the parameter space of interest \cite{Scheel:2025jct_SXScatalog, GW231123_paper}: by contrast, our QNM models have the ability to flexibly fit QNM amplitudes and phases without restriction to any specific physical system. So long as the signal is dominated by damped sinusoids with frequencies from the Kerr spectrum (and even some non-Kerr spectra) our QNM fits should be able to reliably fit all of the signal content. Thus, comparison between the QNM fits we show here and IMR fits can be interpreted as a sanity check of the IMR parameter estimates, especially under the default hypothesis that the signal is from a quasicircular precessing BBH. The tensions we find between QNM and IMR fits highlight the utility of QNM fits as probes of astrophysical parameters in addition to providing tests of general relativity.

The paper is organized as follows. In Sec.~\ref{sec:notation_conventions}, we explain our notational conventions for referring to QNMs and times in the signal. In Sec.~\ref{sec:conditioning} we provide a technical description of the process by which we condition the data before analysis. We also address possible concerns of data quality issues affecting GW231123. In Sec.~\ref{sec:results} we report fitting results from a data-driven perspective, for both Kerr and non-Kerr models. In Sec.~\ref{sec:discussion} we provide discussion and interpretation of the results, and comment on possible origins of systematic error in analysis of GW231123 including NRSur7dq4 and QNM fitting errors. We conclude in Sec.~\ref{sec:conclusion}. We also include a discussion in Appendix~\ref{sec:Appendix_320fits} of QNM fits using the 320 and 321 modes. These models which include the 320 can achieve better goodness-of-fit than the QNM models with the 200 that agree with NRSur7dq4, but they also have seemingly unphysical features which make us disfavor them when compared with the other models in the main text.

\subsection{\label{sec:notation_conventions}Notational conventions}

In this paper, when referring to individual QNMs we follow the conventions of Ref.~\cite{Isi:2021iql_analyzingbhringdowns}. To first order in perturbation theory, individual QNMs can be identified by four indices $(p,\ell,m,n)$. The angular structure of the QNMs is described by spin-weighted spheroidal harmonics with angular indices $\ell$ and $m$. The radial structure of the QNMs is denoted by the index $n$, and is also tied to the lifetime of the QNMs: the longest-lived, ${n=0}$, QNMs are referred to as fundamental QNMs; and faster-decaying, ${n>0}$, QNMs are called overtones. For a given set of $(\ell,m,n)$, when ${m\neq0}$, the sign of $m$ holds the two polarization degrees of freedom; there are also two distinct QNMs which are labeled by an index $p\equiv\mathrm{sgn}[m\,\mathfrak{Re}\left(\omega\right)]$, where $\mathfrak{Re}\left(\omega\right)$ is the real part of the complex QNM frequency $\omega$, and whose phase fronts are either corotating (${p=+}$) or counterrotating (${p=-}$) with the black hole. Solutions with ${m=0}$ are azimuthally symmetric, so that there is no notion of co- vs counter-rotating fronts and the two possible signs of ${\mathfrak{Re}\left(\omega\right)}$ directly encode the two polarization degrees of freedom.

We will frequently make use of the ringdown evolution timescale ${t_M=GM_f^d/c^3}$, which is defined in units of the final detector-frame remnant mass $M_f^d$ when natural units are taken such that ${G=c=1}$. We will use the median detector-frame remnant mass ${M_f^d = 296 ~M_\odot}$ inferred by NRSur7dq4 in order to set ${t_M = 1.46}$~ms. QNM models will be denoted with set notation as comma-separated lists within braces of all simultaneously fitted modes identified by their $\{\ell,m,n\}$ indices; we only consider prograde QNMs in this work, and so ${p=+}$ implicitly throughout. Any QNMs given non-Kerr freedoms in a model will be labeled with underlined text. The time at which a given fit is started will be referred to as $t_\text{start}$, in units of $t_M$ relative to our estimated peak strain time $t_{\rm peak}$ which is given explicitly in Sec.~\ref{sec:conditioning}.

\section{\label{sec:conditioning} Data conditioning and quality}
For all results herein our analysis follows a specific set of steps to condition the data we analyze. These steps are carefully chosen to minimize computational expense, without significantly corrupting parameter estimates. The entire conditioning process is outlined below, and a brief consideration of possible data quality issues is also included.

We start from 4096~s long data segments around the time of the detection in both the Hanford and Livingston interferometers, at sample rates of 16384~Hz. From these data segments, we subtract the 60 Hz AC mains noise line from the data~\cite{Siegel:2024jqd, LIGOScientificNOISE:2019hgc}, a line which overlaps with the central frequencies of the signal. To implement our line subtraction, we use a computationally inexpensive linear algorithm~\cite{LineCleaner_code}. Because the subtraction involves some filter ``warmup'' and tapering, this leaves us with a 4080 s long valid data segment. We then downsample this data segment by a factor of 4 to a sample rate of 4096~Hz using our so-called digital filter. When downsampling, a Tukey window is also applied to the 4080 s data segment, which trims the first and last 10\% of the data. From this downsampled data we select segments for analysis with durations of 0.2 s as motivated by Ref.~\cite{Siegel:2024jqd}. We can analyze relatively short 0.2~s duration segments without losing SNR because of the subtraction of the 60 Hz AC mains line.

Line subtraction for 60 Hz noise was not implemented in production data for GW231123, because the LIGO collaboration's standard nonlinear line subtraction pipeline was found to elevate noise in the sidebands of the line and this might negatively affect continuous gravitational wave searches. Our particular ringdown analysis experiences significant computational and SNR gains when employing line subtraction, and does not suffer from the noise subtraction in the way that other analyses like the continuous wave searches do, especially since our analysis segment is so short that we do not resolve the sidebands as well as those other longer-duration analyses. Our line subtraction algorithm was tested through injecting damped sinusoid signals into real LIGO data from 1000 to 500 seconds before the actual GW231123 signal, and was found to improve the performance of our analysis~\cite{LineCleanerReviewGitlab, LineCleanerInjectionsDCC} (note that these references are to internal LVK documents).

The noise auto-covariance function (ACF) is computed by Fourier transforming a Welch estimate of the PSD. This Welch estimate uses windowed data segments 16 times the duration of the analysis data segment (i.e. 3.2~s), and is computed over the full 4080~s long data segment. While it has been argued previously in Ref.~\cite{Zackay:2019kkv} that long data segments of LIGO noise are subject to PSD estimate drifts, we have not found this empirically to significantly impact our analysis at current SNRs (see e.g. Ref.~\cite{LineCleanerInjectionsDCC}): this may be due to the short duration and higher SNR of our signals of interest.

The geocenter peak strain GPS time in our analysis is ${t_{\rm peak}=1384782888.61823\pm 0.00131~\text{s}}$, with the median value being used throughout. The sky location of our model determines time delays between the data segments of different interferometers, and is selected by using the sample from the NRSur7dq4 posterior which is closest in time to the median estimated peak strain of the most sensitive interferometer. The sky location in radians for our analysis is fixed at a point estimate of ${(\text{ra},~\text{dec},~\text{psi})=(3.80,~0.64,~1.81)}$, when rounded to two decimal places. Note that our $t_{\rm peak}$ is not the same as the time used in Ref.~\cite{GW231123_paper}. In that LVK analysis (and also in another GW231123 ringdown analysis, Ref.~\cite{Wang:2025rvn}), the peak time is estimated using the peak of the complex strain at Earth's location on the celestial sphere, 
\begin{equation}
    {t_{\rm peak}^{\rm LVK}=\arg\max \big[ h_+ ^2(t) + h_\times ^2(t) \big]}.
\end{equation}
This time $t_{\rm peak}^{\rm LVK}$ is not the one typically considered in theoretical studies of ringdown, which more often consider instead an angle-invariant peak strain time over the whole celestial sphere, 
\begin{equation}
{t_{\rm peak}^{\rm invar.}=\arg\max \sum_{\ell m}\,H_{\ell m}(t)^2},
\end{equation}
where $H_{\ell m}(t)$ is related to the complex strain as 
\begin{equation}
    {h_+(t) + ih_\times(t) = \sum_{\ell m} \, _{-2}Y_{\ell m} \, H_{\ell m}(t)},
\end{equation}
and $_{-2}Y_{\ell m}$ are spin-weighted spherical harmonics. Our peak time estimate is obtained by considering the distribution of $t_{\rm peak}^{\rm invar.}$ from 1000 random NRSur7dq4 posterior samples.

There were data quality issues throughout the day of the GW231123 detection~\cite{GW231123_dataquality}, beyond the non-Gaussian noise features deemed to be inconsequential in Ref.~\cite{GW231123_paper}. These additional concerns included notable fluctuations in the binary neutron star observation range throughout the day. The aforementioned injection study which tested our line cleaning algorithm appeared by eye to reliably recover the true parameters of the injected signals in data from this day, suggesting that our analysis is not sensitive to these data quality issues.

See Refs.~\cite{Isi:2021iql_analyzingbhringdowns, Siegel:2024jqd} for more information on our ringdown analysis framework, as well as the data release of this paper~\cite{datarelease_thispaper} for exact analysis settings in configuration files.

\section{\label{sec:results}Results}
Here we report on data analysis results, with a focus on data-driven interpretation and minimal reliance on existing understanding of astrophysical QNM amplitudes in BBH coalescences. In Sec.~\ref{sec:discussion} we then present stricter physical interpretation of these results ex post facto. Given that there is incomplete theoretical knowledge of the regime of validity of QNM fits in BBH coalescences~\cite{Baibhav:2023clw, Berti:2025hly, Giesler:2024hcr, Mitman:2025hgy}, we fit our damped sinusoid models starting over a wide range of times around the NRSur7dq4 $t_{\rm peak}$, both before and after where most works claim the onset of the Kerr perturbative regime to be.  We choose fit start times $t_\text{start}$ in intervals of $2~t_M$ around the estimated peak strain time $t_{\rm peak}$.

A key tenet of our fitting philosophy is that reasonable QNM models which accurately describe the signal should have consistent parameter estimates when fit at different starting times. We expect that good QNM fits to the data should become self-consistent at some time and remain that way until the SNR decreases to the point of the posteriors being uninformative. We assess this self-consistency predominantly visually, by looking for overlapping posteriors from fits at different times, typically considering $90\%$ credible levels (CLs). We also look for amplitude posteriors for individual QNMs that are consistent with being non-zero within at least the $68\%$ CL, as zero amplitude leads to uninformative posteriors for other associated QNM parameters.

When performing model comparison, we choose to use the leave-one-out cross-validation (LOO) \cite{LOO_Paper, LOO_FAQ}.\footnote{A technical detail regarding the LOO: we specifically use the Pareto-smoothed importance sampling estimate of the LOO as defined in Eq.~10 of Ref.~\cite{LOO_Paper}. The shape parameters we recover for the Pareto distribution suggest that we have reliable estimates of the expected log pointwise predictive density for all fitting times where significant model preferences exist.}
For intuition regarding the meaning of the LOO: when using Gaussian likelihood models like ours~\cite{Isi:2021iql_analyzingbhringdowns, Siegel:2024jqd}, the LOO is approximately related to the chi-squared test as ${{\rm LOO} \approx -\frac{1}{2}\chi^2}$ plus a penalty term which depends on the leverage of individual data points~\cite{gelman2013bayesian, hastie2009elements}. Higher LOO values are more preferred. Following Ref.~\cite{mclatchie2024efficient}, LOO differences greater than 4 are very significant. We deem LOO differences of 1 to be our minimum for noteworthy statistical significance. We choose to use the LOO as opposed to the Bayes factor, which is more common in our research field, because the LOO is in general less sensitive to priors and we prefer this behavior. Regardless of which statistical model selection or goodness of fit criterion is chosen, such quantitative measures only form one part of a larger whole in scientific analysis: we use the LOO here merely as a guide in driving our model explorations and supporting our interpretations and conclusions, rather than as an overwhelming figure of merit~\cite{navarro_2018_devildeepbluesea, Cutler:2007mi_PEbias, Cornish:2011ys_stealthbias}.

The first fitting results are presented in Sec.~\ref{sec:results_freedampedsinusoids}, where we consider a general model made up of damped sinusoids that do not have Kerr frequency and damping rate constraints but are forced to be ordered in real frequency to avoid label switching \cite{Buscicchio:2019rir}. This type of ``free'' or ``agnostic'' damped sinusoid model is helpful for developing basic exploratory understanding, before honing in on models with physically motivated frequency spectrum constraints. This free model is also closely related to models we use later which allow deviations from the Kerr spectrum of frequencies and damping rates \cite{Isi:2021iql_analyzingbhringdowns, Gossan:2011ha}.

The free damped sinusoid model has strong statistical preferences for two modes over one, when fitting at start times ranging from before the peak strain to 14~$t_M$, as shown in Fig.~\ref{fig:LOOfig}. The statistical uncertainty of the peak time as given by NRSurd7dq4 is $\pm0.9~ t_M$. Even when accounting for the statistical uncertainty of the peak time, preferences for multimodal fits persist long after the peak strain. If systematic errors in the peak time estimate dominate over the statistical uncertainty, preferences for multimodal fits may correspond to relative fitting times earlier or even later than those we are using as reference. When fitting two free damped sinusoids, the modes are found to have comparable lifetimes and comparable amplitudes. They have frequencies consistent with the prograde ${(\ell,m) = (2,2)}$ and either the $(2,1)$ or $(2,0)$ fundamental modes implied by the mass and spin measurements of NRSur7dq4 (Fig.~\ref{fig:damped_sinusoid_fgamma}).

We then in Sec.~\ref{sec:Kerr_models} fit damped sinusoid models with frequency constraints imposed by first-order Kerr metric perturbation theory. When fitting two-mode Kerr models, we find the best remnant mass and spin agreement with NRSur7dq4 by including the ${(\ell,m) = (2,2)}$ and $(2,0)$ fundamental QNMs. The posteriors of this model overlap visually at the $90~\%$ CL with those of NRSur7dq4 when starting QNM fits from 4~$t_M$ onward (Fig.~\ref{fig:22_20_fgamma}). However, we do find a goodness of fit preference among two-mode models of at least $1\sigma$ for the \{220,~210\} model, until as late as $8~t_M$. This model is also self-consistent over time in both its frequency and damping rate (Fig.~\ref{fig:22_20_fgamma}) and amplitude measurements (Fig.~\ref{fig:amplitude_timescan_fundamental}), although it is in tension with NRSur7dq4.

We also consider three-mode Kerr models. When fitting a \{220,~200,~2$m$1\} model, the remnant mass and spin posteriors overlap at the $90\%$ CL with those of NRSur7dq4 starting as early as $0~t_M$. (Fig.~\ref{fig:22_20_221_fgamma}). Even more strikingly, \{220,~210,~2$m$1\} models make self-consistent remnant mass and spin parameter estimates going back as early at least $-8~t_M$, although these models are in tension with NRSur7dq4 remnant mass and spin estimates. The $m$ index when we just referred to the overtones in these models was left intentionally ambiguous, as any $m$ index of the included overtone in the model will achieve qualitatively similar mass and spin posteriors. However, in terms of goodness of fit, there are goodness of fit preferences at some times for ${m=2}$ over other overtone $m$ indices: restricting ourselves to ${m=2}$ for the overtone, at $-2~t_M$ the \{220,~210,~221\} model is preferred by $\sim 1 \sigma$ over all other models we consider, and at earlier times this model is definitively preferred.

Finally, in Sec.~\ref{sec:TGR_models} we perform a test of general relativity using the Kerr models most preferred by statistical goodness of fit and/or in best agreement with NRSur7dq4. Based on this test, it seems that models with the 210 as opposed to the 200 are consistent with the Kerr metric for GW231123 over a broader range of fitting times, and are capable of recovering the Kerr metric with higher precision at the 90\% CL than 200 models. Comparing with similar fitting times in GW250114~\cite{GW250114_paper1, GW250114_paper2}, the loudest signal to date, we find much better constraints and confirm the Kerr spectrum to $\pm10\%$ using the 210 QNM, and $\pm 40\%$ for the 221 QNM when fit alongside the 210. We report even more precise validation of Kerr when fitting at earlier times. This result, along with some of our observations of goodness of fit and parameter estimate consistency over time for different models, raises questions about whether the NRSur7dq4 model is possibly dominated by systematic error for GW231123, since our analysis seems to have preferences for QNM fits which are in tension with NRSur7dq4. See Table~\ref{tab:TGR-deltas} for a collection of select Kerr constraints.

\subsection{\label{sec:results_freedampedsinusoids}Free damped sinusoid fits}
We fit a model composed of free damped sinusoids that are not constrained by the Kerr frequency spectrum but are forced to be ordered in real frequency in order to avoid degenerate sampling. Each damped sinusoid has two polarizations, and flat priors on the frequency (0 to 150 Hz), damping rate (0 to 250 Hz), phase (0 to 2$\pi$ radians), and amplitude (sampled via the marginalization technique described in Ref.~\cite{amplitude_prior_note}, with ${a_\text{max}=10^{-19}}$).\footnote{The flat prior on the amplitude of each mode is placed on the combined amplitude of both polarizations of the mode, not on each individual polarization's amplitude. See Fig.~3 of Ref.~\cite{Siegel:2023lxl} and Fig.~12 of~\cite{Isi:2022mbx} for discussion of this distinction.} The damping rate and frequency priors are chosen to encompass both the first overtone and ${\ell=3}$ mode parameters suggested by mass and spin measurements of NRSur7dq4.

\begin{figure}[htbp!]
    \centering
    \includegraphics[width=\columnwidth]{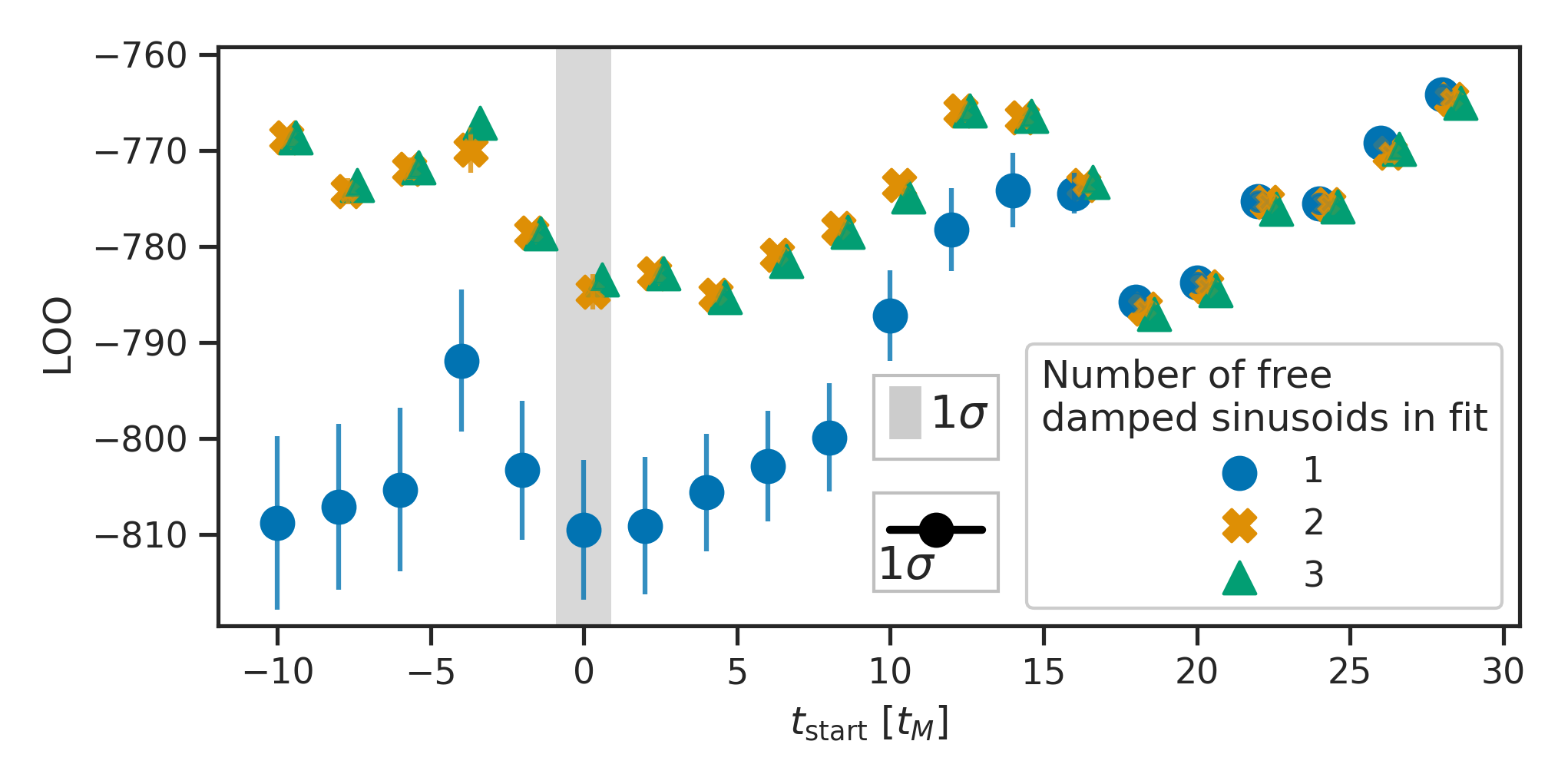}    \includegraphics[width=\columnwidth]{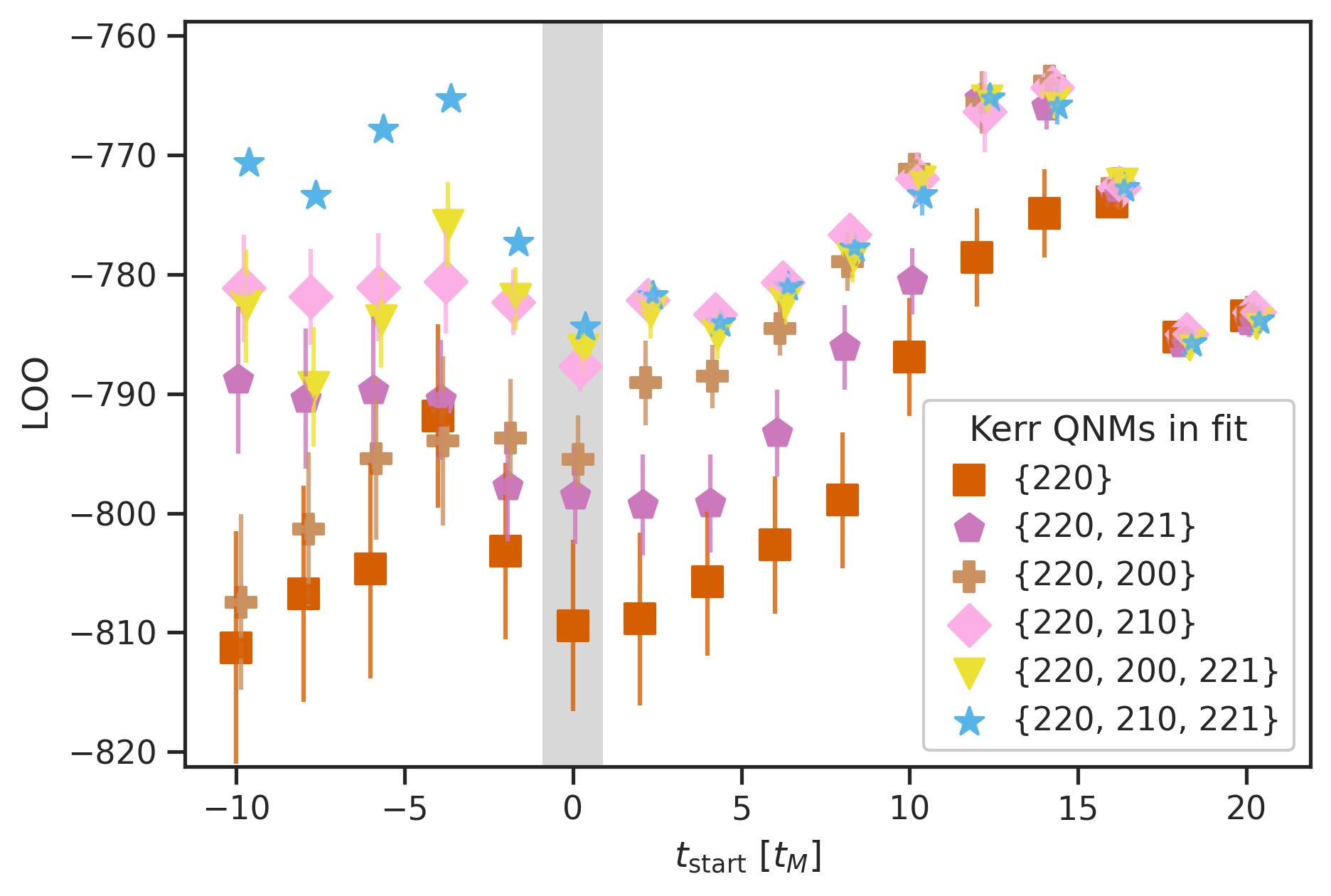}
    \caption{To quantify statistical goodness of fit, we use the Leave-One-Out Cross-Validation (LOO)~\cite{LOO_Paper, LOO_FAQ}, which is approximately related to the chi-squared test as ${{\rm LOO}\approx-\frac{1}{2}\chi^2}$ plus a penalty term which depends on the leverage of individual data points. Following Ref.~\cite{mclatchie2024efficient}, LOO differences greater than 4 are very significant. We deem differences of 1 to be our minimum for noteworthy statistical significance. The error bars indicate differences between LOO of each less-favored model and the top model at a given fit time, and are computed with the \texttt{compare} method in the \textsc{arviz} package \cite{arviz_2019}. Peak strain statistical timing uncertainty from NRSur7dq4 is $\pm 0.9~t_M$, shown as a grey band. \textit{Top:} When fitting free damped sinusoids without Kerr frequency constraints, we find significant preference for two modes over one when the fits start as late as 14~$t_M$, and no preference for three modes over two. \textit{Bottom:} LOO of plausible Kerr models which we found to have the best-constrained non-zero amplitudes. Again, we find significant preference for two modes over one. Within two-mode models, LOO differences in favor of the \{220,~210\} over the \{220,~200\} model have expectation values greater than 1 and statistical significance of at least $1\sigma$ as late as $8~t_M$. When fitting three-mode models with an overtone, there is little LOO improvement if fitting post-peak. However, the \{220,~210,~221\} model is strongly preferred before the peak strain time. This is at odds with the behavior of the free sinusoid models, which had no preference for three-mode fits. The results when replacing the 221 with the 211 or 201 are similar (not shown), although at certain pre-peak times (in particular $6~t_M$) the 221 is still significantly preferred.}
    \label{fig:LOOfig}
\end{figure}

In Fig.~\ref{fig:LOOfig} we show model comparisons using the LOO. There is a statistically significant goodness-of-fit preference for two-mode free damped sinusoid fits over one mode fits when starting the fits as late as 14~$t_M$ after the peak strain. There is no added support for three modes over two modes using these free damped sinusoid fits.

\begin{figure}
    \centering
    \includegraphics[width=\columnwidth]{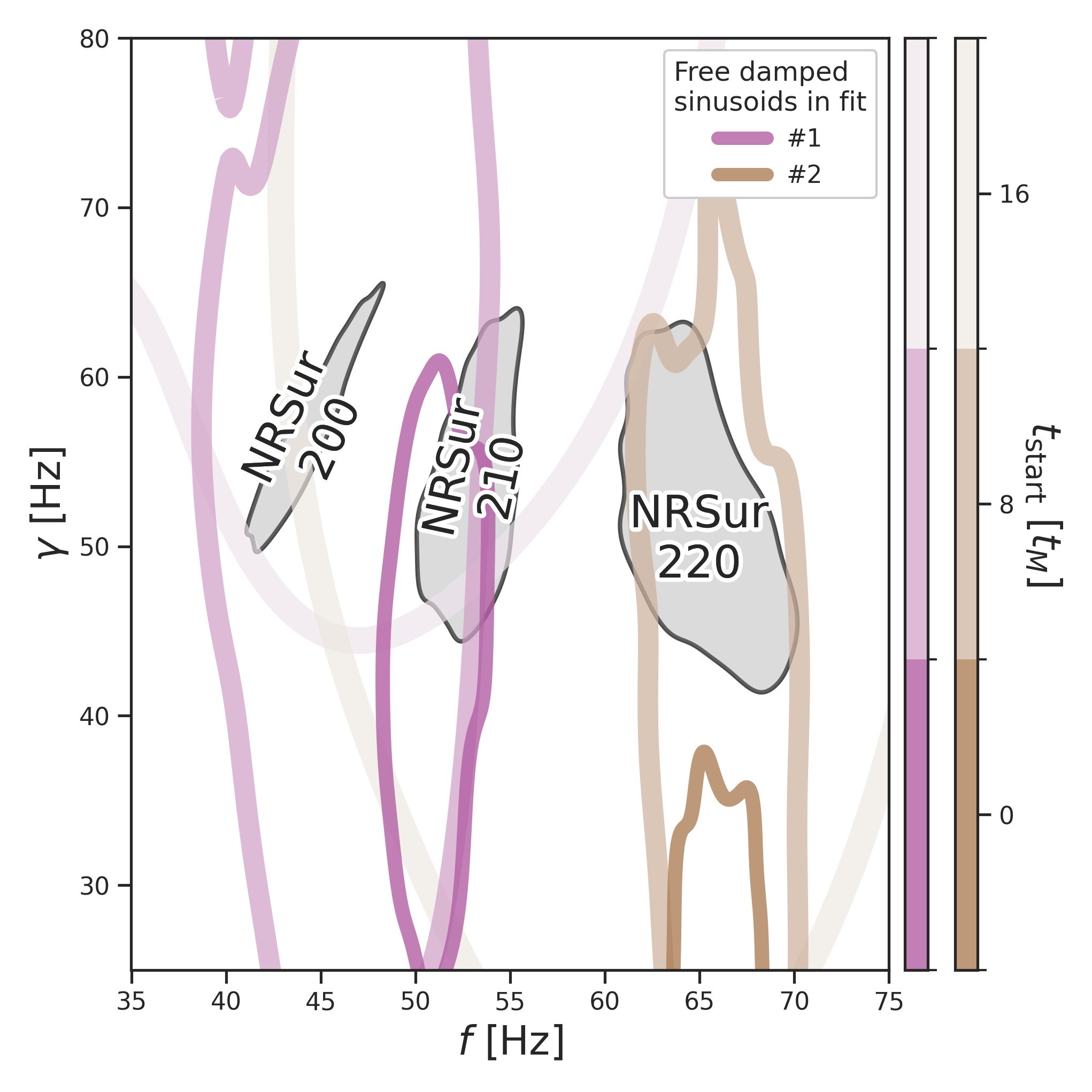}
    \caption{When simultaneously fitting two free damped sinusoids over a wide range of fit start times, the mode frequencies and damping rates remain broadly consistent with the parameters of ${\ell=2}$ fundamental QNMs implied by NRSur7dq4 remnant mass and spin measurements. The frequency and damping rate of modes are shown on the x and y axes respectively. Colors correspond to different fitted sinusoids, and transparency is related to the fit start time as shown in the colorbar. 90\% credible contours shown for all posteriors. The NRSur7dq4 QNM distributions are inferred using the \textsc{qnm} package \cite{qnmpackage_Stein}.}
    \label{fig:damped_sinusoid_fgamma}
\end{figure}

In frequency and damping rate, the two-mode free damped sinusoid fits overlap at the $90\%$ CL with the prograde ${(\ell,m) = (2,2)}$ and either the $(2,1)$ or $(2,0)$ fundamental modes of NRSur7dq4, as shown in Fig.~\ref{fig:damped_sinusoid_fgamma}. Based on this frequency and damping rate behavior as well as the LOO preferences, we are motivated to explore multimodal Kerr QNM fits with at least two prograde fundamental ${\ell=2}$ QNMs, as shown in the next subsection.

Our free damped sinusoid model is not the same as the model used in Ref.~\cite{GW231123_paper}. The model used in the LVK analysis only includes one of the two polarizations, meaning that parameters recovered by that model may not correspond directly to the content of the physical strain. Any free damped sinusoid analysis with the \textsc{pyRing}~\cite{Pyring, Carullo:2019flw, IsiNoHair_GW150914, TestingGR_LIGO_2ndCatalog} code performed for signals from GW231123 and earlier will be subject to this polarization limitation.

The amplitudes of the free damped sinusoids are not shown here. The amplitudes are not as informative as the frequencies and damping rates for this model in our particular analysis. This is because at later fitting times as the SNR decreases, the amplitude uncertainties of our free damped sinusoid fits grow dramatically, likely due to the fact that the priors we use for damping rate and frequency are exceptionally wide.

\subsection{\label{sec:Kerr_models}Kerr QNM fits}
\begin{figure}[htbp!]
    \centering
    \includegraphics[width=\columnwidth]{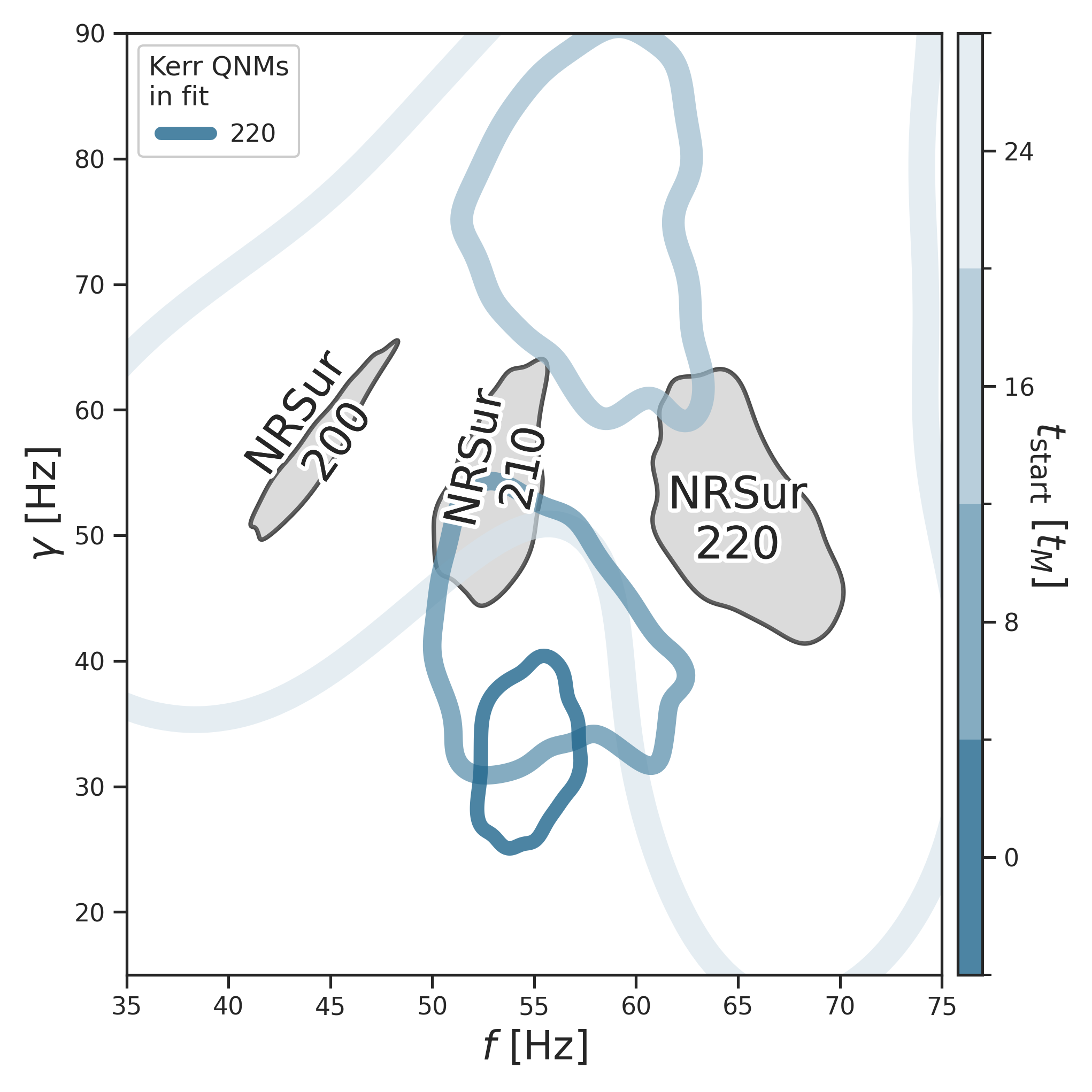}
    \includegraphics[width=\columnwidth]{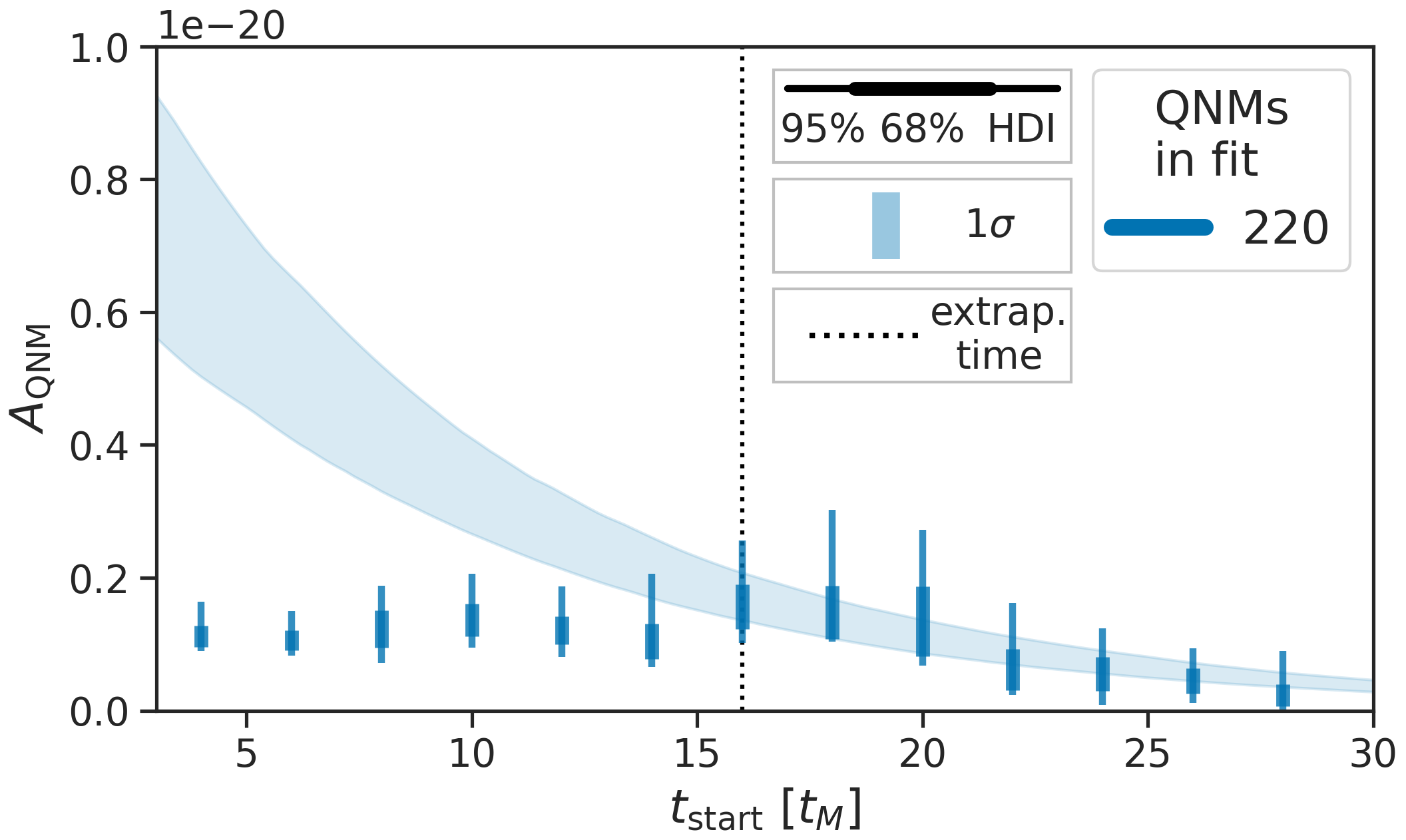}
    \caption{Frequency and damping rate of single-mode Kerr fits, as well as their amplitudes. \textit{Top:}~See Fig.~\ref{fig:damped_sinusoid_fgamma} caption for figure conventions. When fitting a single Kerr mode, consistency with NRSur7dq4 is not found until $22~t_M$, after the point at which single-mode fits are equivalent to multi-mode fits in terms of LOO, as shown in Fig.~\ref{fig:LOOfig}.
    \textit{Bottom:}~QNM amplitudes. Errorbars on each scatterpoint indicate 68\% and 95\% highest density interval (HDI). We extrapolate the amplitude from the fit at 16~$t_M$ (as indicated by dashed line), the earliest time where single-mode fits are equivalent to multi-mode fits in terms of LOO, as shown in Fig.~\ref{fig:LOOfig}. We show the $1\sigma$ uncertainty of the extrapolated exponential decay of the 220 QNM as a colored band. This extrapolation is only consistent with fit times after $16~t_M$. The amplitude of a single-mode Kerr fit is consistent with being non-zero at $2\sigma$ until $28~t_M$.}
    \label{fig:220_fgamma_amplitudetimescan}
\end{figure}

\begin{figure}[htbp!]
    \centering
    \includegraphics[width=\columnwidth]{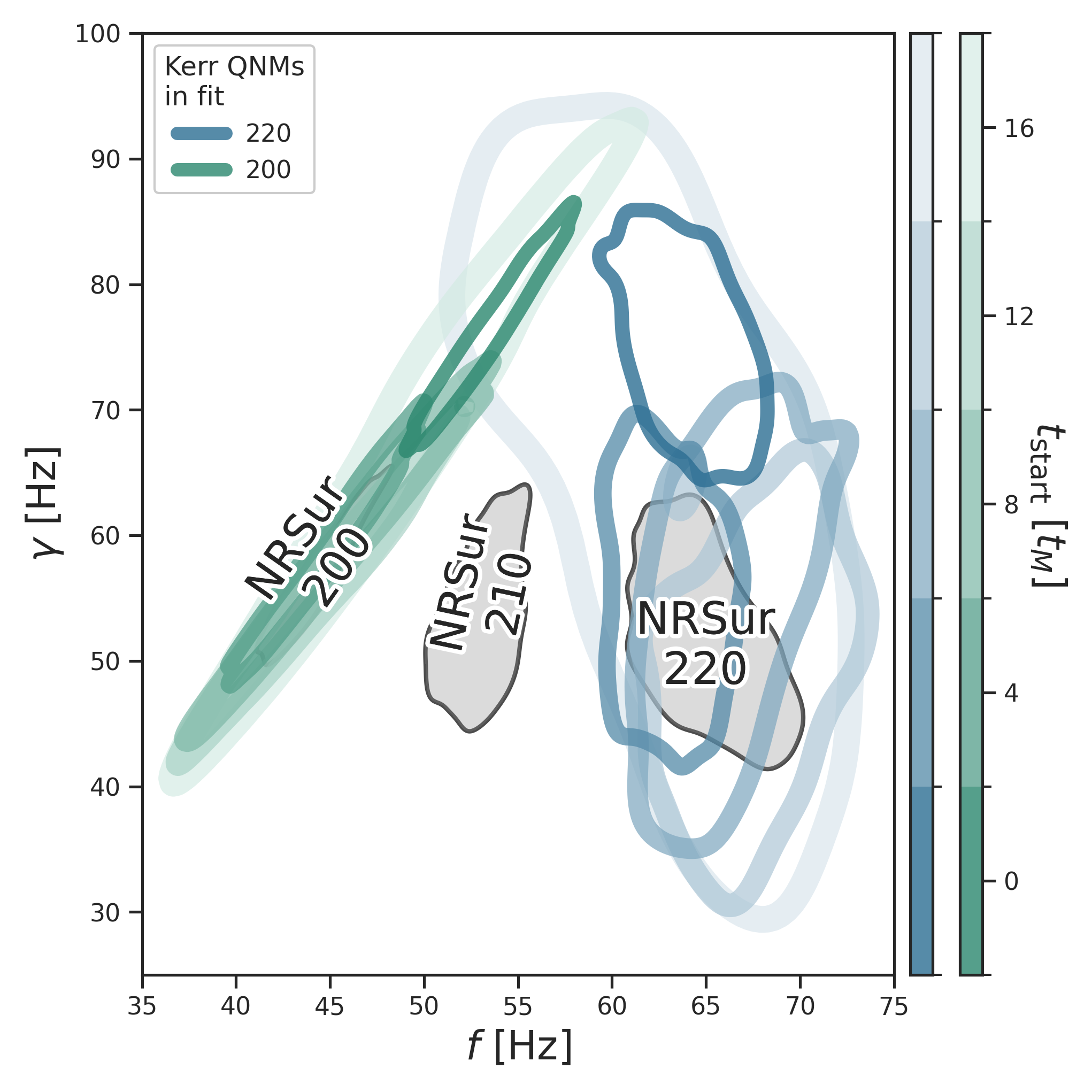}
    \includegraphics[width=\columnwidth]{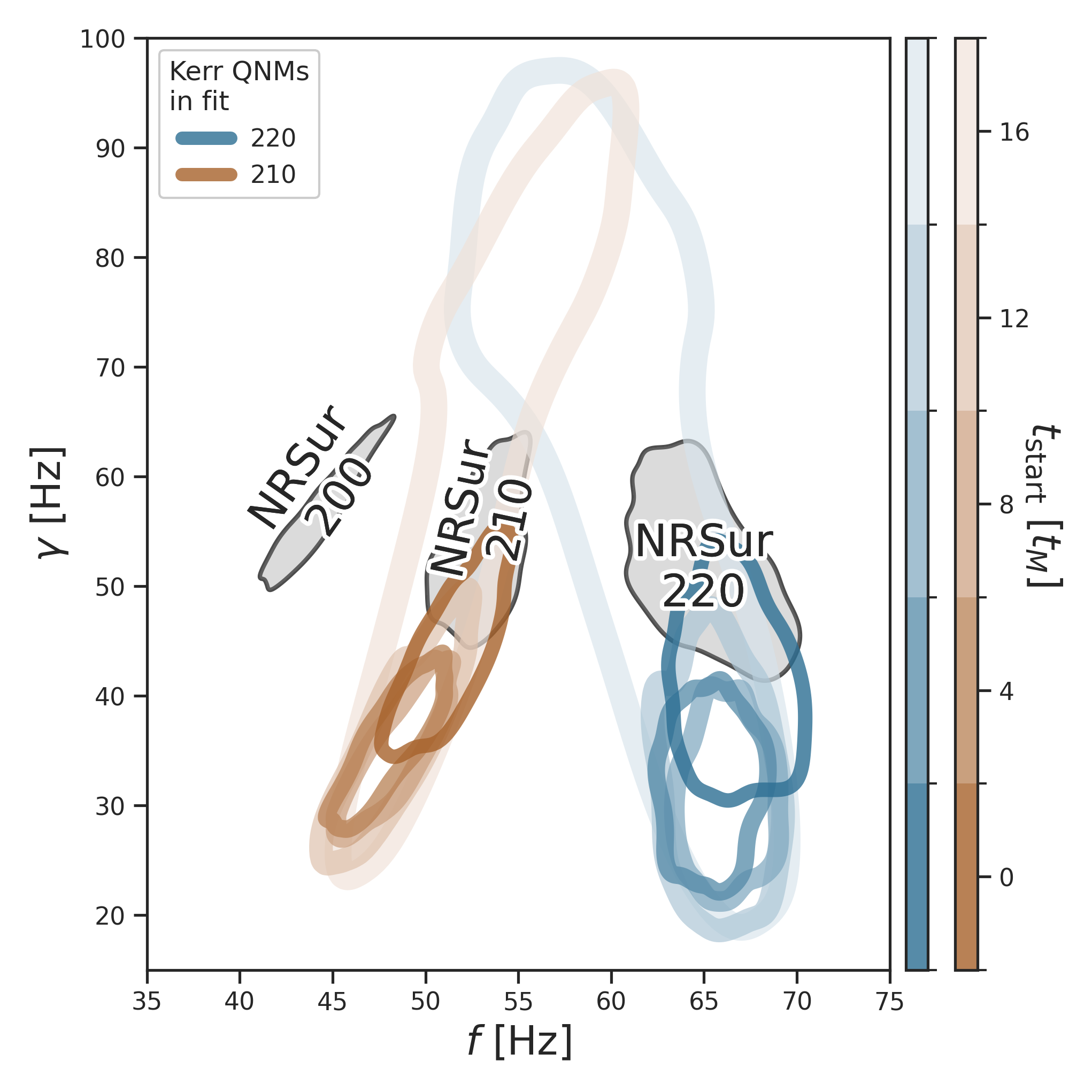}
    \caption{The frequency and damping rate of fits with two fundamental Kerr modes. See Fig.~\ref{fig:damped_sinusoid_fgamma} for figure conventions. \textit{Top:} We find that \{220, 200\} fits are consistent with NRSur7dq4 and also self-consistent over time, starting from 4~$t_M$. We do not find another combination of well-measured fundamental QNMs consistent with NRSur7dq4. \textit{Bottom:} The \{220, 210\} model is similarly self-consistent over time but not in agreement with NRSur7dq4.}
    \label{fig:22_20_fgamma}
\end{figure}

\begin{figure*}
    \centering
    \includegraphics[width=\columnwidth]{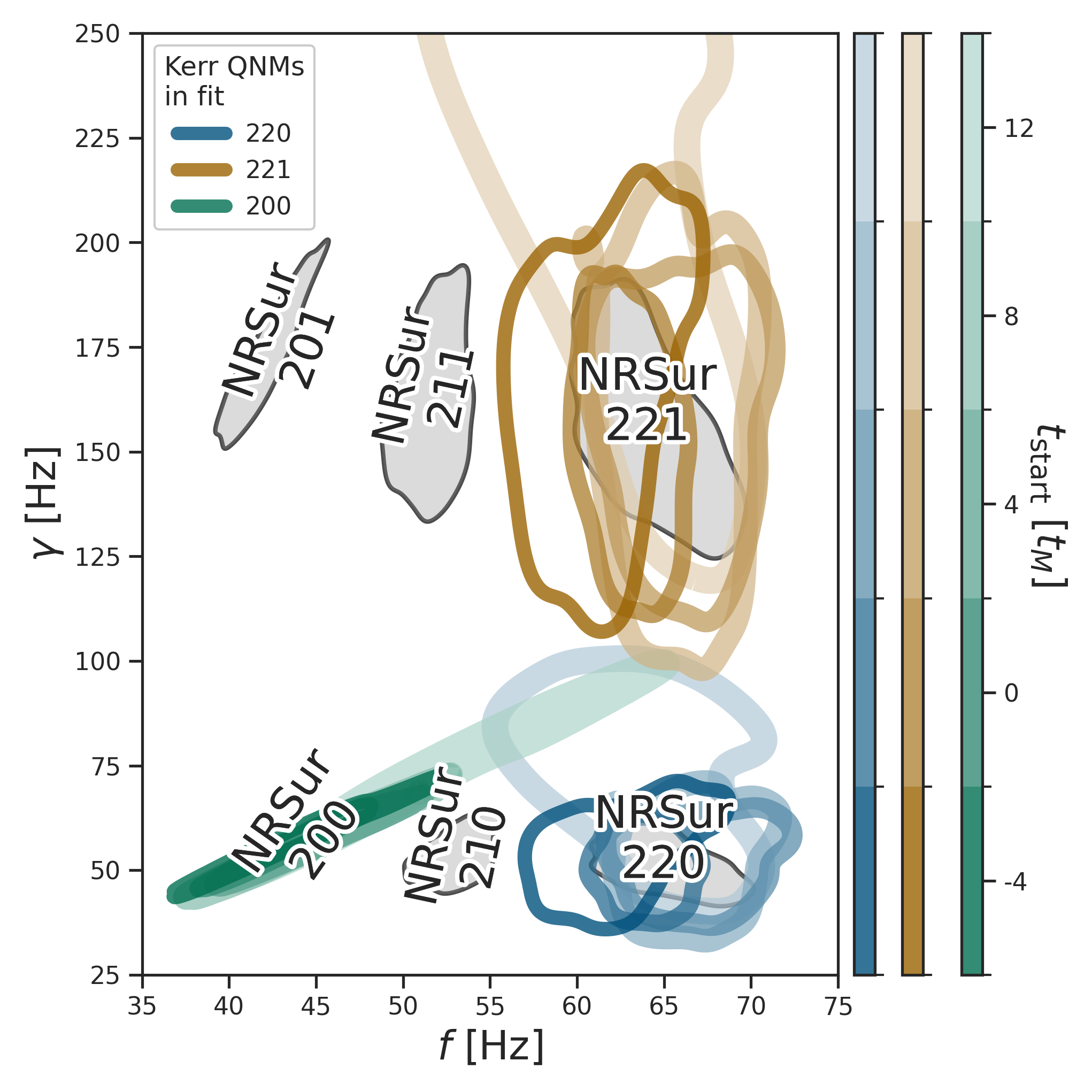}
    \includegraphics[width=\columnwidth]{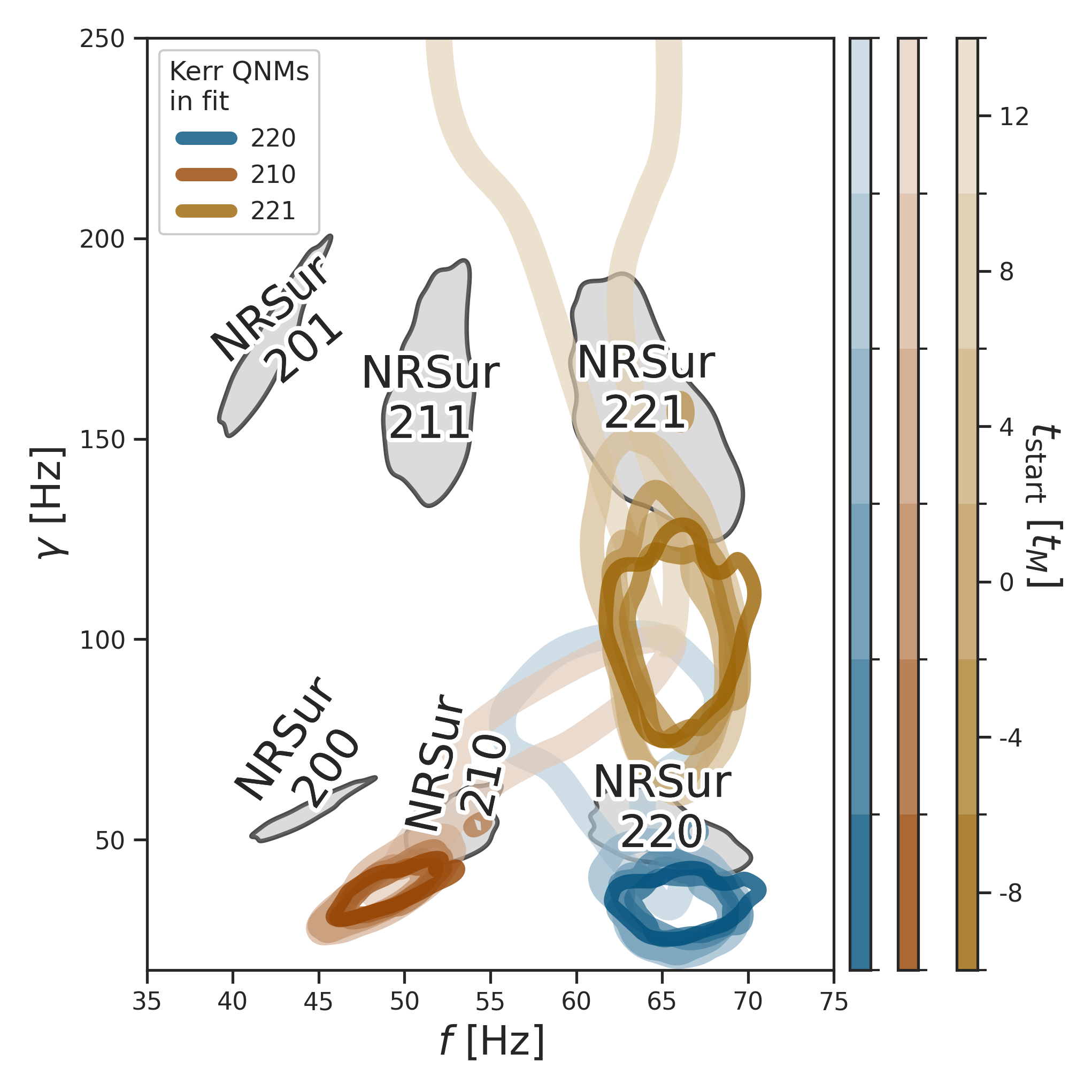}
    \caption{Frequency and damping rate of Kerr models with overtones. See Fig.~\ref{fig:damped_sinusoid_fgamma} caption for figure conventions. \textit{Left:}~We find that \{220, 200, 2$m$1\} fits are consistent with NRSur7dq4 and also self-consistent over time, starting from 0~$t_M$. While the above figure shows only the three-mode model with the 221 as the included overtone, similarly consistent results are obtained with the 211 and 201 QNMS. \textit{Right:}~While \{220, 210, 2$m$1\} fits are inconsistent with NRSur7dq4, the added overtone makes this model self-consistent as early as at least $-8~t_M$ depending on which $m$ index overtone is added.}
    \label{fig:22_20_221_fgamma}
\end{figure*}
We fit a model composed of damped sinusoids constrained to follow the Kerr frequency and damping rate spectrum, limited to QNMs from first-order perturbation theory. We place flat priors on the detector-frame remnant black hole mass (0.5 to 1.5 times the median detector-frame remnant mass inferred by NRSur7dq4, ${M_f^d = 296 ~M_\odot}$) and dimensionless spin (0 to 0.99), mode amplitude (sampled via the marginalization technique described in Ref.~\cite{amplitude_prior_note}, with ${a_\text{max}=10^{-19}}$) and phase (0 to $2\pi$ radians).

As shown in Fig.~\ref{fig:LOOfig}, Kerr models with two QNMs give significantly better fits to the data than a Kerr model with one mode, for the same range of times as in the case of the free damped sinusoid fits. When comparing with NRSur7dq4 mass and spin estimates, a single QNM fit does not agree until late times over 20 $t_M$ after the peak, as shown in Fig.~\ref{fig:220_fgamma_amplitudetimescan}. Also, the expected amplitude decay of these single-mode fits does not arise until at least $16~t_M$, the point at which goodness-of-fit indicates no preference for multimodal fits. The single-mode agreement with NRSur7dq4 comes past the point at which single-mode fits are equivalent in goodness-of-fit to multi-mode fits. Significant non-zero amplitude measurements indicate that a QNM signal persists as late as $28~t_M$. These results are qualitatively similar to those shown in Ref.~\cite{GW231123_paper}. We look for non-zero amplitude constraints because an amplitude of zero leads to uninformative posteriors of other parameters.

For ringdown-dominated signals like GW231123, we expect unbiased QNM and IMR remnant mass and spin posteriors to fully encompass each other, as argued in e.g. Appendix B of Ref.~\cite{Siegel:2023lxl}. As shown in Fig.~\ref{fig:22_20_fgamma}, between 4 and 16 $t_M$ the only two-mode QNM model we find that agrees with the NRSur7dq4 remnant mass and spin is the \{220,~200\}. Another physically plausible model~\cite{Zhu:2023fnf}, \{220,~210\}, has frequency and damping rate posteriors which do not overlap with those of NRSur7dq4 within at least the 90~$\%$ CL, but this model fits the data equally well or slightly better between as late as 8~$t_M$ as shown in Fig.~\ref{fig:LOOfig}. The LOO is greater than 4 in favor of the \{220,~210\} model over the \{220,~200\} model at $0~t_M$ and earlier times, even when taking into account LOO difference uncertainty. At $8~t_M$, the LOO difference between these two models has an expectation of 2 but support for values greater than 4 within statistical uncertainty. At later times, there is no preference between both models.
\begin{figure}
    \centering
    \includegraphics[width=\columnwidth]{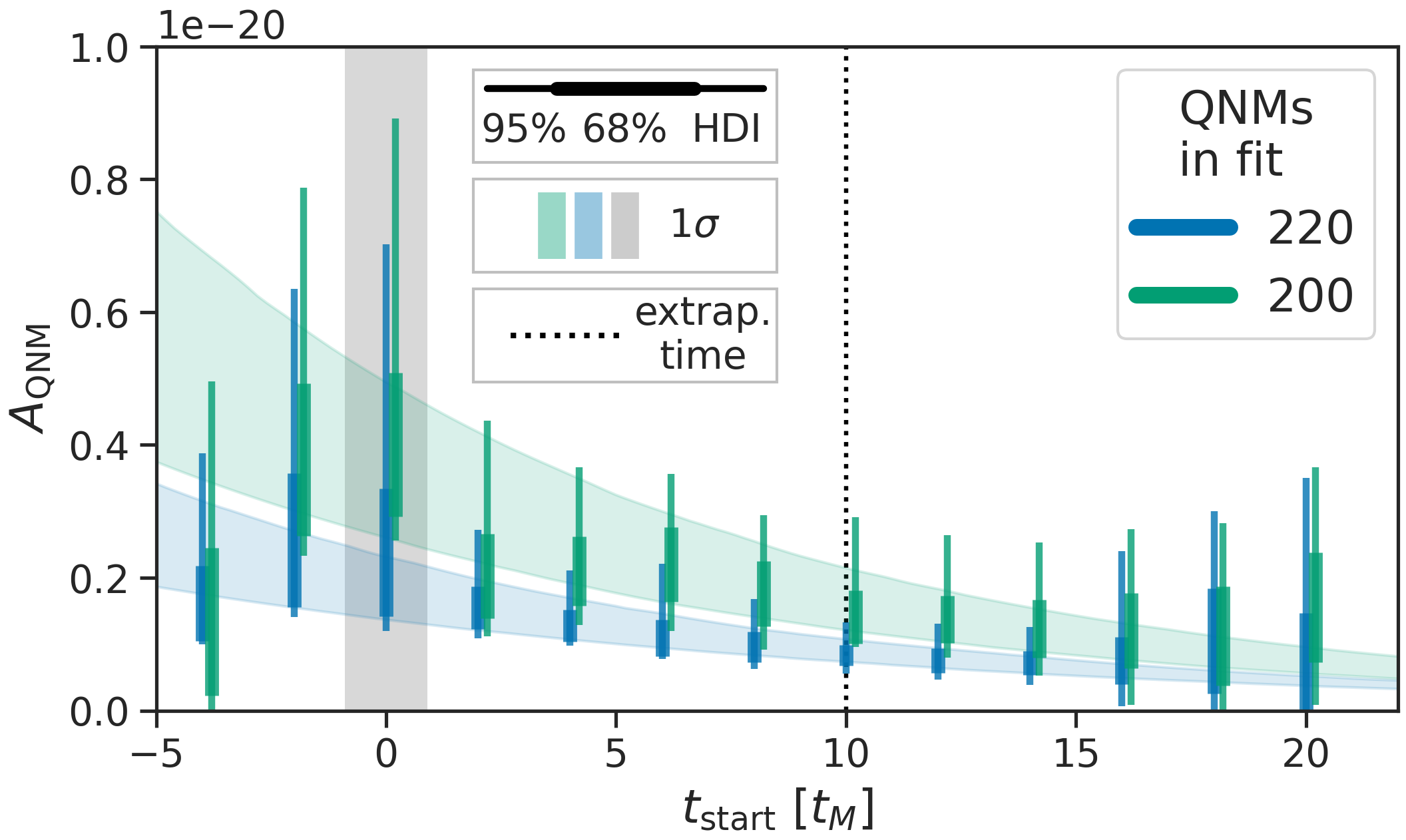}
    \includegraphics[width=\columnwidth]{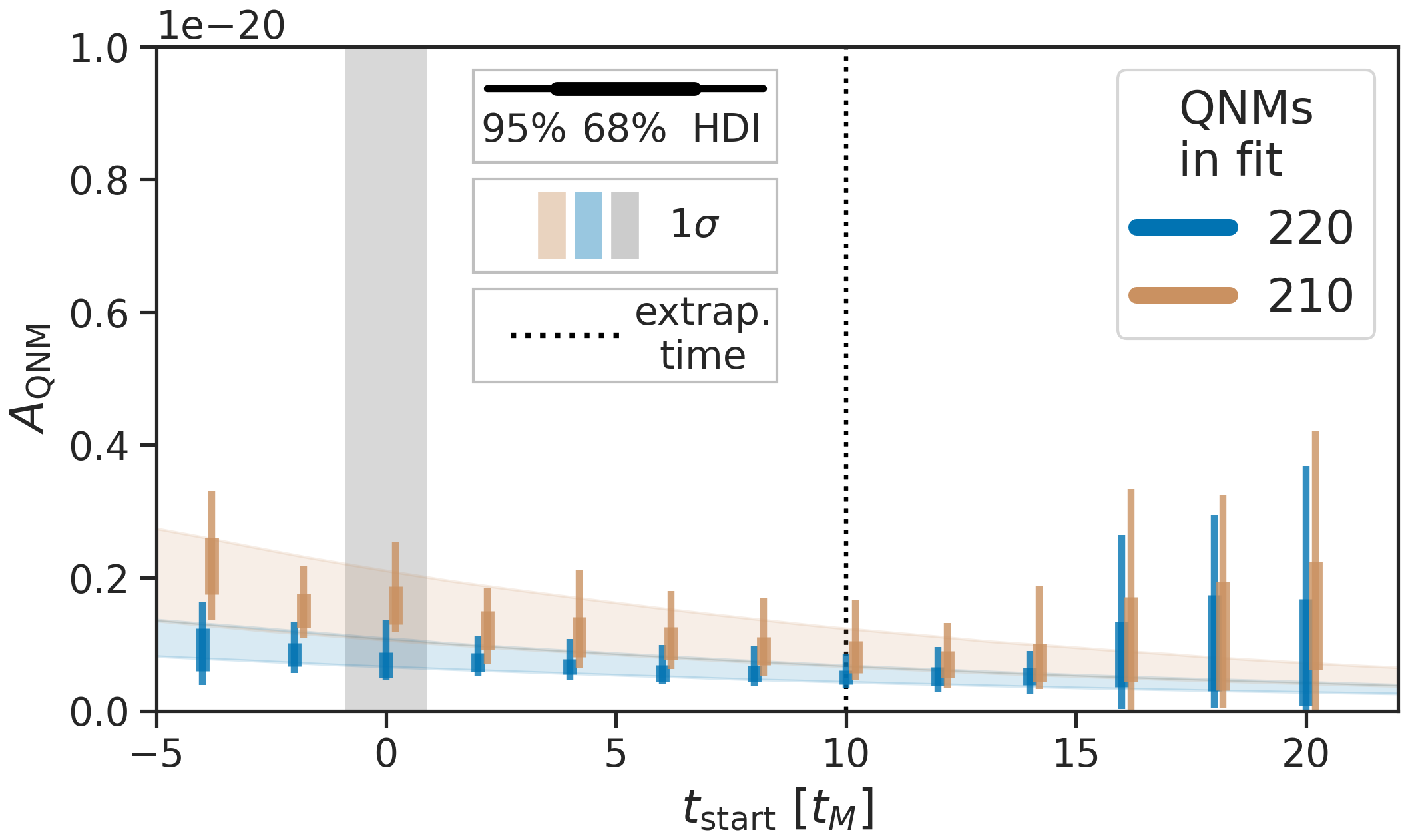}

    \caption{QNM amplitudes. We find multimodal Kerr fits over time to be consistent with their expected decay at least as early as $t_{\rm peak}$. Scatterpoints are staggered horizontally for visibility, but each grouping corresponds to a fit at a single time. The $1\sigma$ peak time uncertainty is shown as a grey band. Extrapolating from the fit at 10 $t_M$ (as indicated by dashed line), we show the $1\sigma$ uncertainty of each fitted QNM as colored bands. \textit{Top:}~Amplitudes of  \{220,~200\} model. The amplitude of the lower-frequency QNM is found to be slightly higher than that of the higher-frequency QNM. \textit{Bottom:}~The \{220,~210\} model produces similar amplitudes, although they are overall smaller. The expected exponential decay extrapolated from late times holds earlier in the signal for the 210 model than for the 200 model. }
    \label{fig:amplitude_timescan_fundamental}
\end{figure}

\begin{figure}
    \centering
    \includegraphics[width=\columnwidth]{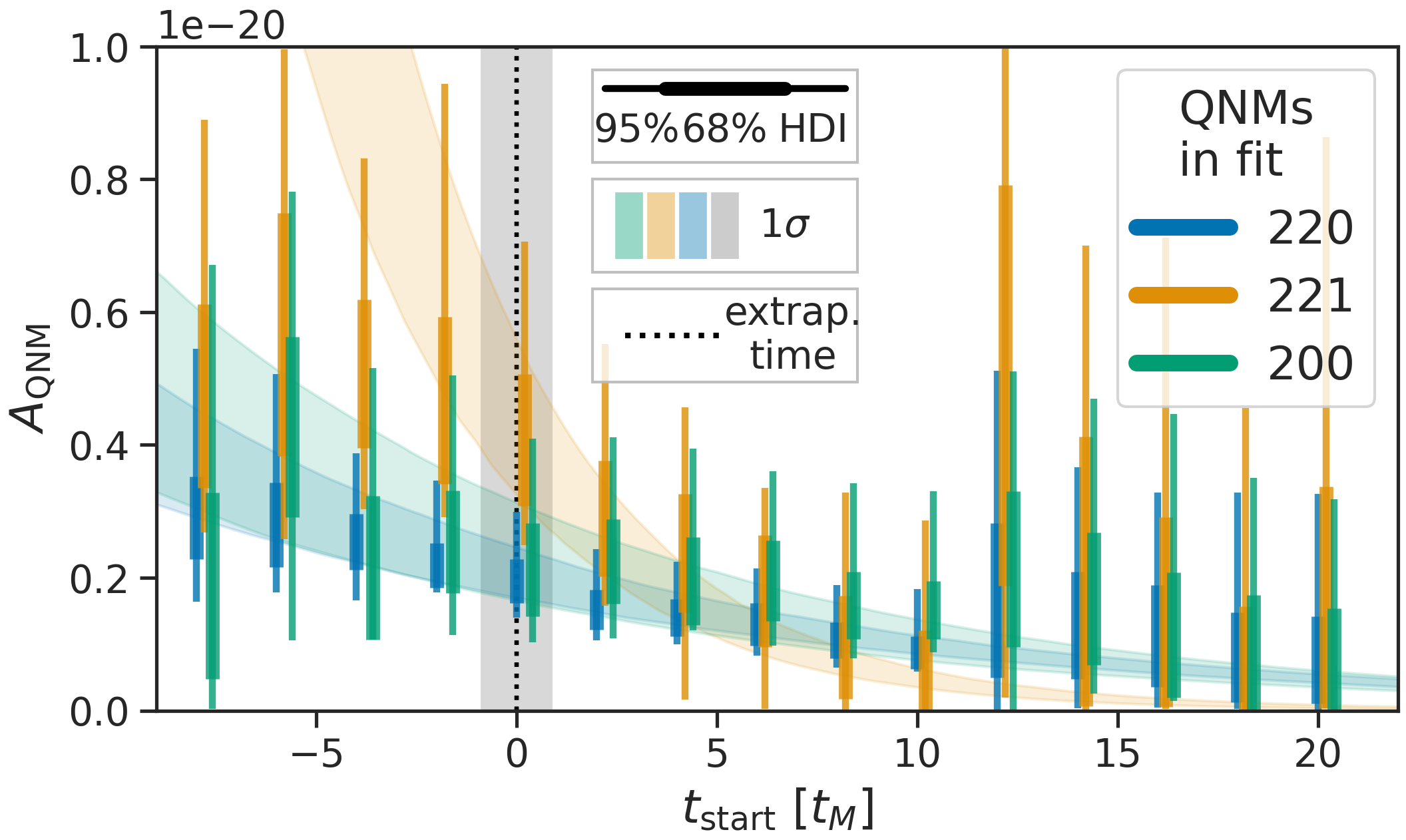}

    \includegraphics[width=\columnwidth]{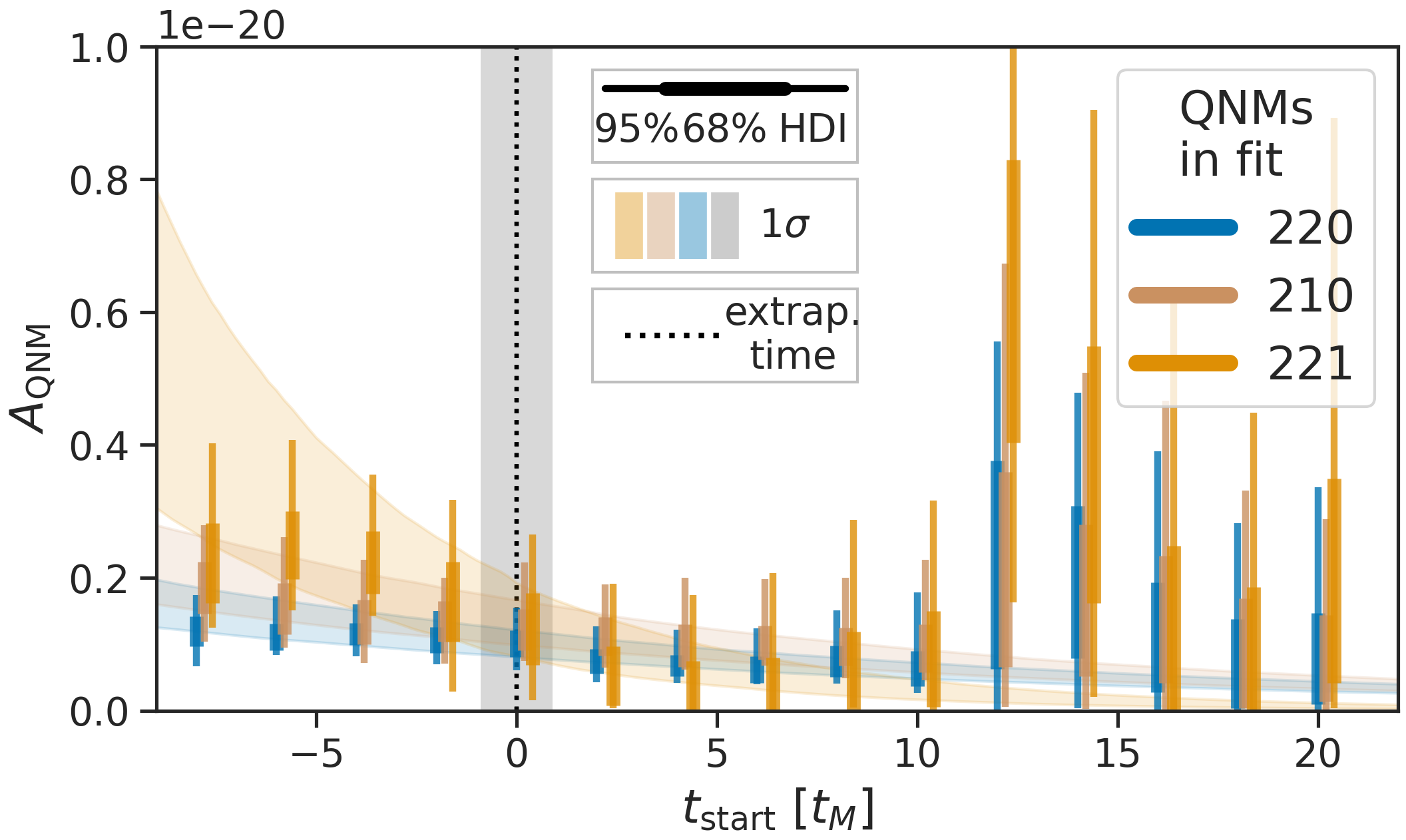}

    \caption{Amplitudes of  \{220,~200,~221\} (top panel) and \{220,~210,~221\} (bottom panel) models. See Fig.~\ref{fig:amplitude_timescan_fundamental} for figure conventions. All three amplitudes are confidently non-zero in the 200~(210) model as late as $4~t_M$~($0~t_M$), and expected decays for the overtone persist from these late times to as early as $-2~t_M$~($-8~t_M$). Qualitatively similar results are obtained with other 2$m$1 overtones instead of the 221. Instead of extrapolating amplitudes from $10~t_M$ as in Fig.~\ref{fig:amplitude_timescan_fundamental}, we choose here to extrapolate from a time where the overtone amplitudes are well-constrained to be non-zero, $0~t_M$, so that the extrapolation is not dominated by noise.}
    \label{fig:amplitude_timescan_overtone}
\end{figure}
As shown in Fig.~\ref{fig:amplitude_timescan_fundamental}, in both the \{220,~200\} and \{220,~210\} models the amplitudes of the two modes are comparable with the lower frequency mode having a slightly larger amplitude. The amplitude decays appear to broadly follow the expected exponential evolution over time.

While there is no goodness-of-fit preference for fitting more than two free damped sinusoids to the data, this does not necessarily guarantee that the same will be true of the Kerr models: and even if three-mode Kerr models are not more preferred by goodness of fit when compared with two-mode models, this does not mean that Kerr models with three or more modes cannot have any explanatory advantages from the physical point of view. Therefore, there is motivation for pursuing at least three-mode Kerr fits.

As shown in Fig.~\ref{fig:22_20_221_fgamma}, when fitting a \{220,~200,~$2m1$\} model ($m$~index intentionally ambiguous, explained immediately below) the mass and spin agrees with NRSur7dq4 as early as 0~$t_M$. This is 4~$t_M$ earlier than was found for the \{220,~200\} model in Fig.~\ref{fig:22_20_fgamma}. The $m$ index has been left ambiguous here because mass and spin agreement can be achieved by adding to the model any one of the $221$, $211$, or $201$ overtones. Furthermore, all three QNMs in these overtone models are well-constrained to have non-zero amplitudes as late as 4~$t_M$, as shown in Fig.~\ref{fig:amplitude_timescan_overtone}. However, at times especially before $t_{\rm peak}$ there are goodness-of-fit preferences overall for the $221$ as opposed to the other overtones in the three-mode models, and we will restrict our attention to the $221$ throughout.

In a similar vein, when fitting a \{220,~210,~221\} model, the frequency and damping rate measurements become self-consistent over time going back at least as far as {$-8~t_M$} (Fig.~\ref{fig:22_20_221_fgamma}). This model also has a significantly preferred LOO over all other models when fitting before the peak strain, which is at odds with the free damped sinusoid models which showed no preference for three-mode fits. The \{220,~210,~221\} model makes confident measurements of all three modes as well, albeit at slightly earlier times than the \{220,~200,~221\} model (Fig.~\ref{fig:amplitude_timescan_overtone}). Note that similar early-time fit consistency was also found in Ref.~\cite{Siegel:2023lxl} for GW190521, another low-frequency signal.

Beyond the Kerr models shown, we also explored two- and three-mode fits with one ${\ell=3}$ mode included, either the 320 or 330. These models confidently constrained the ${\ell=3}$ mode amplitudes to be zero both before and after the peak strain, more than QNMs in the other models we have discussed. Because of this, as well as the fact that the free damped sinusoid models did not explore the ${\ell=3}$ parameter space estimated through NRSur7dq4 measurements, we do not explicitly show these QNM models. We also found the \{220,~221\} model to be a poor fit to the data as shown in Fig.~\ref{fig:LOOfig} and inconsistent with NRSur7dq4 (not shown), and for these reasons we do not consider this model further. Lastly, we also fit four modes, $\{220,~221,~210,~211\}$, but since the $211$ amplitude was not consistently measured to be non-zero even going back to $-10~t_M$ we did not further explore this model. We did not consider retrograde ${p=-}$ Kerr QNM fits, because the free damped sinusoids don't support them when comparing with NRSur7dq4 and the estimated remnant spin being high makes retrograde modes unlikely to be as strongly excited as prograde modes~\cite{QNMsurrogate_Zertuche, Zhu:2023fnf, Mitman:2025hgy}.

\subsection{\label{sec:TGR_models}Beyond-Kerr QNM fits}
\begin{figure*}
    \centering
    \includegraphics[width=\columnwidth]{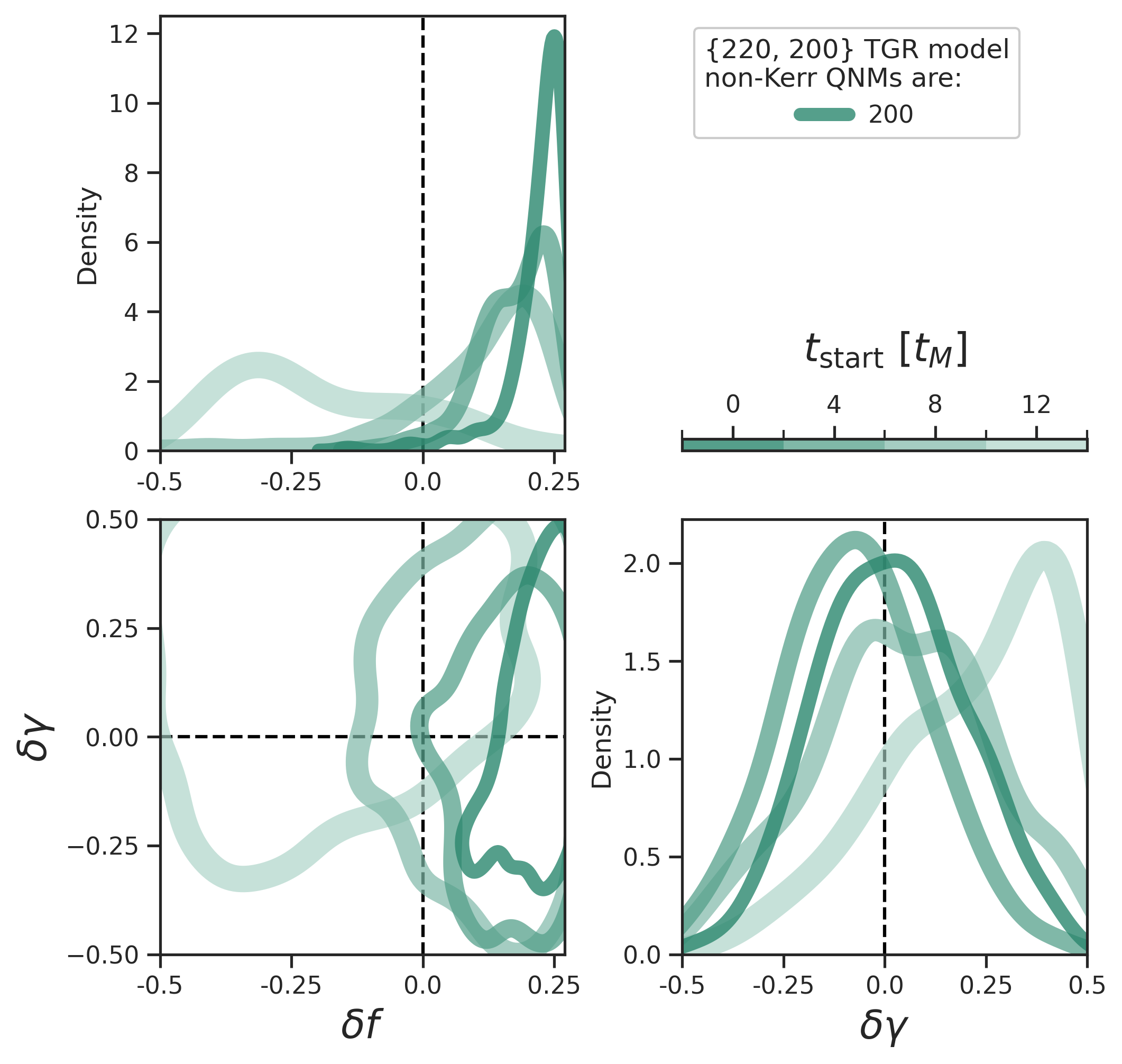}
    \includegraphics[width=\columnwidth]{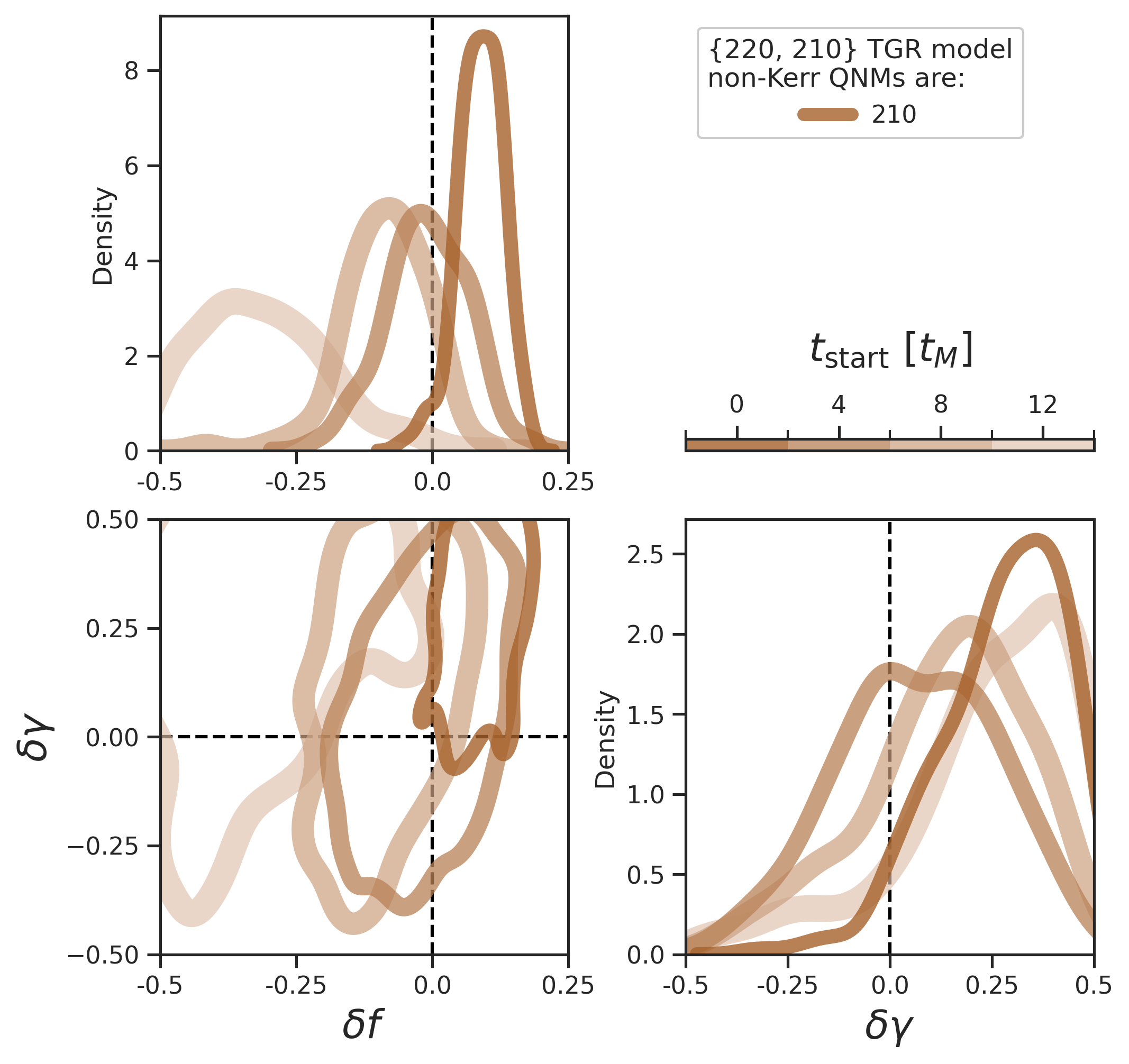}
    \caption{Test of GR (TGR) as parameterized in Eq.~\eqref{eq:TGR_fgamma}, using models with only fundamental modes. Transparency of the posteriors relates to fit start time shown in colorbar, color denotes non-Kerr QNM. Dotted lines indicate GR values. Axis limits correspond to prior boundaries. 90\% credible contours shown. \textit{Left:}~The \{220,~\underline{200}\} model with non-Kerr 200. While the posterior does begin encompassing the Kerr values at 90\% confidence at 8~$t_M$, it is also railing against the upper $\delta f$ prior and so this agreement should be interpreted with caution. At later times it becomes difficult to confidently measure both QNMs in the Kerr model as demonstrated in Fig.~\ref{fig:amplitude_timescan_fundamental}, and there is a clear drift in the TGR posteriors at this time. \textit{Right:}~The \{220,~\underline{210}\} model with non-Kerr 210. The posterior is noticeably more consistent with Kerr than the 200 model through $8~t_M$. This consistency is best achieved at 4~$t_M$, which is near where the Kerr model starts making mass and spin measurements more consistent with itself when fit at later times as shown in Fig.~\ref{fig:22_20_fgamma}.}
    \label{fig:TGR_fundamentals}
\end{figure*}

\begin{figure*}
    \centering
    \includegraphics[width=\columnwidth]{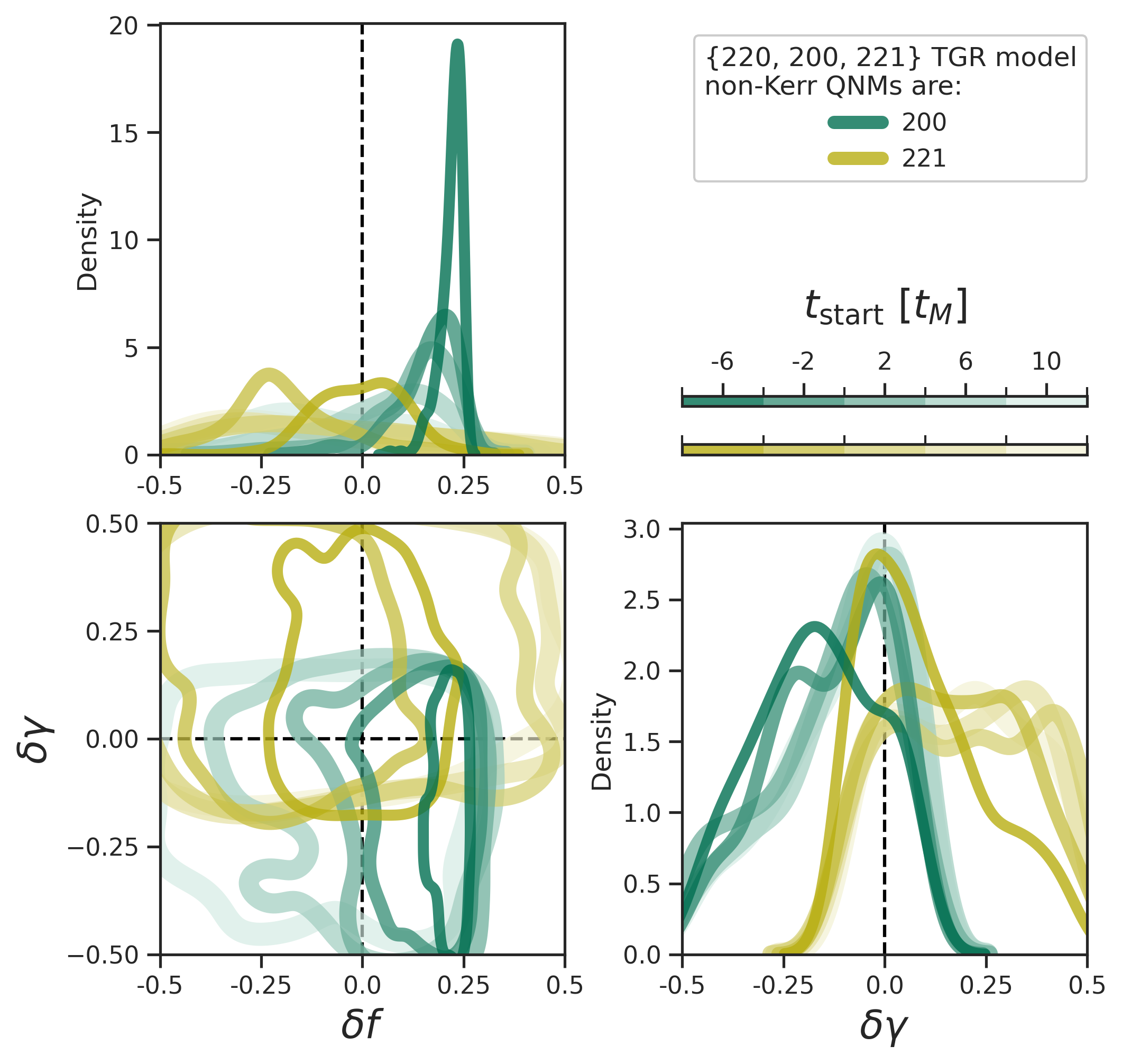}
    \includegraphics[width=\columnwidth]{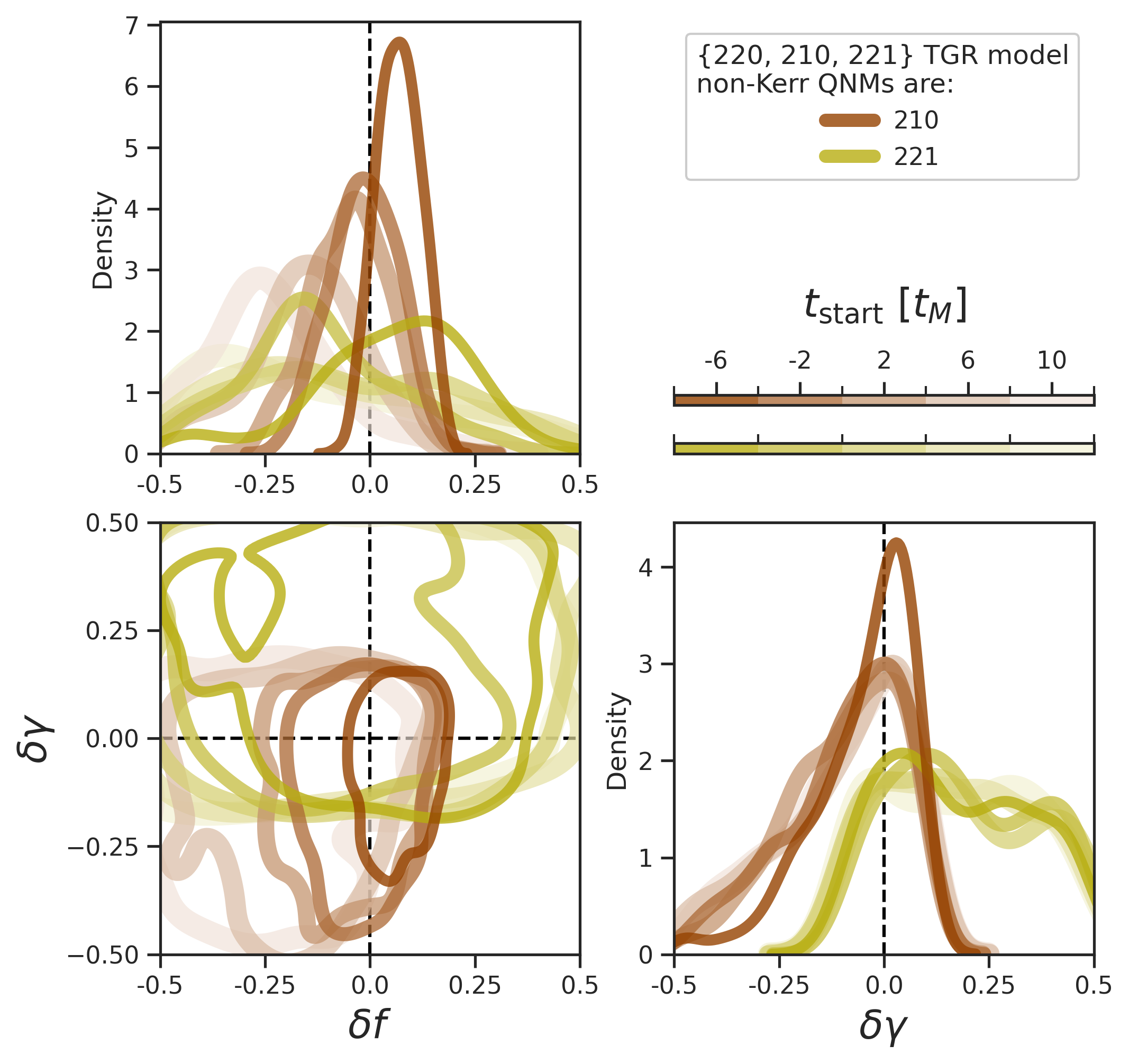}
    \caption{Test of GR (TGR) using models with fundamental and overtone modes, see Fig.~\ref{fig:TGR_fundamentals} for figure conventions. Comparing with Fig.~\ref{fig:TGR_fundamentals}, the overtone models are more consistent with Kerr at all times shown, especially in the case of the \{220,~210,~221\} model. Note that the $\delta \gamma$ parameters in particular have restrictive priors designed to prevent label-switching, which the posteriors consistently rail against. Consistency with Kerr is found at early times before peak strain. \textit{Left:}~The \{220,~\underline{200},~\underline{221}\} model with non-Kerr 200 and 221. While this model does have some support for Kerr at most times shown, the $\delta f_{200}$ posterior is always railing against one of the prior bounds (at 0.27 or -0.5) and thus should be interpreted with caution. At all times from $-6~t_M$ onward, the 221 is in reasonable agreement with Kerr. \textit{Right:}~The \{220,~\underline{210},~\underline{221}\} model with non-Kerr 210 and 221. Kerr is clearly supported at all shown fit times, notably both earlier and later in time than for the fundamental-only fits in in Fig.~\ref{fig:TGR_fundamentals}. At earlier times than those shown, fits start to drift away from Kerr values.}
    \label{fig:TGR_overtones}
\end{figure*}
We perform a test of general relativity (TGR) through a search for deviations from the Kerr frequency and damping rate spectrum. To do so, we fit a TGR model in which a subset of all of the QNMs have extra degrees of freedom $\delta f$ and $\delta \gamma$, such that their frequencies $f$ and damping rates $\gamma$ are given by~\cite{Gossan:2011ha, Isi:2021iql_analyzingbhringdowns}
\begin{align}
\begin{split}
    f_{\text{TGR}} & = f_{\text{Kerr}}\exp(\delta f), \\
    \gamma_{\text{TGR}} & = \gamma_{\text{Kerr}}\exp(\delta \gamma).
    \label{eq:TGR_fgamma}
\end{split}
\end{align}
When referencing specific TGR models in the text, any QNMs with non-Kerr freedoms in the model will be underlined. The default hypothesis of our test is that Kerr should be a good description of the data.

Based on comparisons with NRSur7dq4, we choose to consider the \{220, \underline{200}\} and \{220, \underline{200}, \underline{221}\} TGR models, underlining modes on which the deviation parameter is placed. We also consider TGR models with the \underline{210} instead of the \underline{200}, motivated by goodness of fit from the previous subsection as well as knowledge of precessing BBH coalescences~\cite{Zhu:2023fnf, OShaughnessy:2012iol, Hamilton:2023znn, Nobili:2025ydt}.

We emphasize the measurements of $\delta f$ over $\delta \gamma$, because of their generally greater precision and also because we find the poorer $\delta \gamma$ measurements to be exacerbated  by degeneracies with the remnant mass and spin~\cite{Ghosh:2021mrv, Siegel:2023lxl, Gennari:2023gmx}. TGR parameter priors are chosen to avoid label-switching between QNMs when their Kerr frequencies and damping rates are close in parameter space: this label-switching can cause TGR results to be erroneously inconsistent with GR even in cases where the signal being fit is exactly consistent with GR. A tell-tale sign of such label-switching can be multimodal structure in the TGR parameters. The prior on $\delta f$ goes from -0.5 to +0.25 (+0.27) for the \underline{210} (\underline{200}) in any model where those QNMs have TGR freedoms. Tight $\delta \gamma$ priors with an absolute value of 0.1 are also implemented in whichever single direction would cause overtones and fundamental modes to swap places. When quoting $\delta f$ and $\delta \gamma$ constraints, we will state median values with $\pm 90\%$ HDI uncertainties.
\renewcommand{\arraystretch}{1.3} 
\begin{table}[ht]
  \centering
  \renewcommand{\arraystretch}{1.5} 
  \caption{TGR model constraints on non-Kerr deviations $\delta f$ and $\delta \gamma$ (see Eq.~\eqref{eq:TGR_fgamma}) at selected $t_{\text{start}}$. Each QNM with parameterized deviations from Kerr is indicated with underlined indices. The earliest time for each model in the table is the first time at which $\delta f$ is found to be consistent with 0 at $90\%$ HDI for all non-Kerr QNMs. For comparison with GW250114~\cite{GW250114_paper1,GW250114_paper2}, we include $6~t_M$ when possible. See Sec.~\ref{sec:TGR_models} for further discussion. Many of the Kerr constraints should be interpreted with caution, especially for \underline{200} models which are often railing against priors.}
  \label{tab:TGR-deltas}
  \begin{tabular}{|l|l|l|}
    \hline
    Model & $t_{\text{start}}$ & $(\delta f, \delta \gamma)$, median $\pm90\%$ HDI \\
    \hline
    \{220, \underline{200}\} 
      & $8\,t_M$  
      & \underline{200}:\;$(0.15^{+0.12}_{-0.17},\;0.08^{+0.38}_{-0.34})$ \\
    \hline
    \multirow{2}{*}{\{220, \underline{210}\}} 
      & $2\,t_M$  
      & \underline{210}:\;$(0.05^{+0.08}_{-0.11},\;0.07^{+0.34}_{-0.33})$ \\
      \cline{2-3} & $6\,t_M$  
      & \underline{210}:\;$(-0.03^{+0.10}_{-0.11},\;0.26^{+0.24}_{-0.27})$ \\
    \hline
    \multirow{4}{*}{\{220, \underline{210}, \underline{221}\}} 
      & $-6\,t_M$ 
      & \begin{tabular}[t]{@{}l@{}}\underline{210}:\;($0.06_{-0.08} ^{+0.08}$,\;$-0.03_{-0.20} ^{+0.13}$) \\ \underline{221}:\;($0.07_{-0.26} ^{+0.28}$,\;$0.17_{-0.27} ^{+0.25}$)\end{tabular} \\
      \cline{2-3}
      & $0\,t_M$  
      & \begin{tabular}[t]{@{}l@{}}\underline{210}:\;($-0.08_{-0.16} ^{+0.19}$,\;$-0.09_{-0.28} ^{+0.19}$) \\ \underline{221}:\;($-0.23_{-0.27} ^{+0.24}$,\;$0.21_{-0.25} ^{+0.28}$)\end{tabular} \\
      \cline{2-3}
      & $6\,t_M$  
      & \begin{tabular}[t]{@{}l@{}}\underline{210}:\;($-0.14^{+0.21}_{-0.21}$,\;$-0.06^{+0.16}_{-0.26}$) \\ \underline{221}:\;($-0.12^{+0.44}_{-0.38}$,\;$0.19^{+0.24}_{-0.29}$)\end{tabular} \\
    \hline
  \end{tabular}
\end{table}

\subsubsection{Fundamental modes}

We begin by considering fundamental mode models only. In Fig.~\ref{fig:TGR_fundamentals} we compare the TGR posterior of the \{220,~\underline{200}\} and \{220,~\underline{210}\} models. The \underline{210} model appears more consistent with Kerr over most of the time interval where both modes are confidently measured. Before and up to the peak strain time, the measurements of $\delta f$ and $\delta \gamma$ in both models drift in a manner commensurate with the behavior of Kerr frequency and damping rate measurements shown in Fig.~\ref{fig:22_20_fgamma}. Around 2--4~$t_M$ and later times, the TGR posteriors appear to settle and gradually expand as the SNR decreases. Incidentally, 2--4~$t_M$ corresponds to the time around which Kerr models with overtones start to allow the overtone amplitude to be zero, as in Fig.~\ref{fig:amplitude_timescan_overtone}. At 12~$t_M$, the TGR posteriors then shift again to become broader, and at times after this become uninformative: this final transition seems to correspond to the point in the signal at which the models start to lose resolution of both modes, based on comparison with the Kerr fit behavior shown in Figs.~\ref{fig:22_20_fgamma}~and~\ref{fig:amplitude_timescan_fundamental}.

We first focus on the \{220,~\underline{200}\} TGR model. At $4~t_M$, the earliest time at which we interpret the fundamental mode models to become consistent with their later time fits, the $\delta f_{200}$ posterior is inconsistent with Kerr at the $90\%$ CL. Not only is it inconsistent, but ${\delta f_{200}}$ is railing against the upper prior, and so in principle the GR value may be excluded at even higher CLs: interactions with the prior bounds suggest that CLs should be interpreted with caution. The first time at which the $\delta f_{200}$ posterior gives consistency with GR at the $90\%$ CL is $8~t_M$, where ${\delta f_{200}\,(8~t_M)=0.15 ^{+0.12}
_{-0.17}}$: but again, ${\delta_{f_{200}}}$ is still largely cut off by the prior at positive values, and thus even this Kerr consistency should be interpreted with caution. The Kerr consistency persists when fitting up to $16~t_M$, although the posterior shifts in these later times from railing against the upper prior bound to railing against the lower prior bound and so again interpretations should be made with caution. At $18~t_M$ and later, the  ${\delta f_{200}}$ posterior reverts to the prior.  

By contrast, the \{220,~\underline{210}\} TGR model is not only more in agreement with Kerr but also achieves this agreement at much earlier times. The earliest Kerr-consistent fit start time is $2~t_M$, where we find
${\big(\delta f_{210}\,(2~t_M),~\delta \gamma_{210}\,(2~t_M) \big)=\big(0.05 ^{+0.08}
_{-0.11},~ 0.07 ^{+0.34}
_{-0.33}\big)}$.

Fitting the \{220,~\underline{210}\} TGR model at $6~t_M$ allows for comparison with a similar test of GR for GW250114 reported in Fig.~4 of Ref.~\cite{GW250114_paper1} and Fig.~2 of Ref.~\cite{GW250114_paper2}, which used a \{220,~221\} model. We find ${\big(\delta f_{210}\,(6~t_M),~\delta \gamma_{210}\,(6~t_M) \big)=\big(-0.03 ^{+0.10}
_{-0.11},~ 0.26 ^{+0.24}
_{-0.27}\big)}$, significantly improving on the $\pm30\%$ frequency constraint reported for GW250114 at the equivalent fit start time. The $\delta \gamma_{210}$ posterior is once again somewhat cut off by the upper prior at this time for GW231123 though, and caution in interpretation is warranted.

Fitting the \{220,~\underline{210}\} TGR model between $10~t_M$ and $16~t_M$, the $\delta f_{210}$ posterior drifts to rail against the negative end of the prior, ruling out 0 at the $90\%$ CL from $10~t_M$ to $14~t_M$. At $18~t_M$ and later, the posterior reverts to the prior: this behavior is in line with the evolution of the Kerr fits, which lose resolution of two modes around $16~t_M$ as shown in Figs.~\ref{fig:LOOfig},~\ref{fig:22_20_fgamma},~\ref{fig:amplitude_timescan_fundamental}. The railing may just be a consequence of the decreasing resolution of the two modes with time, or could be a sign of mismodeling.

\subsubsection{Addition of an overtone}

We now consider fitting TGR models with an overtone included. In the previously fit Kerr models of Sec.~\ref{sec:Kerr_models}, overtones are not found to give substantially better fits to the data at times after the peak strain (Fig.~\ref{fig:LOOfig}), but do make remnant mass and spin measurements more consistent over a broader range of times (Fig.~\ref{fig:22_20_221_fgamma}) and have amplitudes confidently measured to be nonzero as late as $\sim4~t_M$ after the peak strain (Fig.~\ref{fig:amplitude_timescan_overtone}). Therefore, we might expect overtones to change the TGR posteriors of the most dominant modes, even if the overtone TGR parameters themselves are not well-constrained.

Fig.~\ref{fig:TGR_overtones} shows representative results of TGR fits with two fundamental modes and one overtone. An interesting feature of these fits is the improved consistency of their TGR parameters with Kerr over a wider range of fitting start times than the fundamental-only models. Notably, the \{220,~\underline{210},~\underline{221}\} model is consistent with Kerr as early as $-6~t_M$. A similar behavior of consistency in very early fits was observed in Ref.~\cite{Siegel:2023lxl} when fitting both Kerr and TGR quasinormal mode models to GW190521, another low-frequency signal. For GW190521, consistent TGR fits seemed to be achievable as early as $-5~t_M$.

The \{220,~\underline{210},~\underline{221}\} TGR model in GW231123 is more consistent overall with Kerr than the \{220,~\underline{200},~\underline{221}\} model, but both have strikingly similar overtone $\delta f$ constraints. Comparing with Fig.~\ref{fig:22_20_221_fgamma}, these shared overtone constraints are perhaps not so surprising: regardless of which Kerr model is fit, the ${(\ell,m)=(2,~2)}$ modes in the Kerr models are fit at the same frequency around 65 Hz. The instrumental noise is lower at higher frequencies in this part of the LIGO band, and correspondingly the ${(\ell,m)=(2,~2)}$ modes are more precisely measured than the lower frequency ${\ell=2\neq m}$ modes, which implies that any improvements in fit when adding TGR freedoms will likely come from the ${\ell=2\neq m}$ modes in the model.

Given that the overtone model closest to Kerr does not seem to be the model in best agreement with NRSur7dq4, interpretation of the TGR results is difficult. We choose to report on Kerr constraints of the \{220,~\underline{210},~\underline{221}\} TGR model at a few time intervals, since this model displays some possible signs of better consistency with Kerr and its $\delta f$ posteriors do not rail against the priors as significantly as the \{220,~\underline{200},~\underline{221}\} model at most fitting times. At $6~t_M$, we find Kerr-consistent frequency constraints of roughly $\pm20\%$ and $\pm40\%$ for the 210 and 221 respectively. These constraints remain Kerr consistent and improve going back in time as early as $-6~t_M$, to $\pm8\%$ and $\pm28\%$ for the 210 and 221 respectively. See Table~\ref{tab:TGR-deltas} for details. Earlier than $-6~t_M$, the \{220,~\underline{210},~\underline{221}\} TGR model begins deviating from Kerr. This time corresponds to an apparent transition in the amplitudes of the Kerr model, as shown in Fig.~\ref{fig:amplitude_timescan_overtone}.

It is possible that the improvement in Kerr consistency of the three-mode models with overtones over two-mode models with only fundamental modes is just a function of the additional parameters of the three-mode models broadening the posteriors. However, given that the Kerr fits with overtones are at least equally preferable if not significantly preferred over fits without overtones in the range of times where Kerr consistent measurements are made (Fig.~\ref{fig:LOOfig}), and the overtone amplitudes are confidently nonzero in most of these times and follow their expected decays (Fig.~\ref{fig:amplitude_timescan_overtone}), it seems plausible that the overtones are meaningfully contributing to the fits and parameter estimates.

\section{\label{sec:discussion}Discussion}
Assessing the best QNM fits to the GW231123 data and determining whether we have successfully validated the QNM frequency and damping rate spectrum of the Kerr metric is complicated not only by our incomplete understanding of the regime of validity of QNM fits in BBH coalescencess but also by the possibility of systematics in the IMR model to which we are comparing our fits, NRSur7dq4. Fully determining the conclusiveness of our results will have to rely on a holistic consideration of statistical quantities, astrophysical expectations, and some educated conjecture regarding the unknowns of QNM analysis from both the data and theory points of view.

\subsection{Motivating choices of QNM fit start times}

A central challenge in the assessment of our fits is determining what should constitute physically meaningful QNM fit start times relative to the peak strain time in BBH coalescence signals. Debate continues on this topic in the literature, centering around fits to NR which hope to subsequently inform actual data analysis. Much of this NR-based work reports some indications of systematic error when fitting damped-sinusoid models close to (but after) the peak strain~\cite{Baibhav:2023clw, Berti:2025hly, Giesler:2024hcr, Mitman:2025hgy}.

However, a key distinction between these NR studies and data analysis of LIGO signals is the noise in both instances. LIGO signals are orders of magnitude quieter than NR, and have a specific noise spectrum morphology which interacts with the signal differently than the imperfectly understood numerical errors associated with most NR simulations in current use. While the findings of systematic errors in QNM fits to NR are important to keep in mind when performing data analysis, we are not aware of any rigorous quantification of these systematic errors. For example, if the extent of these systematics in NR analyses is never more than an error of 1 part in 1000 when fitting after the peak strain (or even before it), that is irrelevant to data analysis of ${\text{SNR}=20}$ LIGO signals where statistical errors of Kerr constraints are $\mathcal{O}(10\%)$.

Furthermore, most NR studies perform maximum likelihood fitting whereas we do fully Bayesian analysis for LIGO data, presenting another complication in making direct comparison. Another issue which further complicates comparison is that our analysis of LIGO data does not estimate the location of the peak strain in the signal using QNM fits, but rather relies on alternate models like NRSur7dq4 to give us these estimates. If the peak strain estimates of these alternate models turn out to be significantly systematically biased, our data analysis may not be able to appeal directly to our theoretical understanding of regimes of validity for QNM fits anyway.
Nevertheless, we are always allowed to empirically explore which regions of the data appear consistent with QNM content.

\subsection{Consistency of QNM fits over time}

A key tenet of our fitting philosophy is that reasonable QNM models which accurately describe the signal should have consistent parameter estimates when fit at different starting times. This philosophy is shared with most NR studies~\cite{Mitman:2025hgy, Giesler:2024hcr, Cheung:2023vki,Clarke:2024lwi}. To confirm consistency between fits at different times, we visually compare credible levels of posteriors, typically the $90\%$ CL but not always: while we could more strictly quantify the level of agreement between posteriors by using a statistical distance, e.g. Kullback-Leibler divergence, there are nontrivial complexities in employing statistical distances~\cite{10.1093/mnras/staa2850_bilbybayesianinference} and we assert that an eye test is sufficiently accurate for our purposes. We find that fits to GW231123 with two fundamental modes are capable of achieving consistent parameter estimates over time, as shown in Figs.~\ref{fig:22_20_fgamma},~\ref{fig:amplitude_timescan_fundamental},~\ref{fig:TGR_fundamentals}. Three-mode fits are capable of achieving this behavior as well, as shown in Figs.~\ref{fig:22_20_221_fgamma},~\ref{fig:amplitude_timescan_overtone},~\ref{fig:TGR_overtones}. This is one indication that our fits are reasonable.

Our two-mode fits confidently give Kerr constraints at some late times close to $10~t_M$, as shown in Fig.~\ref{fig:TGR_fundamentals}. The time $10~t_M$ is often used as approximately where QNM fits should be reliable in general, although we do not assert that this is the earliest possible time for QNM fits to be reliable even in theory. While the two-mode models validate Kerr at the $90\%$~CL at some late times around $10~t_M$, they do not necessarily do so at all late times. This discrepancy in Kerr validation is ameliorated by the inclusion of one overtone in addition to the two fundamental modes in the models, as shown in Fig.~\ref{fig:TGR_overtones}. The overtone is not itself well-measured at these late times as indicated by its amplitude in Fig.~\ref{fig:amplitude_timescan_overtone}, but this is not necessarily a reason to ignore the overtone in the late time fits. The change in posteriors from including the overtone may be an indication of stealth bias in the fundamental mode models~\cite{Cutler:2007mi_PEbias, Cornish:2011ys_stealthbias}. 

Ultimately, we presume that every possible QNM is excited to some non-zero amplitude in astrophysical BBH coalescences, and in principle a physically faithful QNM model would always include every possible QNM even if the contribution of most of the QNMs to the fit was very small, and would also take into account currently ignored effects like the time-dependent ring-up of QNMs~\cite{Chavda:2024awq}. For such types of all-inclusive models, the priors would then determine which QNMs were most significant in the fit. This type of informative-prior driven analysis is a more desirable approach than what is commonly done in our field (and was done in this paper) wherein different numbers of QNMs are added to different models, in part because the prior-driven approach provides a better foundation for model comparison~\cite{LOO_FAQ}. That being said, robust informative priors for QNM fits are currently beyond the reach of our field. Efforts are being undertaken to solve this problem~\cite{Nobili:2025ydt, Mitman:2025hgy, Ghosh:2021mrv, Gennari:2023gmx, QNMsurrogate_Zertuche, Pacilio:2024tdl, Forteza:2022tgq}, but more work is still needed. The lack of informative priors and restricted number of QNMs in our current models is a limitation of the program of black-hole spectroscopy as it was originally envisioned~\cite{Dreyer:2003bv}: even though we may find consistency with the frequencies and damping rates of the Kerr spectrum using our current models, our fits allow such a wide range of amplitudes and phases that they may be settling on astrophysically unlikely or totally unattainable amplitude and phase combinations. An even more ideal improvement over informative priors for QNM models would be full inspiral-merger-ringdown models~\cite{Julie:2024fwy} in different theories of gravity~\cite{Doneva:2023oww, Corman:2022xqg, Corman:2024cdr} which are both well-posed~\cite{Cayuso:2017iqc} and well-motivated, but this seems to be even further from our current reach than informative QNM models. Some works argue that enlarging the model parameter space by adding more and more QNMs will make Bayesian inference unfeasible~\cite{PhysRevD.101.044033}. We contend that if this is an accurate statement, it may only be accurate in the case of models with uninformative priors. Refs.~\cite{Chandra:2025ipu, CalderonBustillo:2020rmh} make qualitatively similar arguments: that being said, we are strictly referring to priors derived from and for direct QNM fitting, while those works seek to use the ringdown stages from existing IMR models which are likely not identical to our proposed priors. In research fields outside of gravitational wave science, it is common to have highly performing models with orders of magnitude more parameters than our current ringdown models~\cite{Efron_Hastie_2021}: if they can do it, perhaps so can we.

\subsection{QNM fits starting before peak strain}

While our GW231123 fits seem to produce consistent parameter estimates over a range of fit start times after the peak strain, they also are consistent in some instances at times before the peak strain. The Kerr constraints at times as early as $-6~t_M$ for the \{220,~\underline{210},~\underline{221}\} model in particular are much tighter than at late times, as shown in Table~\ref{tab:TGR-deltas}, and still support general relativity. Some theoretical studies of extreme and comparable mass binary systems suggest the presence of QNM and other Kerr perturbative content at times near or even before peak strain~\cite{Price:2015gia, Mitman:2025hgy, Giesler:2024hcr, Oshita:2025qmn}, but the early time fit behavior we have observed may not necessarily be physical and deserves scrutiny.

As mentioned above, most theoretical studies of QNM fits are done at very different SNRs from actual data, and with different noise. The specific noise PSD of LIGO whitens very low-frequency BBH signals such that their inspiral content is significantly suppressed, and this may be affecting the ringdown analysis of GW231123 in unanticipated ways. One possibility is that a whitened ringdown signal with no discernible inspiral preceding it in time may still be fit accurately for some time interval before the ringdown signal.

Another possibility is that perturbative content like QNMs exists in the signal even before the peak strain but is obscured by other lower-frequency signal content, and the whitening of the PSD works to reveal this perturbative content which would otherwise be hidden. These explanations are purely speculative however.

It is worth noting that a similar early time fit behavior was also observed in GW190521~\cite{Siegel:2023lxl}, hinting at the possibility that this is a generic feature of low-frequency signals. In future work we aim to perform thorough studies of NR signals in LIGO noise to determine if the fitting behavior we have observed should be expected. The Kerr constraints at these early times are much better than at later times, and so we should make use of them if they are indeed trustworthy. 

\subsection{IMR waveform systematics}

Another possible explanation of the early QNM fit behavior is that we have misidentified the peak strain time and/or the timescale $t_M$ due to NRSur7dq4 systematic error. To assess the systematic error of IMR waveform models, a standard approach is to use the mismatch of the model with NR simulations~\cite{Varma:2019csw_NRsurpaper, GW231123_paper}. Based on the standard method of assessing mismatch and also through empirical observation of fits to NR, when considering highly-spinning NR simulations NRSur7dq4 is the overall best-performing quasicircular precessing BBH IMR model in the GW231123 part of parameter space. Mismatches furthermore indicate that, for many but not all values of parameters in this region and at the SNR of GW231123, we expect NRSur7dq4 to have systematic errors which are subdominant to statistical errors~\cite{GW231123_paper, Varma:2019csw_NRsurpaper}. While in general the mismatches as traditionally computed actually provide a conservative estimate of minimum SNRs at which systematic bias dominates over statistical uncertainty, Ref.~\cite{Thompson:2025hhc} notes that incorrect choices of the number of degrees of freedom in the model may lead to the traditional minimum SNR estimate not being a conservative lower bound, and so the mismatches should be interpreted with caution. We cannot definitively rule out the possibility that the NRSur7dq4 measurement errors are dominated by systematics for GW231123.

A good way to determine if NRSur7dq4 is systematically biased is to find a better fit to the data with a physically plausible model. To this end, we note that the QNM fits with the 200 mode, which produce the best agreement with NRSur7dq4 remnant mass and spin estimates (Figs.~\ref{fig:22_20_fgamma},~\ref{fig:22_20_221_fgamma}), are not as favored by goodness of fit as 210 QNM fits overall (Fig.~\ref{fig:LOOfig}), even though those 210 QNM fits are in tension with NRSur7dq4 parameter estimates. Furthermore, the NRSur7dq4-consistent 200 QNM fits do not seem to be as in agreement with Kerr as the 210 QNM fits (Figs.~\ref{fig:TGR_fundamentals},~\ref{fig:TGR_overtones}).

\subsection{Astrophysical plausibility of 210 and 200 QNMs}

The 210 fits may also constitute a more physically plausible model than the 200 fits. It is known that the 210 can be excited by binary spin-orbit misalignment~\cite{Zhu:2023fnf, OShaughnessy:2012iol, Hamilton:2023znn, Nobili:2025ydt}, and while the 200 is also excited by spin-orbit misalignment it is excited to a lesser extent over the angular two-sphere. To have the 200 be so loud as we find it to be in Figs.~\ref{fig:amplitude_timescan_fundamental},~\ref{fig:amplitude_timescan_overtone}, implies that the 210 should also be similarly loud if not more so over the angular two-sphere: but we do not find strong support for fitting three ${\ell=2}$ damped sinusoids as suggested, e.g., by the behavior of free damped sinsuoid fits in Figs.~\ref{fig:LOOfig},~\ref{fig:damped_sinusoid_fgamma}.

Such a suppression of the 210 would require a rather fine-tuned binary spin configuration. The mechanism by which the 210 could be significantly suppressed is the asymmetric emission of mirror mode QNMs associated with different angular harmonics, as described in e.g. Fig.~3 of Ref.~\cite{Zhu:2023fnf}, combined with a viewing angle closer to the remnant's equator than its poles. Based on current knowledge, this fine-tuned configuration by itself might not even be sufficient to create the observations we have made in GW231123.

\begin{figure}
    \centering
    \includegraphics[width=\columnwidth]{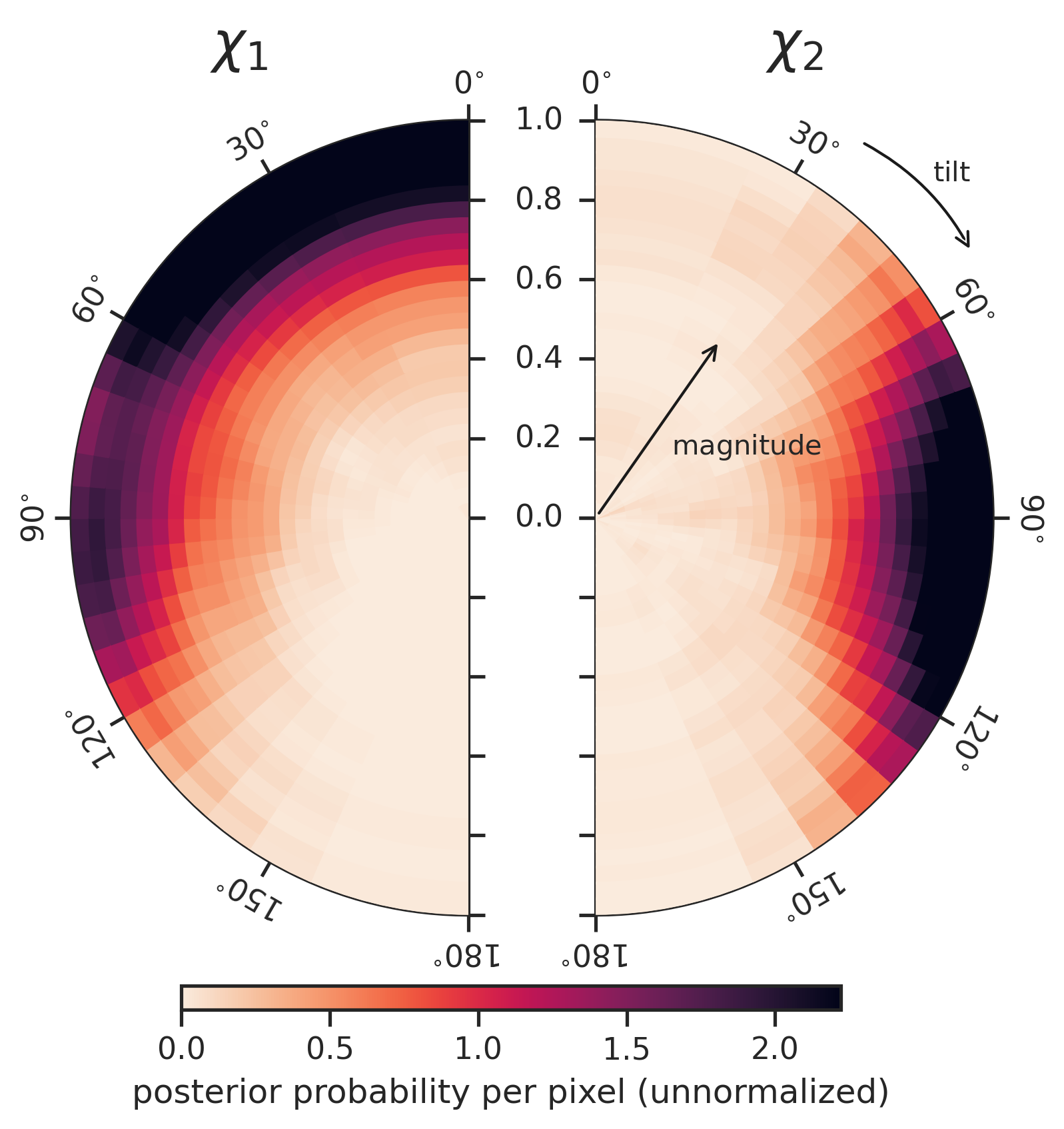}
    \caption{Spin tilts and magnitudes given by NRSur7dq4 for each of the binary components. The bulk of the inferred distribution lies at maximal spins and spin-orbit misalignment of at least the secondary component, with significant support for the primary being misaligned as well. Spin-orbit misalignment is associated with excitation of both prograde and retrograde ${\ell\neq m}$ fundamental QNMs~\cite{Zhu:2023fnf, OShaughnessy:2012iol, Hamilton:2023znn, Nobili:2025ydt} like those fit throughout this work.}
    \label{fig:nrsur_spindisk}
\end{figure}

It is worth noting that NRSur7dq4 finds the most support for binary spins that are not only in the orbital plane but also maximal, as shown in Fig.~\ref{fig:nrsur_spindisk}. It is possible that the aforementioned emission mechanisms are amplified for near-extremally spinning precessing BBHs, a part of parameter space which has not been rigorously studied. Nonetheless, as of now the 210 seems like a more plausible candidate for strong excitation than the 200.

We also comment here on Fig.~10 of Ref.~\cite{GW231123_paper}, which shows a harmonic mode decomposition of the NRSur7dq4 waveform using posterior samples from analysis of GW231123. The text of the accompanying appendix for this figure claims that it demonstrates an inconsistency between the angular harmonic content of NRSur7dq4 and a relatively large observed 200 QNM amplitude. However, direct comparison between this figure and QNM analyses is difficult to make, since the waveforms in the figure are not in the appropriate frame and consider only combined amplitudes over the 2-sphere of the $\pm m$ harmonic modes in this frame. A version of this figure better-suited for direct comparison with our analysis would show the NRSur7dq4 waveform from a specific point on the celestial sphere with angular harmonics defined according to the remnant frame, and the $\pm m$ modes would be kept separate because they can differ dramatically for precessing systems~\cite{Boyle:2014ioa, Zhu:2023fnf}. In its current form, the figure does not provide direct evidence that NRSur7dq4 does not support a relatively large observed 200 amplitude.

\subsection{QNM fit systematics}

One might also worry that our QNM fits are systematically biased rather than the IMR fits. This is possible if the signal we are fitting contains significant content besides damped sinusoids, the data contains significant unmodeled noise features, the data conditioning we perform has corrupted our posteriors~\cite{Siegel:2024jqd}, or the QNM model we fit is misspecified to the actual signal. For a correctly specified QNM model fit to data without significant nonstationary noise features, we expect to see consistent parameter estimates when starting our QNM fits at different times in the signal. We successfully demonstrate such consistent fitting behavior in Figs.~\ref{fig:22_20_fgamma}, \ref{fig:22_20_221_fgamma}, \ref{fig:amplitude_timescan_fundamental}, \ref{fig:amplitude_timescan_overtone}, \ref{fig:TGR_fundamentals}, \ref{fig:TGR_overtones}.

That being said, because the SNR decreases with time and is moderate even at the peak strain, it is only at early times in the signal where definitive preferences between different QNM models appear. It is difficult at present to determine the earliest time at which QNM fits should be considered valid in this signal, so the aforementioned definitive model preferences at early times may not be meaningful. This model selection ambiguity is a general limitation of our very flexible damped sinusoid models: more informative amplitude and phase priors might help better discriminate between different QNM combinations at later fitting times. We note that the post-peak SNR is not so high as to make GW231123 very susceptible to conditioning-induced bias~\cite{Siegel:2024jqd}, and poor data quality affecting our inferences also seems less likely given the discussion in Sec.~\ref{sec:conditioning} as well as Ref.~\cite{GW231123_paper}.

\subsection{Interpretation of Kerr and Beyond-Kerr QNM fits}

\begin{figure}
    \centering
    \includegraphics[width=\columnwidth]{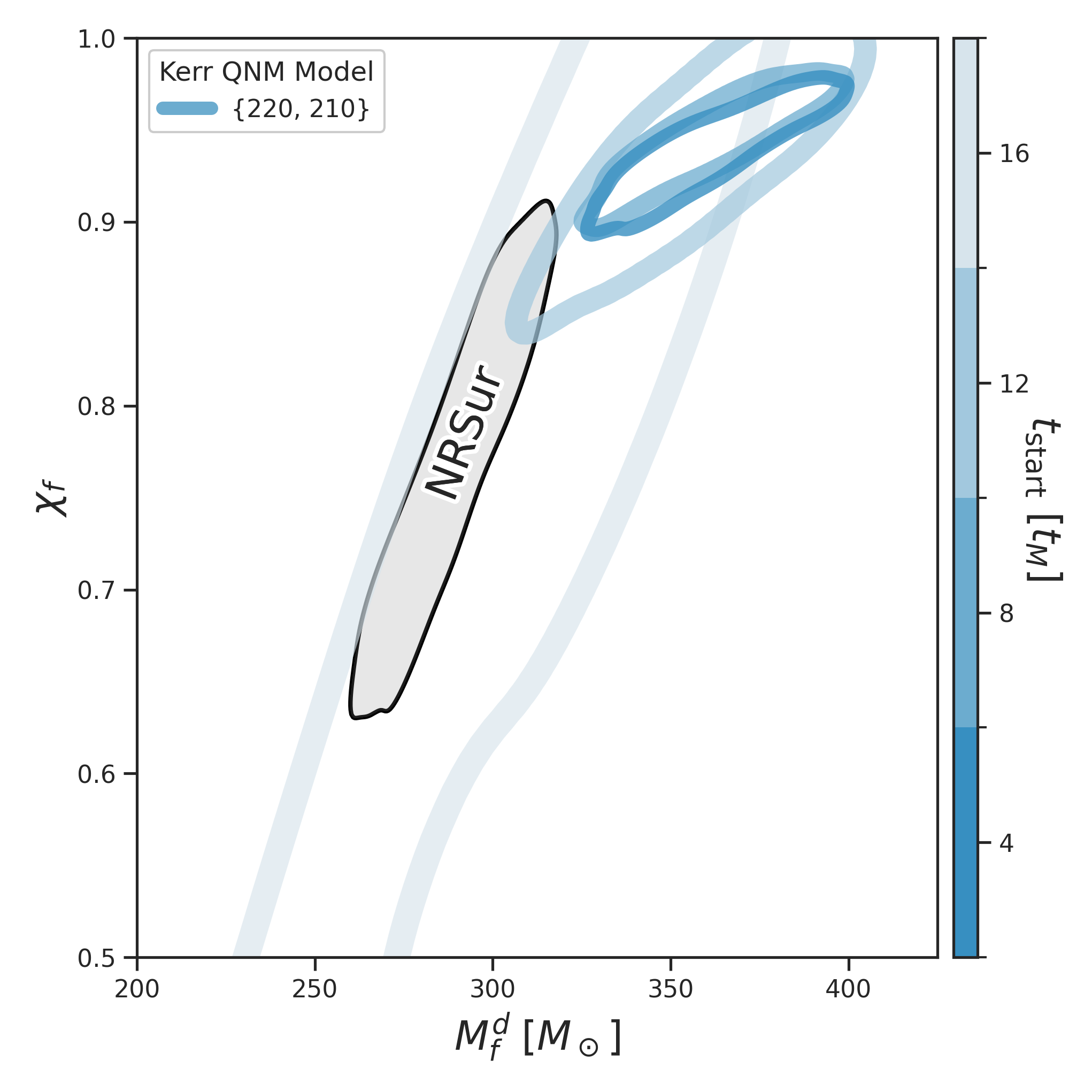}
    \caption{Detector-frame remnant mass~(x-axis) and spin~(y-axis) of NRSur7dq4 compared against the model which may be a more accurate fit, the Kerr 
    \{220,~210\} model. The \{220,~210\} model finds higher remnant spins and masses, implying highly spinning but potentially more spin-aligned binary components than the NRSur7dq4 result shown in Fig.~\ref{fig:nrsur_spindisk}. If other studies consider using our QNM posteriors for parameter estimates of GW231123, we suggest taking a posterior from one of the fits around the times where the Kerr model is tightest and most self-consistent, somewhere between 4 and 12~$t_M$. }
    \label{fig:mchi_210}
\end{figure}

Taking all of the evidence together, there is a case to be made for significant NRSur7dq4 systematic errors affecting inferences of GW231123, although we cannot definitively conclude that this is taking place. Significant systematic error in the IMR parameter estimates would have implications for population and binary formation channel studies using GW231123~\cite{Mandel:2025qnh, Kiroglu:2025vqy, Tong:2025wpz, Paiella:2025qld, Li:2025pyo, Baumgarte:2025syh, Popa:2025dpz, Gottlieb:2025ugy, Delfavero:2025lup, Croon:2025gol, Bartos:2025pkv}. If indeed our 210 QNM fits are more accurate, future studies might consider using the remnant mass and spin estimates of these fits in addition to those of IMR models. The 210 QNM fits imply a higher-spinning and higher-mass remnant than NRSur7dq4 and all other available IMR models~\cite{Estelles:2021gvs, Thompson:2023ase, Ramos-Buades:2023ehm, Varma:2019csw_NRsurpaper, Pratten:2020ceb}. We provide the parameters of the 210 models in the data release of this paper~\cite{datarelease_thispaper}, and for convenience show the mass and spin of the \{220,~210\} model in Fig.~\ref{fig:mchi_210}.

There are alternative BBH dynamics which we cannot conclusively rule out and which may be contributing to the excitation of the QNMs we observe. These dynamics include eccentricity. While we cannot rule out eccentricity, we believe that it is unlikely that eccentricity alone could be responsible for the QNM excitations in GW231123, as eccentricity does not leave a strong imprint in the relative QNM amplitudes unless the eccentricity is so extreme as for the merger to be a nearly head-on collision~\cite{Siegel_eccentricity_unpublished}. Such highly eccentric events are astrophysically unlikely~\cite{Vijaykumar:2024piy}, and also have a detection penalty because their gravitational wave emission intensity can be up to roughly an order of magnitude smaller than their quasicircular counterparts~\cite{Siegel_eccentricity_unpublished, CalderonBustillo:2020xms}.

Gravitational lensing might also be capable of altering the QNM amplitudes, although we are not aware of a lensing study which has shown exactly how the QNM amplitudes and phases might be affected. For example, in the case of millilensing, multiple copies of the signal are overlaid with an amplification, a phase shift, and a time shift~\cite{Liu:2023ikc}. In the regime of the signal dominated by damped sinusoids, added copies of damped sinusoids would produce the same number of observed modes as in the unlensed signal and with the same frequencies, but with different amplitudes and phases, see e.g. Eq.~B.3 of Ref.~\cite{Siegel:2024jqd}.

Regarding our test of general relativity, we note that while the parametrization of Eq.~\eqref{eq:TGR_fgamma} is reasonably motivated by our knowledge of the non-Kerr frequency shifts to QNMs induced in some classes of beyond-GR theories~\cite{Hussain:2022ins, Li:2022pcy}, it may not be optimal for all beyond-GR models~\cite{Crescimbeni:2024sam, Lestingi:2025jyb,Li:2023ulk, Maenaut:2024oci}. Also, when allowing non-Kerr deviations of multiple QNMs which are close in frequency and damping rate, avoiding label-switching is a data analysis challenge. One solution for generic label-switching problems is provided in Ref.~\cite{Buscicchio:2019rir}, which we may consider for future works. And finally, we find that it matters which QNMs in our models are given non-Kerr freedoms. Empirically we find that the method which results in the most precise TGR posteriors is to first look at the Kerr model and leave as-is its best-measured QNM, while adding freedoms to all other QNMs in the model. However, to our knowledge the formalism for this choice has not been rigorously derived in the literature.

\section{\label{sec:conclusion}Conclusion}

By fitting quasinormal modes to GW231123, we confirm strong statistical evidence preferring two-mode fits over one-mode fits for a wide range of fitting times like in Ref.~\cite{GW231123_paper}, as shown in Fig.~\ref{fig:LOOfig}. We confirm that the two modes are similarly long-lived and have comparable amplitudes, as shown in Fig.~\ref{fig:amplitude_timescan_fundamental}. Comparison of the posteriors from our QNM fits with the strain peak time of the overall best-performing inspiral-merger-ringdown model, NRSur7dq4, shows that there is preference for two QNMs until $14~t_M$ after the peak, and evidence of a single QNM as late as  $28~t_M$. Comparison with NRSur7dq4 also identifies the dominant two QNMs as the ${(\ell,m)=(2,2)}$ and $(2,0)$ fundamental prograde modes, as shown in Fig.~\ref{fig:22_20_fgamma}. However, we find that models with the ${(\ell,m)=(2,2)}$ and $(2,1)$ QNMs instead may be better fits to the data and more consistent with the Kerr frequency and damping rate spectrum, as shown in Figs.~\ref{fig:LOOfig},~\ref{fig:TGR_fundamentals}, even though they are in tension with NRSur7dq4 as shown in Fig.~\ref{fig:22_20_fgamma}. The ${(\ell,m)=(2,2)}$ and $(2,1)$ QNM model may also be more physically plausible than the ${(\ell,m)=(2,2)}$ and $(2,0)$ QNM model, if we assume that GW231123 is sourced by a quasicircular precessing BBH coalescence~\cite{Zhu:2023fnf, OShaughnessy:2012iol, Hamilton:2023znn, Nobili:2025ydt}. Lastly, we find confident amplitude measurements and consistent remnant mass and spin inferences when adding an ${(\ell,m,n)=(2,2,1)}$ prograde overtone to our models (Fig.~\ref{fig:amplitude_timescan_overtone}), and the overtone is found to improve validations of the Kerr spectrum as shown in Fig.~\ref{fig:TGR_overtones}.

The fact that we find a QNM model which does not fully agree with NRSur7dq4 (or any of the available IMR models, for that matter) but seems to fit the data better and is more in agreement with the Kerr spectrum raises the possibility of significant systematic errors in the parameter estimates of the NRSur7dq4 model. If NRSur7dq4 is indeed systematically biased and our QNM fits more accurately characterize the remnant mass and spin of GW231123, this has implications for population and binary formation channel studies making use of GW231123. In the data release for this paper~\cite{datarelease_thispaper}, we provide posteriors for the ${(\ell,m)=(2,2)}$ and $(2,1)$ QNM models which may give more accurate estimates of the remnant mass and spin. As shown in Fig.~\ref{fig:mchi_210}, these QNM models suggest that the remnant of GW231123 may have been more highly spinning and heavier than what is found by the inspiral-merger-ringdown models, still implying highly-spinning progenitors but less spin-orbit misalignment. That being said, we cannot definitively rule out the possibility of systematics in our QNM fits, bias caused by non-stationary noise features, or unmodeled physics producing systematic errors.

The possibility of NRSur7dq4 systematics makes interpretation of this signal and its associated constraints of the Kerr metric exceptionally difficult. Nonetheless, if we take the GW231123 QNM model which best fits the signal and is most consistent with the Kerr spectrum, and we compare it at equivalent fitting times relative to the peak strain with QNM fits of GW250114~\cite{GW250114_paper1, GW250114_paper2}, the loudest signal to date, we recover much better constraints of the Kerr metric: our GW231123 fits constrain the Kerr frequencies to within $\pm10\%$ at the $90\%$ credible level (shown in Figs.~\ref{fig:TGR_fundamentals},~\ref{fig:TGR_overtones}, Table~\ref{tab:TGR-deltas}), as opposed to the $\pm30\%$ reported for GW250114. We also report sub-$10\%$ and $30\%$ constraints for one fundamental mode and overtone respectively when fitting at earlier times, although more work is required to definitively confirm whether these constraints are physically meaningful.

The tensions we find between QNM and IMR fits highlight the utility of QNM fits as probes of astrophysical parameters in addition to providing tests of general relativity. The flexibility of QNM models makes them especially useful for probing astrophysical parameters in parts of parameter space where more standard models are known to struggle with systematic errors.

While preparing this manuscript, we became aware of another QNM analysis of GW231123~\cite{Wang:2025rvn}. Ref.~\cite{Wang:2025rvn} finds preference for two modes in the signal, similar to what is reported in Ref.~\cite{GW231123_paper} and our own work, and also reports a test of general relativity. We use different methodologies from both Refs.~\cite{Wang:2025rvn}~and~\cite{GW231123_paper} and do not reach identical conclusions. We consider a broader range of fitting times than both analyses, and we also consider models with combinations of QNMs that are not explored in either work. While all analyses agree on the preference for multimodal fits to the data, we do not interpret this preference in the same way. Ref.~\cite{Wang:2025rvn} broadly reports similar results as contained in the text of Ref.~\cite{GW231123_paper} and concludes that there is a fitting preference for the $\{220,~200\}$ model, whereas our analysis has indications of disfavoring this model in favor of models which use the 210 instead of the 200. Ref.~\cite{Wang:2025rvn} claims that the 200 model is consistent with the Kerr metric, although we do not find the evidence for the Kerr consistency of this model to be as definitive as for the 210 models. Neither Ref.~\cite{GW231123_paper} or Ref.~\cite{Wang:2025rvn} uses the QNM analysis to suggest possible NRSur7dq4 systematics as we do.

In Appendix~\ref{sec:Appendix_320fits}, we discuss QNM fits using the 320 and 321 modes. These models which include the 320 can achieve better goodness-of-fit than the QNM models with the 200 that agree with NRSur7dq4, but they also have seemingly unphysical features which make us disfavor them when compared with the other models in the main text.

\begin{acknowledgments}
We thank Alberto Vecchio and Keefe Mitman for insightful conversation. We also thank Lorenzo Pompili for helpful comments during internal LVK review, as well as Luis Lehner.

This material is based upon work supported by NSF's LIGO Laboratory which is a major facility fully funded by the National Science Foundation.
This research has made use of data or software obtained from the Gravitational Wave Open Science Center (gwosc.org), a service of the LIGO Scientific Collaboration, the Virgo Collaboration, and KAGRA. This material is based upon work supported by NSF's LIGO Laboratory which is a major facility fully funded by the National Science Foundation, as well as the Science and Technology Facilities Council (STFC) of the United Kingdom, the Max-Planck-Society (MPS), and the State of Niedersachsen/Germany for support of the construction of Advanced LIGO and construction and operation of the GEO600 detector. Additional support for Advanced LIGO was provided by the Australian Research Council. Virgo is funded, through the European Gravitational Observatory (EGO), by the French Centre National de Recherche Scientifique (CNRS), the Italian Istituto Nazionale di Fisica Nucleare (INFN) and the Dutch Nikhef, with contributions by institutions from Belgium, Germany, Greece, Hungary, Ireland, Japan, Monaco, Poland, Portugal, Spain. KAGRA is supported by Ministry of Education, Culture, Sports, Science and Technology (MEXT), Japan Society for the Promotion of Science (JSPS) in Japan; National Research Foundation (NRF) and Ministry of Science and ICT (MSIT) in Korea; Academia Sinica (AS) and National Science and Technology Council (NSTC) in Taiwan. This paper carries LIGO document number P2500629.

This research was supported
in part by Perimeter Institute for Theoretical Physics.
Research at Perimeter Institute is supported in part by
the Government of Canada through the Department of
Innovation, Science and Economic Development and by
the Province of Ontario through the Ministry of Colleges
and Universities.

The Flatiron Institute is a division of the Simons Foundation. H.S. was supported by Yuri Levin's Simons Investigator Award 827103 while conducting this research.

\textit{Software:} \textsc{ringdown}~\cite{ringdowncode}, \textsc{arviz}~\cite{arviz_2019}, \textsc{pymc}~\cite{Pymc}, \textsc{seaborn}~\cite{Seaborn}, \textsc{matplotlib}~\cite{matplotlib}, \textsc{jupyter}~\cite{jupyter}, \textsc{numpy}~\cite{numpy}, \textsc{scipy}~\cite{scipy}, \textsc{qnm}~\cite{qnmpackage_Stein},  \textsc{disbatch}~\cite{disbatch}, \textsc{pandas}~\cite{pandas}, \textsc{Python3}~\cite{Python3},
\textsc{ChatGPT}~\cite{openai2024chatgpt} (used for figure plotting code).

\end{acknowledgments}

\appendix

\section{\label{sec:Appendix_320fits} Alternative QNM models better-fitting than \{220,~200\}}
\begin{figure}
    \centering
    \includegraphics[width=\columnwidth]{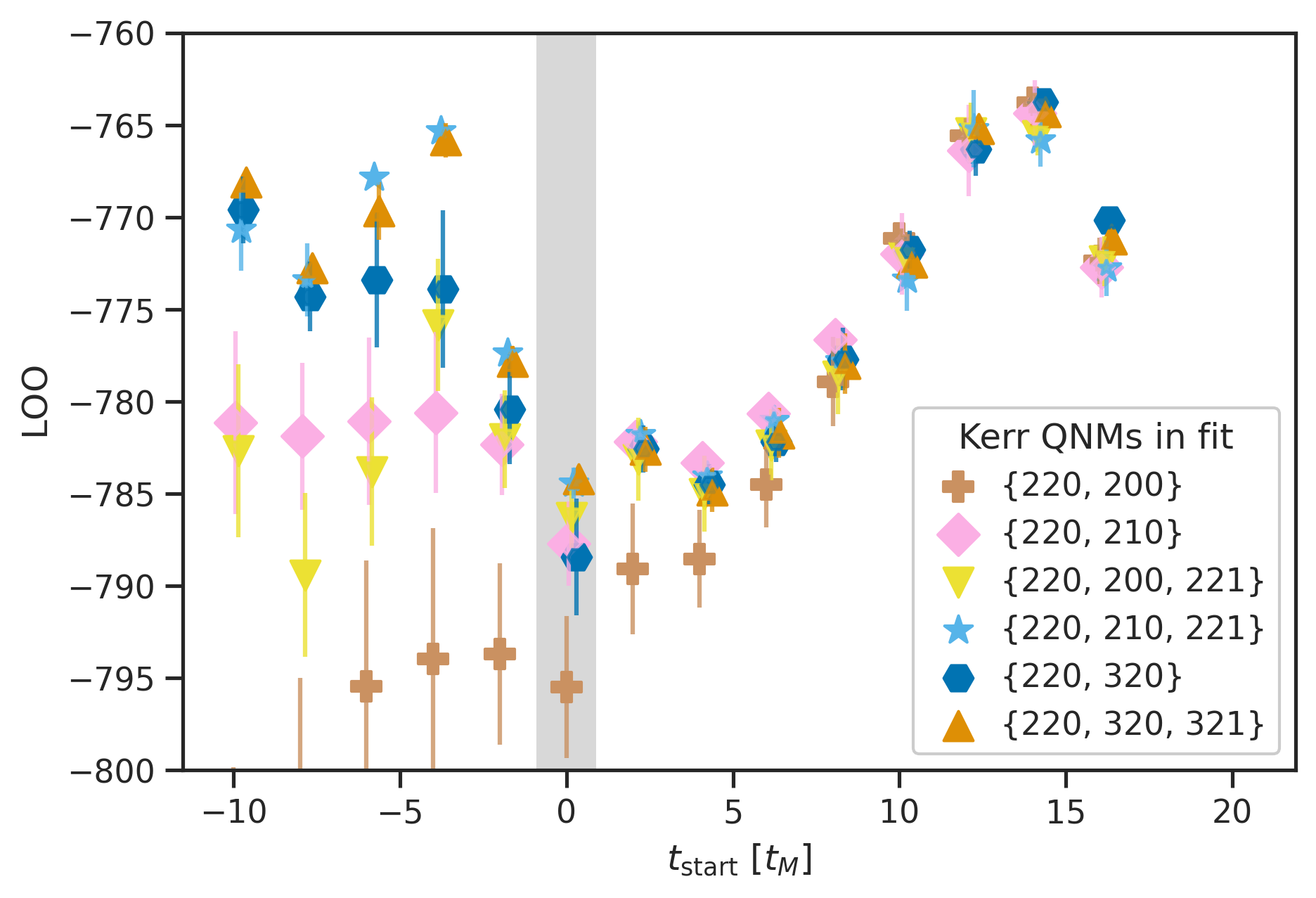}
    \caption{LOO now including fits with the 320. See Fig.~\ref{fig:LOOfig} for figure conventions. The $\{220,~320\}$ model performs as well as $\{220,~210\}$ from $-2~t_M$ onwards, and performs significantly better at earlier times. The $\{220,~320,~321\}$ and $\{220,~210,~221\}$ perform comparably.}
    \label{fig:LOO_320fits}
\end{figure}

The QNM models with 220 and 210 are not the only ones we find that outperform the $\{220,~200\}$ in terms of goodness of fit. We also find that the $\{220,~320\}$ and $\{220,~320,~321\}$ models perform as well as or better than the 200 models and at least as well as the 210 models, as shown in Fig.~\ref{fig:LOO_320fits}.

\begin{figure}
    \centering
    \includegraphics[width=\columnwidth]{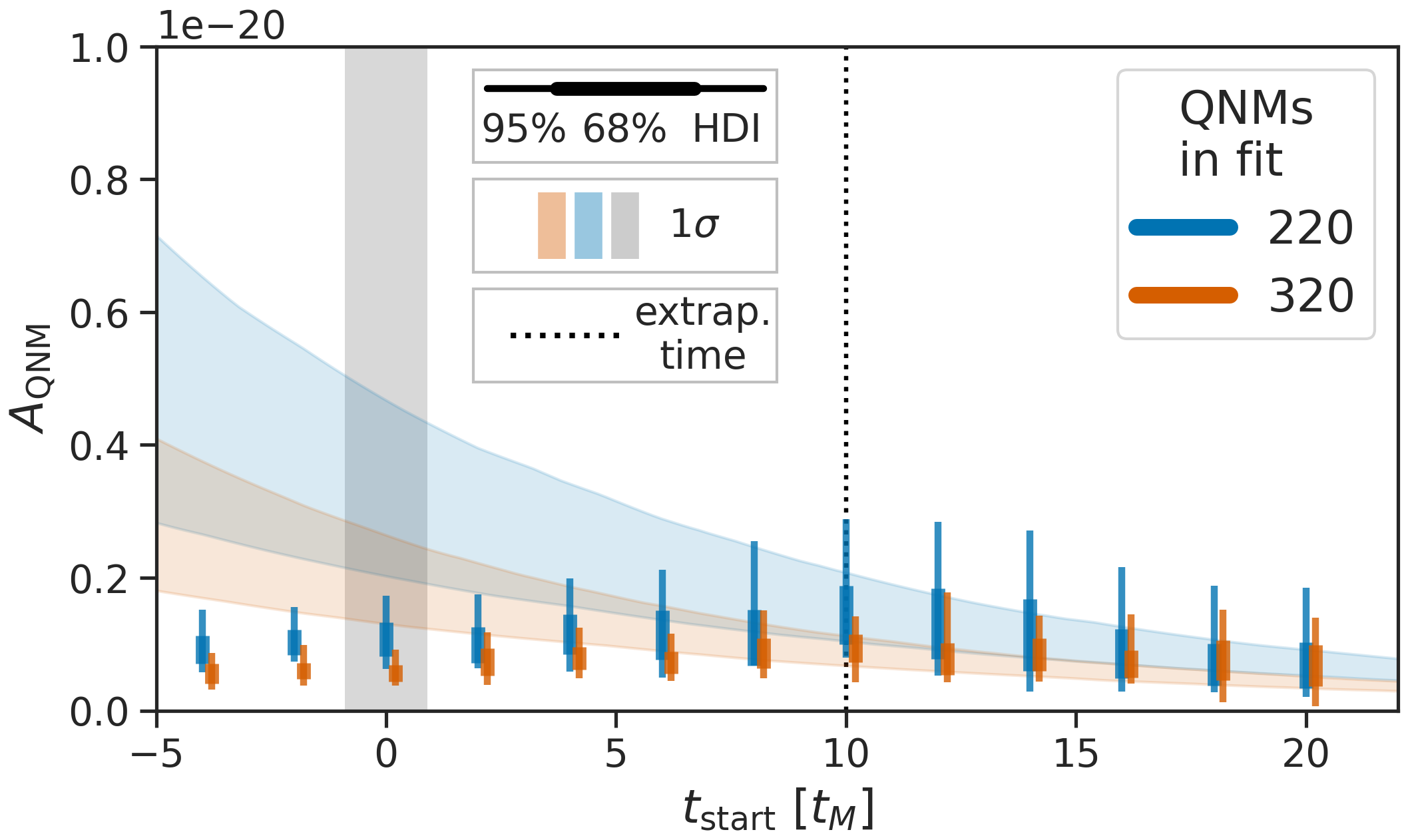}
    \includegraphics[width=\columnwidth]{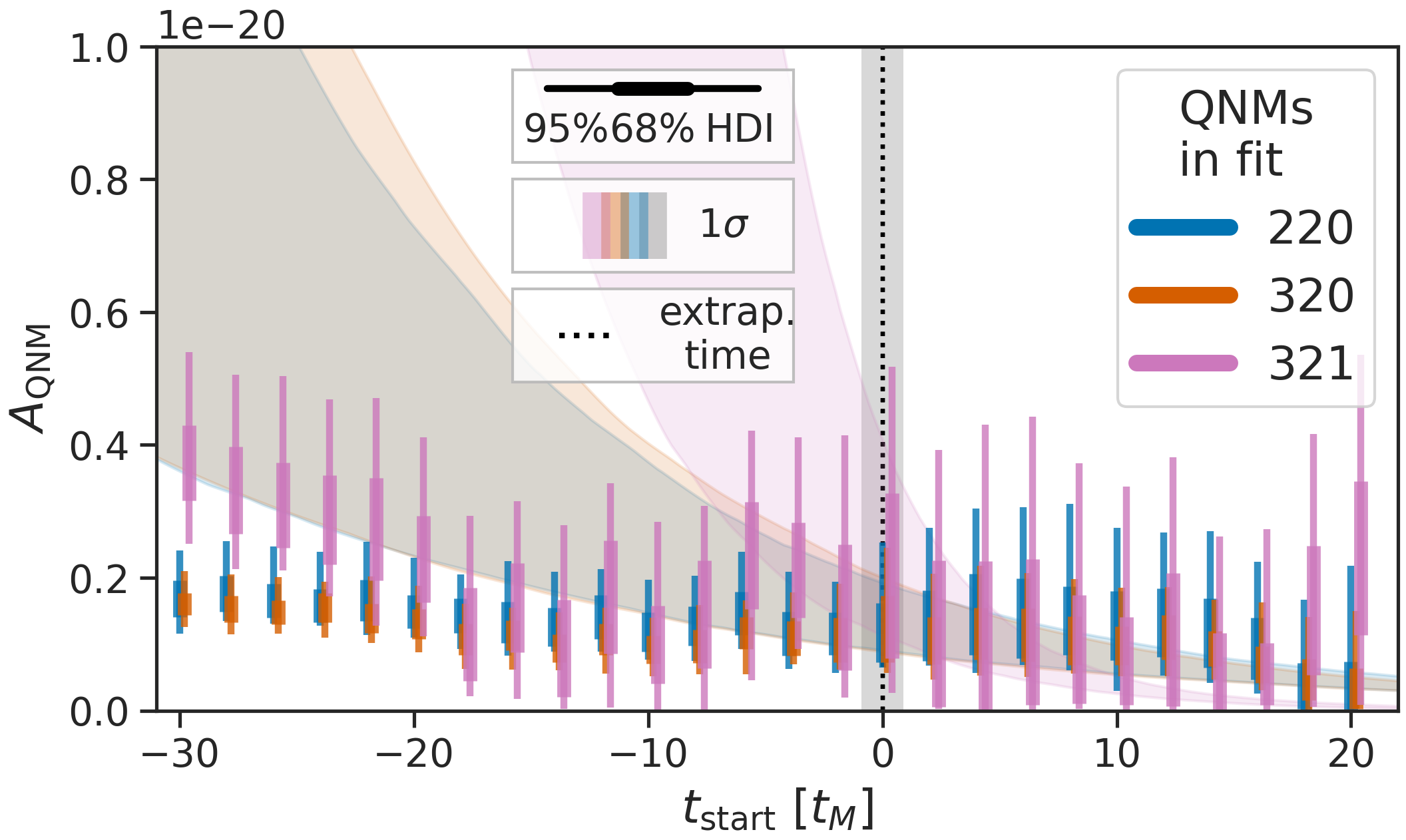}
    \caption{The amplitudes of $\{220,~320\}$ and $\{220,~320,~321\}$ models do not follow physically well-motivated decays as clearly as those of the $\{220,~210,~221\}$ and $\{220,~200,~221\}$ models shown in Figs.~\ref{fig:amplitude_timescan_fundamental} and ~\ref{fig:amplitude_timescan_overtone}, despite the model frequencies being consistent with Kerr as shown in Fig.~\ref{fig:TGR_320}. This is one piece of evidence which favors the models of the main text.}
    \label{fig:A_timescan_320}
\end{figure}

\begin{figure}
    \centering
    \includegraphics[width=\columnwidth]{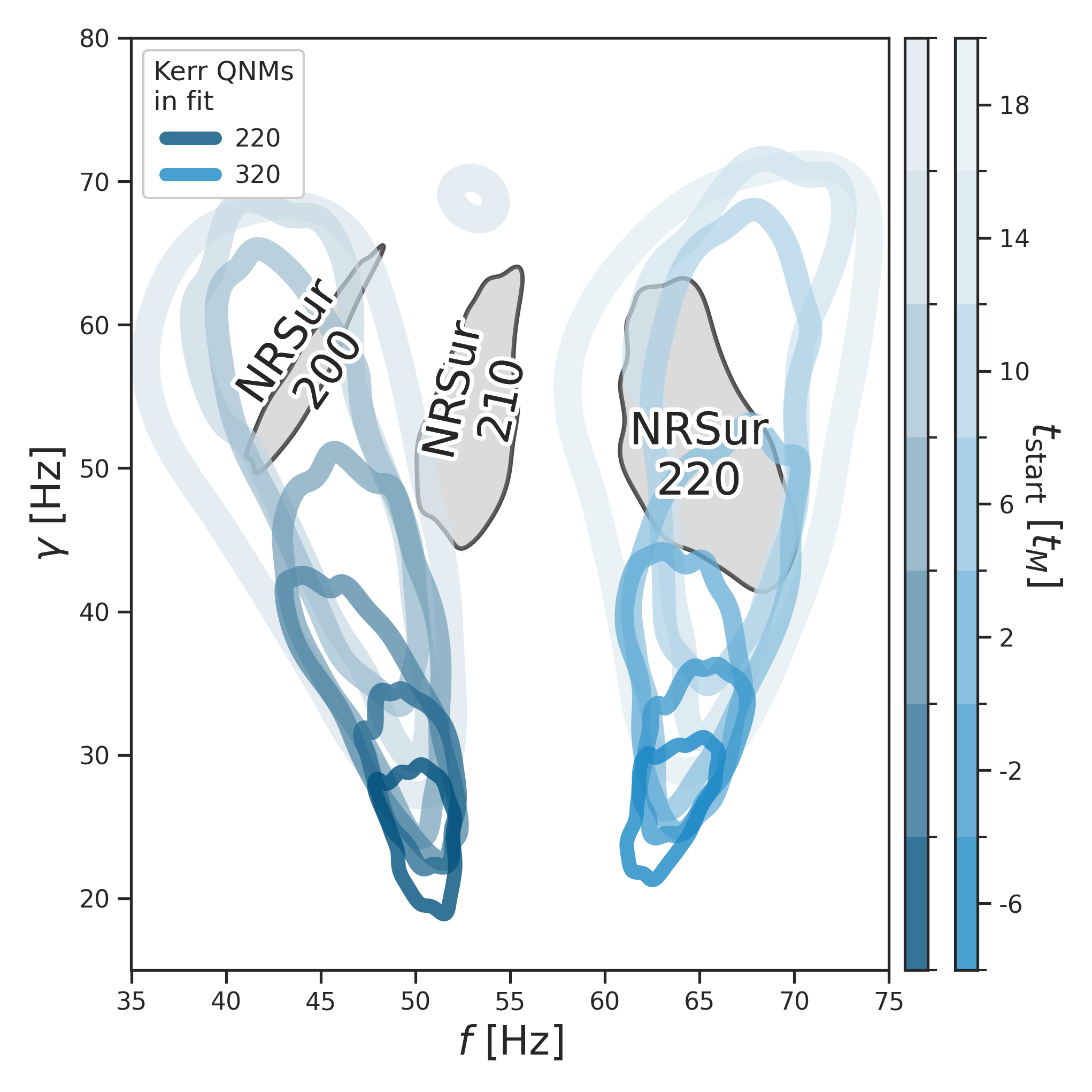}
    \includegraphics[width=\columnwidth]{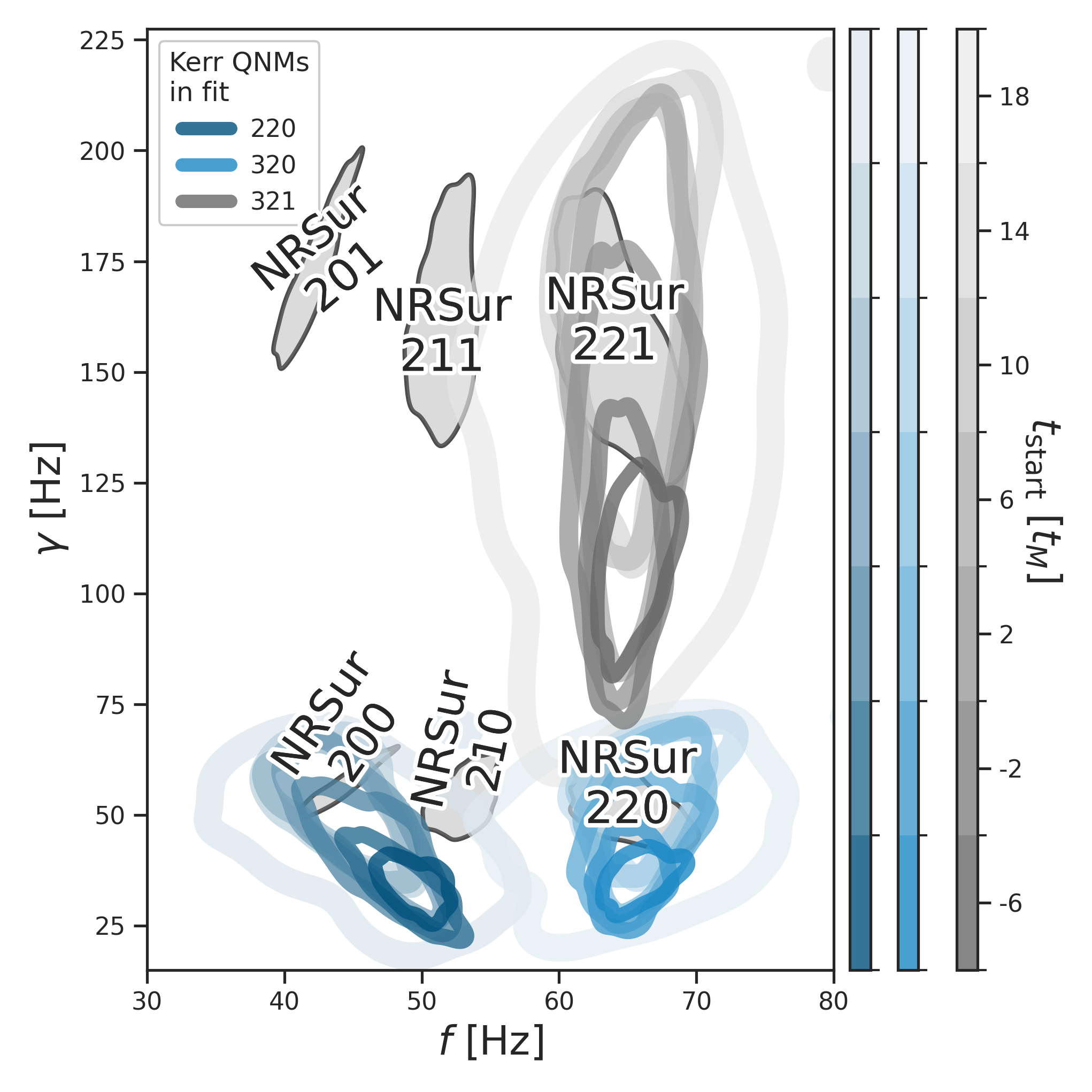}
    \caption{The frequencies of the 320 models are placed roughly in the same locations as those of the 210 models in the main text (Figs.~\ref{fig:22_20_fgamma},~\ref{fig:22_20_221_fgamma}), but for these 320 models the damping rates move more over time. Note that the mass prior for this model is different from those in the main text: it ranges from 250 to 550 $M_\odot$. }
    \label{fig:fgamma_320}
\end{figure}

Despite the promising goodness of fit of the 320 models, we suggest that their parameter estimates evolve over time in physically implausible ways which lead us to prefer the 210 models. First, looking at the amplitudes, we can see that the $\{220,~320\}$ amplitudes do not decay as expected until $10~t_M$, as shown in Fig.~\ref{fig:A_timescan_320}. The frequency and damping rate measurements of the $\{220,~320\}$ model are also shifting over time more than the $\{220,~210\}$ model, as shown in Fig.~\ref{fig:fgamma_320}. By $10~t_M$, there are no longer meaningful goodness of fit preferences between the 210, 200, and 320 models, as shown in Fig.~\ref{fig:LOO_320fits}. So, although at earlier times the $\{220,~320\}$ models may outperform the $\{220,~210\}$ models in terms of goodness of fit, this does not seem to be physically meaningful.

\begin{figure}
    \centering
    \includegraphics[width=\columnwidth]{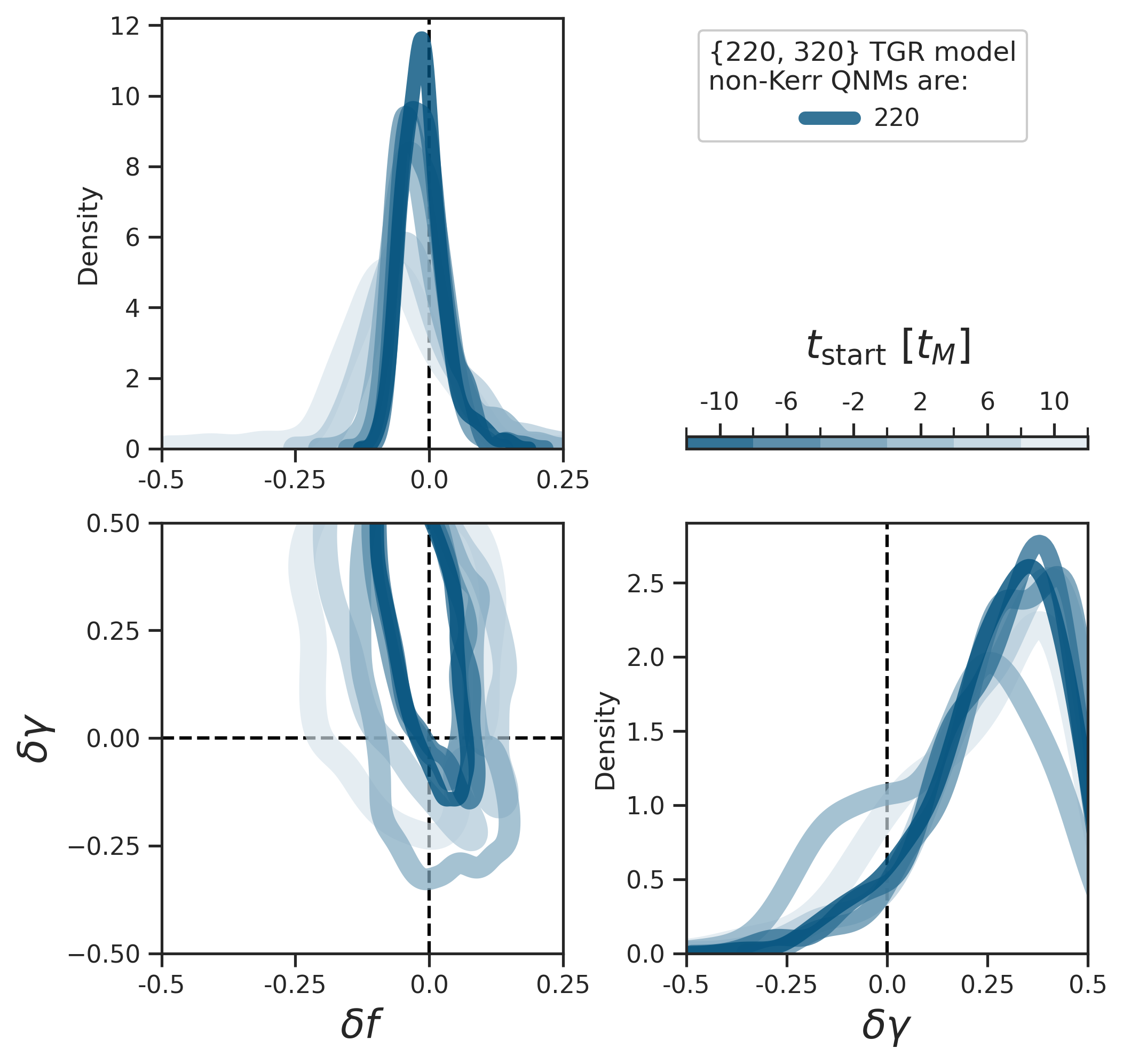}
    \includegraphics[width=\columnwidth]{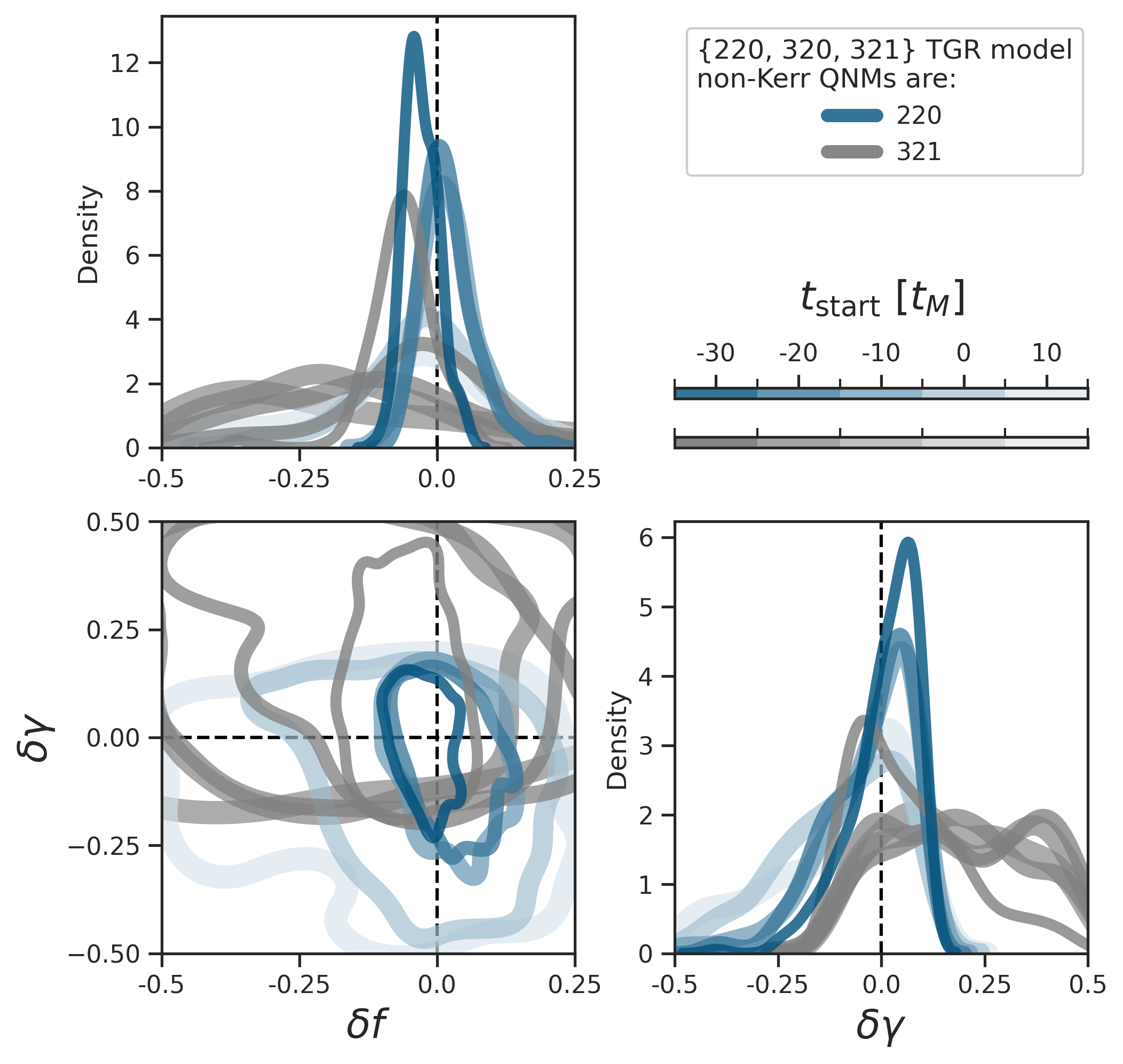}
    \caption{We find the 320 models to be consistent with the Kerr metric at implausibly early times, and over spans of time where the amplitudes of the Kerr models do not decay as expected of QNMs (as shown in Fig.~\ref{fig:A_timescan_320}). Validation of the Kerr metric alone does not seem sufficient to guarantee that the model is physically meaningful, as there seems to be evidence suggesting that this model does not have the physical behavior we expect of QNM models and thus is likely a less accurate description of the signal than those considered in the main text.}
    \label{fig:TGR_320}
\end{figure}

When including a 321 overtone in the model, similar parameter time evolutions occur. Again, the frequencies and damping rates of the modes in the model slowly shift over time, as shown in Fig.~\ref{fig:fgamma_320}. By contrast, the frequency and damping rate measurements of the $\{220,~210,~221\}$ and $\{220,~200,~221\}$ models are more self-consistent over time. Interestingly, when performing our test of the Kerr metric with the $\{220,~320,~321\}$ model, Kerr consistency can be found as early as $-30~t_M$ (Fig.~\ref{fig:TGR_320}) -- despite the fact that the amplitudes are not decaying as damped sinusoids, as shown in Fig.~\ref{fig:A_timescan_320}.

Ultimately, unusual behavior of these QNM fits aside, we appeal to our knowledge of IMR model systematics to claim that these 320 QNM models should not be considered on equal footing with those in the main text. If NRSur7dq4 is indeed systematically biased, its systematic uncertainty is unlikely to be many times that of the statistical uncertainty for GW231123 given the injection studies in Ref.~\cite{GW231123_paper}. If the 320 models were indeed the best description of the signal, it would imply that NRSur7dq4 was strongly dominated by systematic bias, to a degree that seems improbable. The time-dependent parameter behavior we see in these QNM models seems to highlight the difficult nature of our flexible analysis, and shows that consistency with the Kerr frequencies may not by itself be meaningful without consideration of the physicality of the other parameters of the model as motivated by theoretical understanding of black hole ringdown signals. Where the line is drawn as to exactly what constitutes physicality has an element of individual interpretation.



\clearpage
\bibliography{bib}

\providecommand{\noopsort}[1]{}\providecommand{\singleletter}[1]{#1}%
\begin{thebibliography}{133}%
\makeatletter
\providecommand \@ifxundefined [1]{%
 \@ifx{#1\undefined}
}%
\providecommand \@ifnum [1]{%
 \ifnum #1\expandafter \@firstoftwo
 \else \expandafter \@secondoftwo
 \fi
}%
\providecommand \@ifx [1]{%
 \ifx #1\expandafter \@firstoftwo
 \else \expandafter \@secondoftwo
 \fi
}%
\providecommand \natexlab [1]{#1}%
\providecommand \enquote  [1]{``#1''}%
\providecommand \bibnamefont  [1]{#1}%
\providecommand \bibfnamefont [1]{#1}%
\providecommand \citenamefont [1]{#1}%
\providecommand \href@noop [0]{\@secondoftwo}%
\providecommand \href [0]{\begingroup \@sanitize@url \@href}%
\providecommand \@href[1]{\@@startlink{#1}\@@href}%
\providecommand \@@href[1]{\endgroup#1\@@endlink}%
\providecommand \@sanitize@url [0]{\catcode `\\12\catcode `\$12\catcode `\&12\catcode `\#12\catcode `\^12\catcode `\_12\catcode `\%12\relax}%
\providecommand \@@startlink[1]{}%
\providecommand \@@endlink[0]{}%
\providecommand \url  [0]{\begingroup\@sanitize@url \@url }%
\providecommand \@url [1]{\endgroup\@href {#1}{\urlprefix }}%
\providecommand \urlprefix  [0]{URL }%
\providecommand \Eprint [0]{\href }%
\providecommand \doibase [0]{https://doi.org/}%
\providecommand \selectlanguage [0]{\@gobble}%
\providecommand \bibinfo  [0]{\@secondoftwo}%
\providecommand \bibfield  [0]{\@secondoftwo}%
\providecommand \translation [1]{[#1]}%
\providecommand \BibitemOpen [0]{}%
\providecommand \bibitemStop [0]{}%
\providecommand \bibitemNoStop [0]{.\EOS\space}%
\providecommand \EOS [0]{\spacefactor3000\relax}%
\providecommand \BibitemShut  [1]{\csname bibitem#1\endcsname}%
\let\auto@bib@innerbib\@empty
\bibitem [{\citenamefont {Abbott}\ \emph {et~al.}(2025)\citenamefont {Abbott} \emph {et~al.}}]{GW231123_paper}%
  \BibitemOpen
  \bibfield  {author} {\bibinfo {author} {\bibfnamefont {R.}~\bibnamefont {Abbott}} \emph {et~al.} (\bibinfo {collaboration} {LIGO Scientific, VIRGO, KAGRA}),\ }\href@noop {} {\bibfield  {journal} {\bibinfo  {journal} {arXiv eprints}\ } (\bibinfo {year} {2025})},\ \Eprint {https://arxiv.org/abs/2507.08219} {arXiv:2507.08219 [astro-ph.HE]} \BibitemShut {NoStop}%
\bibitem [{\citenamefont {Capote}\ \emph {et~al.}(2025)\citenamefont {Capote} \emph {et~al.}}]{PhysRevD.111.062002_LIGOdetector}%
  \BibitemOpen
  \bibfield  {author} {\bibinfo {author} {\bibfnamefont {E.}~\bibnamefont {Capote}} \emph {et~al.},\ }\href {https://doi.org/10.1103/PhysRevD.111.062002} {\bibfield  {journal} {\bibinfo  {journal} {Phys. Rev. D}\ }\textbf {\bibinfo {volume} {111}},\ \bibinfo {pages} {062002} (\bibinfo {year} {2025})}\BibitemShut {NoStop}%
\bibitem [{\citenamefont {Acernese}\ \emph {et~al.}(2022)\citenamefont {Acernese} \emph {et~al.}}]{Virgo:2022fxr_virgopaper}%
  \BibitemOpen
  \bibfield  {author} {\bibinfo {author} {\bibfnamefont {F.}~\bibnamefont {Acernese}} \emph {et~al.} (\bibinfo {collaboration} {Virgo}),\ }\href@noop {} {\bibfield  {journal} {\bibinfo  {journal} {arXiv eprints}\ } (\bibinfo {year} {2022})},\ \Eprint {https://arxiv.org/abs/2205.01555} {arXiv:2205.01555 [gr-qc]} \BibitemShut {NoStop}%
\bibitem [{\citenamefont {Akutsu}\ \emph {et~al.}(2020)\citenamefont {Akutsu} \emph {et~al.}}]{10.1093/ptep/ptaa125_KAGRApaper}%
  \BibitemOpen
  \bibfield  {author} {\bibinfo {author} {\bibfnamefont {T.}~\bibnamefont {Akutsu}} \emph {et~al.},\ }\href {https://doi.org/10.1093/ptep/ptaa125} {\bibfield  {journal} {\bibinfo  {journal} {Progress of Theoretical and Experimental Physics}\ }\textbf {\bibinfo {volume} {2021}},\ \bibinfo {pages} {05A101} (\bibinfo {year} {2020})},\ \Eprint {https://arxiv.org/abs/https://academic.oup.com/ptep/article-pdf/2021/5/05A101/37974994/ptaa125.pdf} {https://academic.oup.com/ptep/article-pdf/2021/5/05A101/37974994/ptaa125.pdf} \BibitemShut {NoStop}%
\bibitem [{\citenamefont {Aasi}\ \emph {et~al.}(2015)\citenamefont {Aasi} \emph {et~al.}}]{LIGOScientific:2014pky}%
  \BibitemOpen
  \bibfield  {author} {\bibinfo {author} {\bibfnamefont {J.}~\bibnamefont {Aasi}} \emph {et~al.} (\bibinfo {collaboration} {LIGO Scientific}),\ }\href {https://doi.org/10.1088/0264-9381/32/7/074001} {\bibfield  {journal} {\bibinfo  {journal} {Class. Quant. Grav.}\ }\textbf {\bibinfo {volume} {32}},\ \bibinfo {pages} {074001} (\bibinfo {year} {2015})},\ \Eprint {https://arxiv.org/abs/1411.4547} {arXiv:1411.4547 [gr-qc]} \BibitemShut {NoStop}%
\bibitem [{\citenamefont {Abac}\ \emph {et~al.}(2025{\natexlab{a}})\citenamefont {Abac} \emph {et~al.}}]{LIGOScientific:2025slb}%
  \BibitemOpen
  \bibfield  {author} {\bibinfo {author} {\bibfnamefont {A.~G.}\ \bibnamefont {Abac}} \emph {et~al.} (\bibinfo {collaboration} {LIGO Scientific, VIRGO, KAGRA}),\ }\href@noop {} {\bibinfo {title} {{GWTC-4.0: Updating the Gravitational-Wave Transient Catalog with Observations from the First Part of the Fourth LIGO-Virgo-KAGRA Observing Run}}} (\bibinfo {year} {2025}{\natexlab{a}}),\ \Eprint {https://arxiv.org/abs/2508.18082} {arXiv:2508.18082 [gr-qc]} \BibitemShut {NoStop}%
\bibitem [{\citenamefont {Varma}\ \emph {et~al.}(2019)\citenamefont {Varma}, \citenamefont {Field}, \citenamefont {Scheel}, \citenamefont {Blackman}, \citenamefont {Gerosa}, \citenamefont {Stein}, \citenamefont {Kidder},\ and\ \citenamefont {Pfeiffer}}]{Varma:2019csw_NRsurpaper}%
  \BibitemOpen
  \bibfield  {author} {\bibinfo {author} {\bibfnamefont {V.}~\bibnamefont {Varma}}, \bibinfo {author} {\bibfnamefont {S.~E.}\ \bibnamefont {Field}}, \bibinfo {author} {\bibfnamefont {M.~A.}\ \bibnamefont {Scheel}}, \bibinfo {author} {\bibfnamefont {J.}~\bibnamefont {Blackman}}, \bibinfo {author} {\bibfnamefont {D.}~\bibnamefont {Gerosa}}, \bibinfo {author} {\bibfnamefont {L.~C.}\ \bibnamefont {Stein}}, \bibinfo {author} {\bibfnamefont {L.~E.}\ \bibnamefont {Kidder}},\ and\ \bibinfo {author} {\bibfnamefont {H.~P.}\ \bibnamefont {Pfeiffer}},\ }\href {https://doi.org/10.1103/PhysRevResearch.1.033015} {\bibfield  {journal} {\bibinfo  {journal} {Phys. Rev. Research.}\ }\textbf {\bibinfo {volume} {1}},\ \bibinfo {pages} {033015} (\bibinfo {year} {2019})},\ \Eprint {https://arxiv.org/abs/1905.09300} {arXiv:1905.09300 [gr-qc]} \BibitemShut {NoStop}%
\bibitem [{\citenamefont {Talbot}\ \emph {et~al.}(2025)\citenamefont {Talbot} \emph {et~al.}}]{Talbot:2025vth}%
  \BibitemOpen
  \bibfield  {author} {\bibinfo {author} {\bibfnamefont {C.}~\bibnamefont {Talbot}} \emph {et~al.},\ }\href@noop {} {\bibinfo {title} {{Inference with finite time series II: the window strikes back}}} (\bibinfo {year} {2025}),\ \Eprint {https://arxiv.org/abs/2508.11091} {arXiv:2508.11091 [gr-qc]} \BibitemShut {NoStop}%
\bibitem [{\citenamefont {Isi}\ and\ \citenamefont {Farr}(2021)}]{Isi:2021iql_analyzingbhringdowns}%
  \BibitemOpen
  \bibfield  {author} {\bibinfo {author} {\bibfnamefont {M.}~\bibnamefont {Isi}}\ and\ \bibinfo {author} {\bibfnamefont {W.~M.}\ \bibnamefont {Farr}},\ }\href@noop {} {\bibfield  {journal} {\bibinfo  {journal} {arXiv eprints}\ } (\bibinfo {year} {2021})},\ \Eprint {https://arxiv.org/abs/2107.05609} {arXiv:2107.05609 [gr-qc]} \BibitemShut {NoStop}%
\bibitem [{\citenamefont {Siegel}\ \emph {et~al.}(2025)\citenamefont {Siegel}, \citenamefont {Isi},\ and\ \citenamefont {Farr}}]{Siegel:2024jqd}%
  \BibitemOpen
  \bibfield  {author} {\bibinfo {author} {\bibfnamefont {H.}~\bibnamefont {Siegel}}, \bibinfo {author} {\bibfnamefont {M.}~\bibnamefont {Isi}},\ and\ \bibinfo {author} {\bibfnamefont {W.~M.}\ \bibnamefont {Farr}},\ }\href {https://doi.org/10.1103/PhysRevD.111.044070} {\bibfield  {journal} {\bibinfo  {journal} {Phys. Rev. D}\ }\textbf {\bibinfo {volume} {111}},\ \bibinfo {pages} {044070} (\bibinfo {year} {2025})},\ \Eprint {https://arxiv.org/abs/2410.02704} {arXiv:2410.02704 [gr-qc]} \BibitemShut {NoStop}%
\bibitem [{\citenamefont {Abac}\ \emph {et~al.}(2025{\natexlab{b}})\citenamefont {Abac} \emph {et~al.}}]{GW250114_paper1}%
  \BibitemOpen
  \bibfield  {author} {\bibinfo {author} {\bibfnamefont {A.~G.}\ \bibnamefont {Abac}} \emph {et~al.} (\bibinfo {collaboration} {KAGRA, Virgo, LIGO Scientific}),\ }\href {https://doi.org/10.1103/kw5g-d732} {\bibfield  {journal} {\bibinfo  {journal} {Phys. Rev. Lett.}\ }\textbf {\bibinfo {volume} {135}},\ \bibinfo {pages} {111403} (\bibinfo {year} {2025}{\natexlab{b}})},\ \Eprint {https://arxiv.org/abs/2509.08054} {arXiv:2509.08054 [gr-qc]} \BibitemShut {NoStop}%
\bibitem [{GW2(2025)}]{GW250114_paper2}%
  \BibitemOpen
  \href@noop {} {\bibfield  {journal} {\bibinfo  {journal} {arXiv eprints}\ } (\bibinfo {year} {2025})},\ \Eprint {https://arxiv.org/abs/2509.08099} {arXiv:2509.08099 [gr-qc]} \BibitemShut {NoStop}%
\bibitem [{\citenamefont {Teukolsky}(1973)}]{Teukolsky:1973ha}%
  \BibitemOpen
  \bibfield  {author} {\bibinfo {author} {\bibfnamefont {S.~A.}\ \bibnamefont {Teukolsky}},\ }\href {https://doi.org/10.1086/152444} {\bibfield  {journal} {\bibinfo  {journal} {Astrophys. J.}\ }\textbf {\bibinfo {volume} {185}},\ \bibinfo {pages} {635} (\bibinfo {year} {1973})}\BibitemShut {NoStop}%
\bibitem [{\citenamefont {Dreyer}\ \emph {et~al.}(2004{\natexlab{a}})\citenamefont {Dreyer}, \citenamefont {Kelly}, \citenamefont {Krishnan}, \citenamefont {Finn}, \citenamefont {Garrison},\ and\ \citenamefont {Lopez-Aleman}}]{testgr_ringdown_2}%
  \BibitemOpen
  \bibfield  {author} {\bibinfo {author} {\bibfnamefont {O.}~\bibnamefont {Dreyer}}, \bibinfo {author} {\bibfnamefont {B.~J.}\ \bibnamefont {Kelly}}, \bibinfo {author} {\bibfnamefont {B.}~\bibnamefont {Krishnan}}, \bibinfo {author} {\bibfnamefont {L.~S.}\ \bibnamefont {Finn}}, \bibinfo {author} {\bibfnamefont {D.}~\bibnamefont {Garrison}},\ and\ \bibinfo {author} {\bibfnamefont {R.}~\bibnamefont {Lopez-Aleman}},\ }\href {https://doi.org/10.1088/0264-9381/21/4/003} {\bibfield  {journal} {\bibinfo  {journal} {Class. Quant. Grav.}\ }\textbf {\bibinfo {volume} {21}},\ \bibinfo {pages} {787} (\bibinfo {year} {2004}{\natexlab{a}})},\ \Eprint {https://arxiv.org/abs/gr-qc/0309007} {arXiv:gr-qc/0309007} \BibitemShut {NoStop}%
\bibitem [{\citenamefont {Gossan}\ \emph {et~al.}(2012{\natexlab{a}})\citenamefont {Gossan}, \citenamefont {Veitch},\ and\ \citenamefont {Sathyaprakash}}]{testgr_ringdownbayesian}%
  \BibitemOpen
  \bibfield  {author} {\bibinfo {author} {\bibfnamefont {S.}~\bibnamefont {Gossan}}, \bibinfo {author} {\bibfnamefont {J.}~\bibnamefont {Veitch}},\ and\ \bibinfo {author} {\bibfnamefont {B.~S.}\ \bibnamefont {Sathyaprakash}},\ }\href {https://doi.org/10.1103/PhysRevD.85.124056} {\bibfield  {journal} {\bibinfo  {journal} {Phys. Rev. D}\ }\textbf {\bibinfo {volume} {85}},\ \bibinfo {pages} {124056} (\bibinfo {year} {2012}{\natexlab{a}})},\ \Eprint {https://arxiv.org/abs/1111.5819} {arXiv:1111.5819 [gr-qc]} \BibitemShut {NoStop}%
\bibitem [{\citenamefont {Berti}\ \emph {et~al.}(2015)\citenamefont {Berti} \emph {et~al.}}]{testgr_overview}%
  \BibitemOpen
  \bibfield  {author} {\bibinfo {author} {\bibfnamefont {E.}~\bibnamefont {Berti}} \emph {et~al.},\ }\href {https://doi.org/10.1088/0264-9381/32/24/243001} {\bibfield  {journal} {\bibinfo  {journal} {Class. Quant. Grav.}\ }\textbf {\bibinfo {volume} {32}},\ \bibinfo {pages} {243001} (\bibinfo {year} {2015})},\ \Eprint {https://arxiv.org/abs/1501.07274} {arXiv:1501.07274 [gr-qc]} \BibitemShut {NoStop}%
\bibitem [{\citenamefont {Zhu}\ \emph {et~al.}(2025)\citenamefont {Zhu} \emph {et~al.}}]{Zhu:2023fnf}%
  \BibitemOpen
  \bibfield  {author} {\bibinfo {author} {\bibfnamefont {H.}~\bibnamefont {Zhu}} \emph {et~al.},\ }\href {https://doi.org/10.1103/PhysRevD.111.064052} {\bibfield  {journal} {\bibinfo  {journal} {Phys. Rev. D}\ }\textbf {\bibinfo {volume} {111}},\ \bibinfo {pages} {064052} (\bibinfo {year} {2025})},\ \Eprint {https://arxiv.org/abs/2312.08588} {arXiv:2312.08588 [gr-qc]} \BibitemShut {NoStop}%
\bibitem [{\citenamefont {Kamaretsos}\ \emph {et~al.}(2012)\citenamefont {Kamaretsos}, \citenamefont {Hannam},\ and\ \citenamefont {Sathyaprakash}}]{Kamaretsos:2012bs}%
  \BibitemOpen
  \bibfield  {author} {\bibinfo {author} {\bibfnamefont {I.}~\bibnamefont {Kamaretsos}}, \bibinfo {author} {\bibfnamefont {M.}~\bibnamefont {Hannam}},\ and\ \bibinfo {author} {\bibfnamefont {B.}~\bibnamefont {Sathyaprakash}},\ }\href {https://doi.org/10.1103/PhysRevLett.109.141102} {\bibfield  {journal} {\bibinfo  {journal} {Phys. Rev. Lett.}\ }\textbf {\bibinfo {volume} {109}},\ \bibinfo {pages} {141102} (\bibinfo {year} {2012})},\ \Eprint {https://arxiv.org/abs/1207.0399} {arXiv:1207.0399 [gr-qc]} \BibitemShut {NoStop}%
\bibitem [{\citenamefont {Berti}\ \emph {et~al.}(2025)\citenamefont {Berti} \emph {et~al.}}]{Berti:2025hly}%
  \BibitemOpen
  \bibfield  {author} {\bibinfo {author} {\bibfnamefont {E.}~\bibnamefont {Berti}} \emph {et~al.},\ }\href@noop {} {\bibfield  {journal} {\bibinfo  {journal} {arXiv eprints}\ } (\bibinfo {year} {2025})},\ \Eprint {https://arxiv.org/abs/2505.23895} {arXiv:2505.23895 [gr-qc]} \BibitemShut {NoStop}%
\bibitem [{\citenamefont {Carullo}\ \emph {et~al.}(2019)\citenamefont {Carullo}, \citenamefont {Del~Pozzo},\ and\ \citenamefont {Veitch}}]{Carullo:2019flw}%
  \BibitemOpen
  \bibfield  {author} {\bibinfo {author} {\bibfnamefont {G.}~\bibnamefont {Carullo}}, \bibinfo {author} {\bibfnamefont {W.}~\bibnamefont {Del~Pozzo}},\ and\ \bibinfo {author} {\bibfnamefont {J.}~\bibnamefont {Veitch}},\ }\href {https://doi.org/10.1103/PhysRevD.99.123029} {\bibfield  {journal} {\bibinfo  {journal} {Phys. Rev. D}\ }\textbf {\bibinfo {volume} {99}},\ \bibinfo {pages} {123029} (\bibinfo {year} {2019})},\ \bibinfo {note} {[Erratum: Phys.Rev.D 100, 089903 (2019)]},\ \Eprint {https://arxiv.org/abs/1902.07527} {arXiv:1902.07527 [gr-qc]} \BibitemShut {NoStop}%
\bibitem [{\citenamefont {Isi}\ \emph {et~al.}(2019)\citenamefont {Isi}, \citenamefont {Giesler}, \citenamefont {Farr}, \citenamefont {Scheel},\ and\ \citenamefont {Teukolsky}}]{IsiNoHair_GW150914}%
  \BibitemOpen
  \bibfield  {author} {\bibinfo {author} {\bibfnamefont {M.}~\bibnamefont {Isi}}, \bibinfo {author} {\bibfnamefont {M.}~\bibnamefont {Giesler}}, \bibinfo {author} {\bibfnamefont {W.~M.}\ \bibnamefont {Farr}}, \bibinfo {author} {\bibfnamefont {M.~A.}\ \bibnamefont {Scheel}},\ and\ \bibinfo {author} {\bibfnamefont {S.~A.}\ \bibnamefont {Teukolsky}},\ }\href {https://link.aps.org/doi/10.1103/PhysRevLett.123.111102} {\bibfield  {journal} {\bibinfo  {journal} {Phys. Rev. Lett.}\ }\textbf {\bibinfo {volume} {123}},\ \bibinfo {pages} {111102} (\bibinfo {year} {2019})}\BibitemShut {NoStop}%
\bibitem [{\citenamefont {Isi}\ \emph {et~al.}(2021)\citenamefont {Isi}, \citenamefont {Farr}, \citenamefont {Giesler}, \citenamefont {Scheel},\ and\ \citenamefont {Teukolsky}}]{Isi_BHArea}%
  \BibitemOpen
  \bibfield  {author} {\bibinfo {author} {\bibfnamefont {M.}~\bibnamefont {Isi}}, \bibinfo {author} {\bibfnamefont {W.~M.}\ \bibnamefont {Farr}}, \bibinfo {author} {\bibfnamefont {M.}~\bibnamefont {Giesler}}, \bibinfo {author} {\bibfnamefont {M.~A.}\ \bibnamefont {Scheel}},\ and\ \bibinfo {author} {\bibfnamefont {S.~A.}\ \bibnamefont {Teukolsky}},\ }\href {https://link.aps.org/doi/10.1103/PhysRevLett.127.011103} {\bibfield  {journal} {\bibinfo  {journal} {Phys. Rev. Lett.}\ }\textbf {\bibinfo {volume} {127}},\ \bibinfo {pages} {011103} (\bibinfo {year} {2021})}\BibitemShut {NoStop}%
\bibitem [{\citenamefont {Abbott}\ \emph {et~al.}(2021)\citenamefont {Abbott} \emph {et~al.}}]{LIGOScientific:2021sio_TGRGwtc-3}%
  \BibitemOpen
  \bibfield  {author} {\bibinfo {author} {\bibfnamefont {R.}~\bibnamefont {Abbott}} \emph {et~al.} (\bibinfo {collaboration} {LIGO Scientific, VIRGO, KAGRA}),\ }\href@noop {} {\bibfield  {journal} {\bibinfo  {journal} {arXiv eprints}\ } (\bibinfo {year} {2021})},\ \Eprint {https://arxiv.org/abs/2112.06861} {arXiv:2112.06861 [gr-qc]} \BibitemShut {NoStop}%
\bibitem [{\citenamefont {Isi}\ and\ \citenamefont {Farr}(2022)}]{Isi_revisitGW150914}%
  \BibitemOpen
  \bibfield  {author} {\bibinfo {author} {\bibfnamefont {M.}~\bibnamefont {Isi}}\ and\ \bibinfo {author} {\bibfnamefont {W.~M.}\ \bibnamefont {Farr}},\ }\Eprint {https://arxiv.org/abs/2202.02941} {arXiv:2202.02941 [gr-qc]}  (\bibinfo {year} {2022})\BibitemShut {NoStop}%
\bibitem [{\citenamefont {Finch}\ and\ \citenamefont {Moore}(2022)}]{FinchMoore_GW150914}%
  \BibitemOpen
  \bibfield  {author} {\bibinfo {author} {\bibfnamefont {E.}~\bibnamefont {Finch}}\ and\ \bibinfo {author} {\bibfnamefont {C.~J.}\ \bibnamefont {Moore}},\ }\href {https://doi.org/10.1103/PhysRevD.106.043005} {\bibfield  {journal} {\bibinfo  {journal} {Phys. Rev. D}\ }\textbf {\bibinfo {volume} {106}},\ \bibinfo {pages} {043005} (\bibinfo {year} {2022})}\BibitemShut {NoStop}%
\bibitem [{\citenamefont {Cotesta}\ \emph {et~al.}(2022)\citenamefont {Cotesta}, \citenamefont {Carullo}, \citenamefont {Berti},\ and\ \citenamefont {Cardoso}}]{Cotesta_ringdownGW150914}%
  \BibitemOpen
  \bibfield  {author} {\bibinfo {author} {\bibfnamefont {R.}~\bibnamefont {Cotesta}}, \bibinfo {author} {\bibfnamefont {G.}~\bibnamefont {Carullo}}, \bibinfo {author} {\bibfnamefont {E.}~\bibnamefont {Berti}},\ and\ \bibinfo {author} {\bibfnamefont {V.}~\bibnamefont {Cardoso}},\ }\href {https://link.aps.org/doi/10.1103/PhysRevLett.129.111102} {\bibfield  {journal} {\bibinfo  {journal} {Phys. Rev. Lett.}\ }\textbf {\bibinfo {volume} {129}},\ \bibinfo {pages} {111102} (\bibinfo {year} {2022})}\BibitemShut {NoStop}%
\bibitem [{\citenamefont {Isi}\ and\ \citenamefont {Farr}(2023)}]{Isi:2023nif}%
  \BibitemOpen
  \bibfield  {author} {\bibinfo {author} {\bibfnamefont {M.}~\bibnamefont {Isi}}\ and\ \bibinfo {author} {\bibfnamefont {W.~M.}\ \bibnamefont {Farr}},\ }\href {https://doi.org/10.1103/PhysRevLett.131.169001} {\bibfield  {journal} {\bibinfo  {journal} {Phys. Rev. Lett.}\ }\textbf {\bibinfo {volume} {131}},\ \bibinfo {pages} {169001} (\bibinfo {year} {2023})},\ \Eprint {https://arxiv.org/abs/2310.13869} {arXiv:2310.13869 [astro-ph.HE]} \BibitemShut {NoStop}%
\bibitem [{\citenamefont {Siegel}\ \emph {et~al.}(2023)\citenamefont {Siegel}, \citenamefont {Isi},\ and\ \citenamefont {Farr}}]{Siegel:2023lxl}%
  \BibitemOpen
  \bibfield  {author} {\bibinfo {author} {\bibfnamefont {H.}~\bibnamefont {Siegel}}, \bibinfo {author} {\bibfnamefont {M.}~\bibnamefont {Isi}},\ and\ \bibinfo {author} {\bibfnamefont {W.~M.}\ \bibnamefont {Farr}},\ }\href {https://doi.org/10.1103/PhysRevD.108.064008} {\bibfield  {journal} {\bibinfo  {journal} {Phys. Rev. D}\ }\textbf {\bibinfo {volume} {108}},\ \bibinfo {pages} {064008} (\bibinfo {year} {2023})},\ \Eprint {https://arxiv.org/abs/2307.11975} {arXiv:2307.11975 [gr-qc]} \BibitemShut {NoStop}%
\bibitem [{\citenamefont {Capano}\ \emph {et~al.}(2023)\citenamefont {Capano}, \citenamefont {Cabero}, \citenamefont {Westerweck}, \citenamefont {Abedi}, \citenamefont {Kastha}, \citenamefont {Nitz}, \citenamefont {Wang}, \citenamefont {Nielsen},\ and\ \citenamefont {Krishnan}}]{Capano:2021etf}%
  \BibitemOpen
  \bibfield  {author} {\bibinfo {author} {\bibfnamefont {C.~D.}\ \bibnamefont {Capano}}, \bibinfo {author} {\bibfnamefont {M.}~\bibnamefont {Cabero}}, \bibinfo {author} {\bibfnamefont {J.}~\bibnamefont {Westerweck}}, \bibinfo {author} {\bibfnamefont {J.}~\bibnamefont {Abedi}}, \bibinfo {author} {\bibfnamefont {S.}~\bibnamefont {Kastha}}, \bibinfo {author} {\bibfnamefont {A.~H.}\ \bibnamefont {Nitz}}, \bibinfo {author} {\bibfnamefont {Y.-F.}\ \bibnamefont {Wang}}, \bibinfo {author} {\bibfnamefont {A.~B.}\ \bibnamefont {Nielsen}},\ and\ \bibinfo {author} {\bibfnamefont {B.}~\bibnamefont {Krishnan}},\ }\href {https://doi.org/10.1103/PhysRevLett.131.221402} {\bibfield  {journal} {\bibinfo  {journal} {Phys. Rev. Lett.}\ }\textbf {\bibinfo {volume} {131}},\ \bibinfo {pages} {221402} (\bibinfo {year} {2023})},\ \Eprint {https://arxiv.org/abs/2105.05238} {arXiv:2105.05238 [gr-qc]} \BibitemShut {NoStop}%
\bibitem [{\citenamefont {Capano}\ \emph {et~al.}(2024)\citenamefont {Capano}, \citenamefont {Abedi}, \citenamefont {Kastha}, \citenamefont {Nitz}, \citenamefont {Westerweck}, \citenamefont {Wang}, \citenamefont {Cabero}, \citenamefont {Nielsen},\ and\ \citenamefont {Krishnan}}]{Capano:2022zqm}%
  \BibitemOpen
  \bibfield  {author} {\bibinfo {author} {\bibfnamefont {C.~D.}\ \bibnamefont {Capano}}, \bibinfo {author} {\bibfnamefont {J.}~\bibnamefont {Abedi}}, \bibinfo {author} {\bibfnamefont {S.}~\bibnamefont {Kastha}}, \bibinfo {author} {\bibfnamefont {A.~H.}\ \bibnamefont {Nitz}}, \bibinfo {author} {\bibfnamefont {J.}~\bibnamefont {Westerweck}}, \bibinfo {author} {\bibfnamefont {Y.-F.}\ \bibnamefont {Wang}}, \bibinfo {author} {\bibfnamefont {M.}~\bibnamefont {Cabero}}, \bibinfo {author} {\bibfnamefont {A.~B.}\ \bibnamefont {Nielsen}},\ and\ \bibinfo {author} {\bibfnamefont {B.}~\bibnamefont {Krishnan}},\ }\href {https://doi.org/10.1088/1361-6382/ad84ae} {\bibfield  {journal} {\bibinfo  {journal} {Class. Quant. Grav.}\ }\textbf {\bibinfo {volume} {41}},\ \bibinfo {pages} {245009} (\bibinfo {year} {2024})},\ \Eprint {https://arxiv.org/abs/2209.00640} {arXiv:2209.00640 [gr-qc]} \BibitemShut {NoStop}%
\bibitem [{\citenamefont {Brito}\ \emph {et~al.}(2018)\citenamefont {Brito}, \citenamefont {Buonanno},\ and\ \citenamefont {Raymond}}]{Brito:2018rfr}%
  \BibitemOpen
  \bibfield  {author} {\bibinfo {author} {\bibfnamefont {R.}~\bibnamefont {Brito}}, \bibinfo {author} {\bibfnamefont {A.}~\bibnamefont {Buonanno}},\ and\ \bibinfo {author} {\bibfnamefont {V.}~\bibnamefont {Raymond}},\ }\href {https://doi.org/10.1103/PhysRevD.98.084038} {\bibfield  {journal} {\bibinfo  {journal} {Phys. Rev. D}\ }\textbf {\bibinfo {volume} {98}},\ \bibinfo {pages} {084038} (\bibinfo {year} {2018})},\ \Eprint {https://arxiv.org/abs/1805.00293} {arXiv:1805.00293 [gr-qc]} \BibitemShut {NoStop}%
\bibitem [{\citenamefont {Pompili}\ \emph {et~al.}(2025)\citenamefont {Pompili}, \citenamefont {Maggio}, \citenamefont {Silva},\ and\ \citenamefont {Buonanno}}]{Pompili:2025cdc}%
  \BibitemOpen
  \bibfield  {author} {\bibinfo {author} {\bibfnamefont {L.}~\bibnamefont {Pompili}}, \bibinfo {author} {\bibfnamefont {E.}~\bibnamefont {Maggio}}, \bibinfo {author} {\bibfnamefont {H.~O.}\ \bibnamefont {Silva}},\ and\ \bibinfo {author} {\bibfnamefont {A.}~\bibnamefont {Buonanno}},\ }\href {https://doi.org/10.1103/ng8w-98sz} {\bibfield  {journal} {\bibinfo  {journal} {Phys. Rev. D}\ }\textbf {\bibinfo {volume} {111}},\ \bibinfo {pages} {124040} (\bibinfo {year} {2025})},\ \Eprint {https://arxiv.org/abs/2504.10130} {arXiv:2504.10130 [gr-qc]} \BibitemShut {NoStop}%
\bibitem [{\citenamefont {Ma}\ \emph {et~al.}(2022)\citenamefont {Ma}, \citenamefont {Mitman}, \citenamefont {Sun}, \citenamefont {Deppe}, \citenamefont {H{\'e}bert}, \citenamefont {Kidder}, \citenamefont {Moxon}, \citenamefont {Throwe}, \citenamefont {Vu},\ and\ \citenamefont {Chen}}]{Ma:2022wpv}%
  \BibitemOpen
  \bibfield  {author} {\bibinfo {author} {\bibfnamefont {S.}~\bibnamefont {Ma}}, \bibinfo {author} {\bibfnamefont {K.}~\bibnamefont {Mitman}}, \bibinfo {author} {\bibfnamefont {L.}~\bibnamefont {Sun}}, \bibinfo {author} {\bibfnamefont {N.}~\bibnamefont {Deppe}}, \bibinfo {author} {\bibfnamefont {F.}~\bibnamefont {H{\'e}bert}}, \bibinfo {author} {\bibfnamefont {L.~E.}\ \bibnamefont {Kidder}}, \bibinfo {author} {\bibfnamefont {J.}~\bibnamefont {Moxon}}, \bibinfo {author} {\bibfnamefont {W.}~\bibnamefont {Throwe}}, \bibinfo {author} {\bibfnamefont {N.~L.}\ \bibnamefont {Vu}},\ and\ \bibinfo {author} {\bibfnamefont {Y.}~\bibnamefont {Chen}},\ }\href {https://doi.org/10.1103/PhysRevD.106.084036} {\bibfield  {journal} {\bibinfo  {journal} {Phys. Rev. D}\ }\textbf {\bibinfo {volume} {106}},\ \bibinfo {pages} {084036} (\bibinfo {year} {2022})},\ \Eprint {https://arxiv.org/abs/2207.10870} {arXiv:2207.10870 [gr-qc]} \BibitemShut {NoStop}%
\bibitem [{\citenamefont {Lu}\ \emph {et~al.}(2025)\citenamefont {Lu}, \citenamefont {Ma}, \citenamefont {Piccinni}, \citenamefont {Sun},\ and\ \citenamefont {Finch}}]{Lu:2025mwp}%
  \BibitemOpen
  \bibfield  {author} {\bibinfo {author} {\bibfnamefont {N.}~\bibnamefont {Lu}}, \bibinfo {author} {\bibfnamefont {S.}~\bibnamefont {Ma}}, \bibinfo {author} {\bibfnamefont {O.~J.}\ \bibnamefont {Piccinni}}, \bibinfo {author} {\bibfnamefont {L.}~\bibnamefont {Sun}},\ and\ \bibinfo {author} {\bibfnamefont {E.}~\bibnamefont {Finch}},\ }\href {https://doi.org/10.1103/h6jy-sd94} {\bibfield  {journal} {\bibinfo  {journal} {Phys. Rev. D}\ }\textbf {\bibinfo {volume} {112}},\ \bibinfo {pages} {064047} (\bibinfo {year} {2025})},\ \Eprint {https://arxiv.org/abs/2505.18560} {arXiv:2505.18560 [gr-qc]} \BibitemShut {NoStop}%
\bibitem [{\citenamefont {Abbott}\ \emph {et~al.}(2016)\citenamefont {Abbott} \emph {et~al.}}]{GW150914_detection}%
  \BibitemOpen
  \bibfield  {author} {\bibinfo {author} {\bibfnamefont {B.~P.}\ \bibnamefont {Abbott}} \emph {et~al.} (\bibinfo {collaboration} {LIGO Scientific, Virgo}),\ }\href {https://doi.org/10.1103/PhysRevLett.116.061102} {\bibfield  {journal} {\bibinfo  {journal} {Phys. Rev. Lett.}\ }\textbf {\bibinfo {volume} {116}},\ \bibinfo {pages} {061102} (\bibinfo {year} {2016})},\ \Eprint {https://arxiv.org/abs/1602.03837} {arXiv:1602.03837 [gr-qc]} \BibitemShut {NoStop}%
\bibitem [{\citenamefont {Bailes}\ \emph {et~al.}(2021)\citenamefont {Bailes} \emph {et~al.}}]{GWAstro_Review}%
  \BibitemOpen
  \bibfield  {author} {\bibinfo {author} {\bibfnamefont {M.}~\bibnamefont {Bailes}} \emph {et~al.},\ }\href {https://doi.org/10.1038/s42254-021-00303-8} {\bibfield  {journal} {\bibinfo  {journal} {Nature Rev. Phys.}\ }\textbf {\bibinfo {volume} {3}},\ \bibinfo {pages} {344} (\bibinfo {year} {2021})}\BibitemShut {NoStop}%
\bibitem [{\citenamefont {Flanagan}\ and\ \citenamefont {Hughes}(2005)}]{Flanagan:2005yc}%
  \BibitemOpen
  \bibfield  {author} {\bibinfo {author} {\bibfnamefont {E.~E.}\ \bibnamefont {Flanagan}}\ and\ \bibinfo {author} {\bibfnamefont {S.~A.}\ \bibnamefont {Hughes}},\ }\href {https://doi.org/10.1088/1367-2630/7/1/204} {\bibfield  {journal} {\bibinfo  {journal} {New J. Phys.}\ }\textbf {\bibinfo {volume} {7}},\ \bibinfo {pages} {204} (\bibinfo {year} {2005})},\ \Eprint {https://arxiv.org/abs/gr-qc/0501041} {arXiv:gr-qc/0501041} \BibitemShut {NoStop}%
\bibitem [{\citenamefont {Dreyer}\ \emph {et~al.}(2004{\natexlab{b}})\citenamefont {Dreyer}, \citenamefont {Kelly}, \citenamefont {Krishnan}, \citenamefont {Finn}, \citenamefont {Garrison},\ and\ \citenamefont {Lopez-Aleman}}]{Dreyer:2003bv}%
  \BibitemOpen
  \bibfield  {author} {\bibinfo {author} {\bibfnamefont {O.}~\bibnamefont {Dreyer}}, \bibinfo {author} {\bibfnamefont {B.~J.}\ \bibnamefont {Kelly}}, \bibinfo {author} {\bibfnamefont {B.}~\bibnamefont {Krishnan}}, \bibinfo {author} {\bibfnamefont {L.~S.}\ \bibnamefont {Finn}}, \bibinfo {author} {\bibfnamefont {D.}~\bibnamefont {Garrison}},\ and\ \bibinfo {author} {\bibfnamefont {R.}~\bibnamefont {Lopez-Aleman}},\ }\href {https://doi.org/10.1088/0264-9381/21/4/003} {\bibfield  {journal} {\bibinfo  {journal} {Class. Quant. Grav.}\ }\textbf {\bibinfo {volume} {21}},\ \bibinfo {pages} {787} (\bibinfo {year} {2004}{\natexlab{b}})},\ \Eprint {https://arxiv.org/abs/gr-qc/0309007} {arXiv:gr-qc/0309007} \BibitemShut {NoStop}%
\bibitem [{\citenamefont {Isi}\ \emph {et~al.}(2024)\citenamefont {Isi}, \citenamefont {Siegel}, \citenamefont {Farr}, \citenamefont {Khusid}, \citenamefont {Hussain},\ and\ \citenamefont {Udall}}]{ringdowncode}%
  \BibitemOpen
  \bibfield  {author} {\bibinfo {author} {\bibfnamefont {M.}~\bibnamefont {Isi}}, \bibinfo {author} {\bibfnamefont {H.}~\bibnamefont {Siegel}}, \bibinfo {author} {\bibfnamefont {W.~M.}\ \bibnamefont {Farr}}, \bibinfo {author} {\bibfnamefont {N.}~\bibnamefont {Khusid}}, \bibinfo {author} {\bibfnamefont {A.}~\bibnamefont {Hussain}},\ and\ \bibinfo {author} {\bibfnamefont {R.}~\bibnamefont {Udall}},\ }\href {https://doi.org/10.5281/zenodo.13892015} {\bibinfo {title} {maxisi/ringdown: v1.0.0}} (\bibinfo {year} {2024})\BibitemShut {NoStop}%
\bibitem [{\citenamefont {O'Shaughnessy}\ \emph {et~al.}(2013)\citenamefont {O'Shaughnessy}, \citenamefont {London}, \citenamefont {Healy},\ and\ \citenamefont {Shoemaker}}]{OShaughnessy:2012iol}%
  \BibitemOpen
  \bibfield  {author} {\bibinfo {author} {\bibfnamefont {R.}~\bibnamefont {O'Shaughnessy}}, \bibinfo {author} {\bibfnamefont {L.}~\bibnamefont {London}}, \bibinfo {author} {\bibfnamefont {J.}~\bibnamefont {Healy}},\ and\ \bibinfo {author} {\bibfnamefont {D.}~\bibnamefont {Shoemaker}},\ }\href {https://doi.org/10.1103/PhysRevD.87.044038} {\bibfield  {journal} {\bibinfo  {journal} {Phys. Rev. D}\ }\textbf {\bibinfo {volume} {87}},\ \bibinfo {pages} {044038} (\bibinfo {year} {2013})},\ \Eprint {https://arxiv.org/abs/1209.3712} {arXiv:1209.3712 [gr-qc]} \BibitemShut {NoStop}%
\bibitem [{\citenamefont {Hamilton}\ \emph {et~al.}(2023)\citenamefont {Hamilton}, \citenamefont {London},\ and\ \citenamefont {Hannam}}]{Hamilton:2023znn}%
  \BibitemOpen
  \bibfield  {author} {\bibinfo {author} {\bibfnamefont {E.}~\bibnamefont {Hamilton}}, \bibinfo {author} {\bibfnamefont {L.}~\bibnamefont {London}},\ and\ \bibinfo {author} {\bibfnamefont {M.}~\bibnamefont {Hannam}},\ }\href {https://doi.org/10.1103/PhysRevD.107.104035} {\bibfield  {journal} {\bibinfo  {journal} {Phys. Rev. D}\ }\textbf {\bibinfo {volume} {107}},\ \bibinfo {pages} {104035} (\bibinfo {year} {2023})},\ \Eprint {https://arxiv.org/abs/2301.06558} {arXiv:2301.06558 [gr-qc]} \BibitemShut {NoStop}%
\bibitem [{\citenamefont {Nobili}\ \emph {et~al.}(2025)\citenamefont {Nobili}, \citenamefont {Bhagwat}, \citenamefont {Pacilio},\ and\ \citenamefont {Gerosa}}]{Nobili:2025ydt}%
  \BibitemOpen
  \bibfield  {author} {\bibinfo {author} {\bibfnamefont {F.}~\bibnamefont {Nobili}}, \bibinfo {author} {\bibfnamefont {S.}~\bibnamefont {Bhagwat}}, \bibinfo {author} {\bibfnamefont {C.}~\bibnamefont {Pacilio}},\ and\ \bibinfo {author} {\bibfnamefont {D.}~\bibnamefont {Gerosa}},\ }\href {https://doi.org/10.1103/cl3k-3xt2} {\bibfield  {journal} {\bibinfo  {journal} {Phys. Rev. D}\ }\textbf {\bibinfo {volume} {112}},\ \bibinfo {pages} {044058} (\bibinfo {year} {2025})},\ \Eprint {https://arxiv.org/abs/2504.17021} {arXiv:2504.17021 [gr-qc]} \BibitemShut {NoStop}%
\bibitem [{\citenamefont {Estell{\'e}s}\ \emph {et~al.}(2022)\citenamefont {Estell{\'e}s}, \citenamefont {Colleoni}, \citenamefont {Garc{\'\i}a-Quir{\'o}s}, \citenamefont {Husa}, \citenamefont {Keitel}, \citenamefont {Mateu-Lucena}, \citenamefont {Planas},\ and\ \citenamefont {Ramos-Buades}}]{Estelles:2021gvs}%
  \BibitemOpen
  \bibfield  {author} {\bibinfo {author} {\bibfnamefont {H.}~\bibnamefont {Estell{\'e}s}}, \bibinfo {author} {\bibfnamefont {M.}~\bibnamefont {Colleoni}}, \bibinfo {author} {\bibfnamefont {C.}~\bibnamefont {Garc{\'\i}a-Quir{\'o}s}}, \bibinfo {author} {\bibfnamefont {S.}~\bibnamefont {Husa}}, \bibinfo {author} {\bibfnamefont {D.}~\bibnamefont {Keitel}}, \bibinfo {author} {\bibfnamefont {M.}~\bibnamefont {Mateu-Lucena}}, \bibinfo {author} {\bibfnamefont {M.~d.~L.}\ \bibnamefont {Planas}},\ and\ \bibinfo {author} {\bibfnamefont {A.}~\bibnamefont {Ramos-Buades}},\ }\href {https://doi.org/10.1103/PhysRevD.105.084040} {\bibfield  {journal} {\bibinfo  {journal} {Phys. Rev. D}\ }\textbf {\bibinfo {volume} {105}},\ \bibinfo {pages} {084040} (\bibinfo {year} {2022})},\ \Eprint {https://arxiv.org/abs/2105.05872} {arXiv:2105.05872 [gr-qc]} \BibitemShut {NoStop}%
\bibitem [{\citenamefont {Thompson}\ \emph {et~al.}(2024)\citenamefont {Thompson}, \citenamefont {Hamilton}, \citenamefont {London}, \citenamefont {Ghosh}, \citenamefont {Kolitsidou}, \citenamefont {Hoy},\ and\ \citenamefont {Hannam}}]{Thompson:2023ase}%
  \BibitemOpen
  \bibfield  {author} {\bibinfo {author} {\bibfnamefont {J.~E.}\ \bibnamefont {Thompson}}, \bibinfo {author} {\bibfnamefont {E.}~\bibnamefont {Hamilton}}, \bibinfo {author} {\bibfnamefont {L.}~\bibnamefont {London}}, \bibinfo {author} {\bibfnamefont {S.}~\bibnamefont {Ghosh}}, \bibinfo {author} {\bibfnamefont {P.}~\bibnamefont {Kolitsidou}}, \bibinfo {author} {\bibfnamefont {C.}~\bibnamefont {Hoy}},\ and\ \bibinfo {author} {\bibfnamefont {M.}~\bibnamefont {Hannam}},\ }\href {https://doi.org/10.1103/PhysRevD.109.063012} {\bibfield  {journal} {\bibinfo  {journal} {Phys. Rev. D}\ }\textbf {\bibinfo {volume} {109}},\ \bibinfo {pages} {063012} (\bibinfo {year} {2024})},\ \Eprint {https://arxiv.org/abs/2312.10025} {arXiv:2312.10025 [gr-qc]} \BibitemShut {NoStop}%
\bibitem [{\citenamefont {Ramos-Buades}\ \emph {et~al.}(2023)\citenamefont {Ramos-Buades}, \citenamefont {Buonanno}, \citenamefont {Estell{\'e}s}, \citenamefont {Khalil}, \citenamefont {Mihaylov}, \citenamefont {Ossokine}, \citenamefont {Pompili},\ and\ \citenamefont {Shiferaw}}]{Ramos-Buades:2023ehm}%
  \BibitemOpen
  \bibfield  {author} {\bibinfo {author} {\bibfnamefont {A.}~\bibnamefont {Ramos-Buades}}, \bibinfo {author} {\bibfnamefont {A.}~\bibnamefont {Buonanno}}, \bibinfo {author} {\bibfnamefont {H.}~\bibnamefont {Estell{\'e}s}}, \bibinfo {author} {\bibfnamefont {M.}~\bibnamefont {Khalil}}, \bibinfo {author} {\bibfnamefont {D.~P.}\ \bibnamefont {Mihaylov}}, \bibinfo {author} {\bibfnamefont {S.}~\bibnamefont {Ossokine}}, \bibinfo {author} {\bibfnamefont {L.}~\bibnamefont {Pompili}},\ and\ \bibinfo {author} {\bibfnamefont {M.}~\bibnamefont {Shiferaw}},\ }\href {https://doi.org/10.1103/PhysRevD.108.124037} {\bibfield  {journal} {\bibinfo  {journal} {Phys. Rev. D}\ }\textbf {\bibinfo {volume} {108}},\ \bibinfo {pages} {124037} (\bibinfo {year} {2023})},\ \Eprint {https://arxiv.org/abs/2303.18046} {arXiv:2303.18046 [gr-qc]} \BibitemShut {NoStop}%
\bibitem [{\citenamefont {Pratten}\ \emph {et~al.}(2021)\citenamefont {Pratten} \emph {et~al.}}]{Pratten:2020ceb}%
  \BibitemOpen
  \bibfield  {author} {\bibinfo {author} {\bibfnamefont {G.}~\bibnamefont {Pratten}} \emph {et~al.},\ }\href {https://doi.org/10.1103/PhysRevD.103.104056} {\bibfield  {journal} {\bibinfo  {journal} {Phys. Rev. D}\ }\textbf {\bibinfo {volume} {103}},\ \bibinfo {pages} {104056} (\bibinfo {year} {2021})},\ \Eprint {https://arxiv.org/abs/2004.06503} {arXiv:2004.06503 [gr-qc]} \BibitemShut {NoStop}%
\bibitem [{\citenamefont {Colleoni}\ \emph {et~al.}(2025)\citenamefont {Colleoni}, \citenamefont {Vidal}, \citenamefont {Garc{\'\i}a-Quir{\'o}s}, \citenamefont {Ak{\c{c}}ay},\ and\ \citenamefont {Bera}}]{Colleoni:2024knd}%
  \BibitemOpen
  \bibfield  {author} {\bibinfo {author} {\bibfnamefont {M.}~\bibnamefont {Colleoni}}, \bibinfo {author} {\bibfnamefont {F.~A.~R.}\ \bibnamefont {Vidal}}, \bibinfo {author} {\bibfnamefont {C.}~\bibnamefont {Garc{\'\i}a-Quir{\'o}s}}, \bibinfo {author} {\bibfnamefont {S.}~\bibnamefont {Ak{\c{c}}ay}},\ and\ \bibinfo {author} {\bibfnamefont {S.}~\bibnamefont {Bera}},\ }\href {https://doi.org/10.1103/PhysRevD.111.104019} {\bibfield  {journal} {\bibinfo  {journal} {Phys. Rev. D}\ }\textbf {\bibinfo {volume} {111}},\ \bibinfo {pages} {104019} (\bibinfo {year} {2025})},\ \Eprint {https://arxiv.org/abs/2412.16721} {arXiv:2412.16721 [gr-qc]} \BibitemShut {NoStop}%
\bibitem [{\citenamefont {Mac~Uilliam}\ \emph {et~al.}(2024)\citenamefont {Mac~Uilliam}, \citenamefont {Akcay},\ and\ \citenamefont {Thompson}}]{MacUilliam:2024oif}%
  \BibitemOpen
  \bibfield  {author} {\bibinfo {author} {\bibfnamefont {J.}~\bibnamefont {Mac~Uilliam}}, \bibinfo {author} {\bibfnamefont {S.}~\bibnamefont {Akcay}},\ and\ \bibinfo {author} {\bibfnamefont {J.~E.}\ \bibnamefont {Thompson}},\ }\href {https://doi.org/10.1103/PhysRevD.109.084077} {\bibfield  {journal} {\bibinfo  {journal} {Phys. Rev. D}\ }\textbf {\bibinfo {volume} {109}},\ \bibinfo {pages} {084077} (\bibinfo {year} {2024})},\ \Eprint {https://arxiv.org/abs/2402.06781} {arXiv:2402.06781 [gr-qc]} \BibitemShut {NoStop}%
\bibitem [{\citenamefont {Dhani}\ \emph {et~al.}(2024)\citenamefont {Dhani}, \citenamefont {V\"olkel}, \citenamefont {Buonanno}, \citenamefont {Estelles}, \citenamefont {Gair}, \citenamefont {Pfeiffer}, \citenamefont {Pompili},\ and\ \citenamefont {Toubiana}}]{Dhani:2024jja}%
  \BibitemOpen
  \bibfield  {author} {\bibinfo {author} {\bibfnamefont {A.}~\bibnamefont {Dhani}}, \bibinfo {author} {\bibfnamefont {S.}~\bibnamefont {V\"olkel}}, \bibinfo {author} {\bibfnamefont {A.}~\bibnamefont {Buonanno}}, \bibinfo {author} {\bibfnamefont {H.}~\bibnamefont {Estelles}}, \bibinfo {author} {\bibfnamefont {J.}~\bibnamefont {Gair}}, \bibinfo {author} {\bibfnamefont {H.~P.}\ \bibnamefont {Pfeiffer}}, \bibinfo {author} {\bibfnamefont {L.}~\bibnamefont {Pompili}},\ and\ \bibinfo {author} {\bibfnamefont {A.}~\bibnamefont {Toubiana}},\ }\href@noop {} {\  (\bibinfo {year} {2024})},\ \Eprint {https://arxiv.org/abs/2404.05811} {arXiv:2404.05811 [gr-qc]} \BibitemShut {NoStop}%
\bibitem [{\citenamefont {Mandel}(2025)}]{Mandel:2025qnh}%
  \BibitemOpen
  \bibfield  {author} {\bibinfo {author} {\bibfnamefont {I.}~\bibnamefont {Mandel}},\ }\href@noop {} {\bibfield  {journal} {\bibinfo  {journal} {arXiv eprints}\ } (\bibinfo {year} {2025})},\ \Eprint {https://arxiv.org/abs/2509.05885} {arXiv:2509.05885 [astro-ph.HE]} \BibitemShut {NoStop}%
\bibitem [{\citenamefont {K{\i}ro{\u{g}}lu}\ \emph {et~al.}(2025)\citenamefont {K{\i}ro{\u{g}}lu}, \citenamefont {Kremer},\ and\ \citenamefont {Rasio}}]{Kiroglu:2025vqy}%
  \BibitemOpen
  \bibfield  {author} {\bibinfo {author} {\bibfnamefont {F.}~\bibnamefont {K{\i}ro{\u{g}}lu}}, \bibinfo {author} {\bibfnamefont {K.}~\bibnamefont {Kremer}},\ and\ \bibinfo {author} {\bibfnamefont {F.~A.}\ \bibnamefont {Rasio}},\ }\href@noop {} {\bibfield  {journal} {\bibinfo  {journal} {arXiv eprints}\ } (\bibinfo {year} {2025})},\ \Eprint {https://arxiv.org/abs/2509.05415} {arXiv:2509.05415 [astro-ph.HE]} \BibitemShut {NoStop}%
\bibitem [{\citenamefont {Tong}\ \emph {et~al.}(2025)\citenamefont {Tong} \emph {et~al.}}]{Tong:2025wpz}%
  \BibitemOpen
  \bibfield  {author} {\bibinfo {author} {\bibfnamefont {H.}~\bibnamefont {Tong}} \emph {et~al.},\ }\href@noop {} {\bibfield  {journal} {\bibinfo  {journal} {arXiv eprints}\ } (\bibinfo {year} {2025})},\ \Eprint {https://arxiv.org/abs/2509.04151} {arXiv:2509.04151 [astro-ph.HE]} \BibitemShut {NoStop}%
\bibitem [{\citenamefont {Paiella}\ \emph {et~al.}(2025)\citenamefont {Paiella}, \citenamefont {Ugolini}, \citenamefont {Spera}, \citenamefont {Branchesi},\ and\ \citenamefont {Sedda}}]{Paiella:2025qld}%
  \BibitemOpen
  \bibfield  {author} {\bibinfo {author} {\bibfnamefont {L.}~\bibnamefont {Paiella}}, \bibinfo {author} {\bibfnamefont {C.}~\bibnamefont {Ugolini}}, \bibinfo {author} {\bibfnamefont {M.}~\bibnamefont {Spera}}, \bibinfo {author} {\bibfnamefont {M.}~\bibnamefont {Branchesi}},\ and\ \bibinfo {author} {\bibfnamefont {M.~A.}\ \bibnamefont {Sedda}},\ }\href@noop {} {\bibfield  {journal} {\bibinfo  {journal} {arXiv eprints}\ } (\bibinfo {year} {2025})},\ \Eprint {https://arxiv.org/abs/2509.10609} {arXiv:2509.10609 [astro-ph.GA]} \BibitemShut {NoStop}%
\bibitem [{\citenamefont {Li}\ and\ \citenamefont {Fan}(2025)}]{Li:2025pyo}%
  \BibitemOpen
  \bibfield  {author} {\bibinfo {author} {\bibfnamefont {G.-P.}\ \bibnamefont {Li}}\ and\ \bibinfo {author} {\bibfnamefont {X.-L.}\ \bibnamefont {Fan}},\ }\href@noop {} {\bibfield  {journal} {\bibinfo  {journal} {arXiv eprints}\ } (\bibinfo {year} {2025})},\ \Eprint {https://arxiv.org/abs/2509.08298} {arXiv:2509.08298 [astro-ph.HE]} \BibitemShut {NoStop}%
\bibitem [{\citenamefont {Baumgarte}\ and\ \citenamefont {Shapiro}(2025)}]{Baumgarte:2025syh}%
  \BibitemOpen
  \bibfield  {author} {\bibinfo {author} {\bibfnamefont {T.~W.}\ \bibnamefont {Baumgarte}}\ and\ \bibinfo {author} {\bibfnamefont {S.~L.}\ \bibnamefont {Shapiro}},\ }\href@noop {} {\bibfield  {journal} {\bibinfo  {journal} {arXiv eprints}\ } (\bibinfo {year} {2025})},\ \Eprint {https://arxiv.org/abs/2509.04574} {arXiv:2509.04574 [astro-ph.HE]} \BibitemShut {NoStop}%
\bibitem [{\citenamefont {Popa}\ and\ \citenamefont {de~Mink}(2025)}]{Popa:2025dpz}%
  \BibitemOpen
  \bibfield  {author} {\bibinfo {author} {\bibfnamefont {S.~A.}\ \bibnamefont {Popa}}\ and\ \bibinfo {author} {\bibfnamefont {S.~E.}\ \bibnamefont {de~Mink}},\ }\href@noop {} {\bibfield  {journal} {\bibinfo  {journal} {arXiv eprints}\ } (\bibinfo {year} {2025})},\ \Eprint {https://arxiv.org/abs/2509.00154} {arXiv:2509.00154 [astro-ph.HE]} \BibitemShut {NoStop}%
\bibitem [{\citenamefont {Gottlieb}\ \emph {et~al.}(2025)\citenamefont {Gottlieb}, \citenamefont {Metzger}, \citenamefont {Issa}, \citenamefont {Li}, \citenamefont {Renzo},\ and\ \citenamefont {Isi}}]{Gottlieb:2025ugy}%
  \BibitemOpen
  \bibfield  {author} {\bibinfo {author} {\bibfnamefont {O.}~\bibnamefont {Gottlieb}}, \bibinfo {author} {\bibfnamefont {B.~D.}\ \bibnamefont {Metzger}}, \bibinfo {author} {\bibfnamefont {D.}~\bibnamefont {Issa}}, \bibinfo {author} {\bibfnamefont {S.~E.}\ \bibnamefont {Li}}, \bibinfo {author} {\bibfnamefont {M.}~\bibnamefont {Renzo}},\ and\ \bibinfo {author} {\bibfnamefont {M.}~\bibnamefont {Isi}},\ }\href@noop {} {\bibfield  {journal} {\bibinfo  {journal} {arXiv eprints}\ } (\bibinfo {year} {2025})},\ \Eprint {https://arxiv.org/abs/2508.15887} {arXiv:2508.15887 [astro-ph.HE]} \BibitemShut {NoStop}%
\bibitem [{\citenamefont {Delfavero}\ \emph {et~al.}(2025)\citenamefont {Delfavero}, \citenamefont {Ray}, \citenamefont {Cook}, \citenamefont {Nathaniel}, \citenamefont {McKernan}, \citenamefont {Ford}, \citenamefont {Postiglione}, \citenamefont {McPike},\ and\ \citenamefont {O'Shaughnessy}}]{Delfavero:2025lup}%
  \BibitemOpen
  \bibfield  {author} {\bibinfo {author} {\bibfnamefont {V.}~\bibnamefont {Delfavero}}, \bibinfo {author} {\bibfnamefont {S.}~\bibnamefont {Ray}}, \bibinfo {author} {\bibfnamefont {H.~E.}\ \bibnamefont {Cook}}, \bibinfo {author} {\bibfnamefont {K.}~\bibnamefont {Nathaniel}}, \bibinfo {author} {\bibfnamefont {B.}~\bibnamefont {McKernan}}, \bibinfo {author} {\bibfnamefont {K.~E.~S.}\ \bibnamefont {Ford}}, \bibinfo {author} {\bibfnamefont {J.}~\bibnamefont {Postiglione}}, \bibinfo {author} {\bibfnamefont {E.}~\bibnamefont {McPike}},\ and\ \bibinfo {author} {\bibfnamefont {R.}~\bibnamefont {O'Shaughnessy}},\ }\href@noop {} {\bibfield  {journal} {\bibinfo  {journal} {arXiv eprints}\ } (\bibinfo {year} {2025})},\ \Eprint {https://arxiv.org/abs/2508.13412} {arXiv:2508.13412 [gr-qc]} \BibitemShut {NoStop}%
\bibitem [{\citenamefont {Croon}\ \emph {et~al.}(2025)\citenamefont {Croon}, \citenamefont {Sakstein},\ and\ \citenamefont {Gerosa}}]{Croon:2025gol}%
  \BibitemOpen
  \bibfield  {author} {\bibinfo {author} {\bibfnamefont {D.}~\bibnamefont {Croon}}, \bibinfo {author} {\bibfnamefont {J.}~\bibnamefont {Sakstein}},\ and\ \bibinfo {author} {\bibfnamefont {D.}~\bibnamefont {Gerosa}},\ }\href@noop {} {\bibfield  {journal} {\bibinfo  {journal} {arXiv eprints}\ } (\bibinfo {year} {2025})},\ \Eprint {https://arxiv.org/abs/2508.10088} {arXiv:2508.10088 [astro-ph.HE]} \BibitemShut {NoStop}%
\bibitem [{\citenamefont {Bartos}\ and\ \citenamefont {Haiman}(2025)}]{Bartos:2025pkv}%
  \BibitemOpen
  \bibfield  {author} {\bibinfo {author} {\bibfnamefont {I.}~\bibnamefont {Bartos}}\ and\ \bibinfo {author} {\bibfnamefont {Z.}~\bibnamefont {Haiman}},\ }\href@noop {} {\bibfield  {journal} {\bibinfo  {journal} {arXiv eprints}\ } (\bibinfo {year} {2025})},\ \Eprint {https://arxiv.org/abs/2508.08558} {arXiv:2508.08558 [astro-ph.HE]} \BibitemShut {NoStop}%
\bibitem [{\citenamefont {Abbott}\ \emph {et~al.}(2020{\natexlab{a}})\citenamefont {Abbott} \emph {et~al.}}]{LIGOScientific:2020iuh}%
  \BibitemOpen
  \bibfield  {author} {\bibinfo {author} {\bibfnamefont {R.}~\bibnamefont {Abbott}} \emph {et~al.} (\bibinfo {collaboration} {LIGO Scientific, Virgo}),\ }\href {https://doi.org/10.1103/PhysRevLett.125.101102} {\bibfield  {journal} {\bibinfo  {journal} {Phys. Rev. Lett.}\ }\textbf {\bibinfo {volume} {125}},\ \bibinfo {pages} {101102} (\bibinfo {year} {2020}{\natexlab{a}})},\ \Eprint {https://arxiv.org/abs/2009.01075} {arXiv:2009.01075 [gr-qc]} \BibitemShut {NoStop}%
\bibitem [{\citenamefont {Abbott}\ \emph {et~al.}(2020{\natexlab{b}})\citenamefont {Abbott} \emph {et~al.}}]{LIGOScientific:2020ufj}%
  \BibitemOpen
  \bibfield  {author} {\bibinfo {author} {\bibfnamefont {R.}~\bibnamefont {Abbott}} \emph {et~al.} (\bibinfo {collaboration} {LIGO Scientific, Virgo}),\ }\href {https://doi.org/10.3847/2041-8213/aba493} {\bibfield  {journal} {\bibinfo  {journal} {Astrophys. J. Lett.}\ }\textbf {\bibinfo {volume} {900}},\ \bibinfo {pages} {L13} (\bibinfo {year} {2020}{\natexlab{b}})},\ \Eprint {https://arxiv.org/abs/2009.01190} {arXiv:2009.01190 [astro-ph.HE]} \BibitemShut {NoStop}%
\bibitem [{\citenamefont {Abbott}\ \emph {et~al.}(2023)\citenamefont {Abbott} \emph {et~al.}}]{KAGRA:2021vkt}%
  \BibitemOpen
  \bibfield  {author} {\bibinfo {author} {\bibfnamefont {R.}~\bibnamefont {Abbott}} \emph {et~al.} (\bibinfo {collaboration} {KAGRA, VIRGO, LIGO Scientific}),\ }\href {https://doi.org/10.1103/PhysRevX.13.041039} {\bibfield  {journal} {\bibinfo  {journal} {Phys. Rev. X}\ }\textbf {\bibinfo {volume} {13}},\ \bibinfo {pages} {041039} (\bibinfo {year} {2023})},\ \Eprint {https://arxiv.org/abs/2111.03606} {arXiv:2111.03606 [gr-qc]} \BibitemShut {NoStop}%
\bibitem [{\citenamefont {Scheel}\ \emph {et~al.}(2025)\citenamefont {Scheel} \emph {et~al.}}]{Scheel:2025jct_SXScatalog}%
  \BibitemOpen
  \bibfield  {author} {\bibinfo {author} {\bibfnamefont {M.~A.}\ \bibnamefont {Scheel}} \emph {et~al.},\ }\href@noop {} {\bibfield  {journal} {\bibinfo  {journal} {arXiv eprints}\ } (\bibinfo {year} {2025})},\ \Eprint {https://arxiv.org/abs/2505.13378} {arXiv:2505.13378 [gr-qc]} \BibitemShut {NoStop}%
\bibitem [{\citenamefont {Abbott}\ \emph {et~al.}(2020{\natexlab{c}})\citenamefont {Abbott} \emph {et~al.}}]{LIGOScientificNOISE:2019hgc}%
  \BibitemOpen
  \bibfield  {author} {\bibinfo {author} {\bibfnamefont {B.~P.}\ \bibnamefont {Abbott}} \emph {et~al.} (\bibinfo {collaboration} {LIGO Scientific, Virgo}),\ }\href {https://doi.org/10.1088/1361-6382/ab685e} {\bibfield  {journal} {\bibinfo  {journal} {Class. Quant. Grav.}\ }\textbf {\bibinfo {volume} {37}},\ \bibinfo {pages} {055002} (\bibinfo {year} {2020}{\natexlab{c}})},\ \Eprint {https://arxiv.org/abs/1908.11170} {arXiv:1908.11170 [gr-qc]} \BibitemShut {NoStop}%
\bibitem [{\citenamefont {Farr}(2024)}]{LineCleaner_code}%
  \BibitemOpen
  \bibfield  {author} {\bibinfo {author} {\bibfnamefont {W.~M.}\ \bibnamefont {Farr}},\ }\href {https://github.com/farr/LineCleaner/blob/main-pdf/note/linecleaner.pdf} {\bibinfo {title} {farr/linecleaner}} (\bibinfo {year} {2024})\BibitemShut {NoStop}%
\bibitem [{\citenamefont {Siegel}\ and\ \citenamefont {Farr}(2025)}]{LineCleanerReviewGitlab}%
  \BibitemOpen
  \bibfield  {author} {\bibinfo {author} {\bibfnamefont {H.}~\bibnamefont {Siegel}}\ and\ \bibinfo {author} {\bibfnamefont {W.~M.}\ \bibnamefont {Farr}},\ }\href {https://git.ligo.org/will-farr/line-cleaner-review} {\bibinfo {title} {Line cleaner review gitlab}} (\bibinfo {year} {2025})\BibitemShut {NoStop}%
\bibitem [{\citenamefont {Siegel}(2025)}]{LineCleanerInjectionsDCC}%
  \BibitemOpen
  \bibfield  {author} {\bibinfo {author} {\bibfnamefont {H.}~\bibnamefont {Siegel}},\ }\href {https://dcc.ligo.org/G2501032} {\bibinfo {title} {Line cleaner injection study dcc slides}} (\bibinfo {year} {2025})\BibitemShut {NoStop}%
\bibitem [{\citenamefont {Zackay}\ \emph {et~al.}(2021)\citenamefont {Zackay}, \citenamefont {Venumadhav}, \citenamefont {Roulet}, \citenamefont {Dai},\ and\ \citenamefont {Zaldarriaga}}]{Zackay:2019kkv}%
  \BibitemOpen
  \bibfield  {author} {\bibinfo {author} {\bibfnamefont {B.}~\bibnamefont {Zackay}}, \bibinfo {author} {\bibfnamefont {T.}~\bibnamefont {Venumadhav}}, \bibinfo {author} {\bibfnamefont {J.}~\bibnamefont {Roulet}}, \bibinfo {author} {\bibfnamefont {L.}~\bibnamefont {Dai}},\ and\ \bibinfo {author} {\bibfnamefont {M.}~\bibnamefont {Zaldarriaga}},\ }\href {https://doi.org/10.1103/PhysRevD.104.063034} {\bibfield  {journal} {\bibinfo  {journal} {Phys. Rev. D}\ }\textbf {\bibinfo {volume} {104}},\ \bibinfo {pages} {063034} (\bibinfo {year} {2021})},\ \Eprint {https://arxiv.org/abs/1908.05644} {arXiv:1908.05644 [astro-ph.IM]} \BibitemShut {NoStop}%
\bibitem [{\citenamefont {Wang}\ \emph {et~al.}(2025)\citenamefont {Wang}, \citenamefont {Tang}, \citenamefont {Li},\ and\ \citenamefont {Fan}}]{Wang:2025rvn}%
  \BibitemOpen
  \bibfield  {author} {\bibinfo {author} {\bibfnamefont {H.-T.}\ \bibnamefont {Wang}}, \bibinfo {author} {\bibfnamefont {S.-P.}\ \bibnamefont {Tang}}, \bibinfo {author} {\bibfnamefont {P.-C.}\ \bibnamefont {Li}},\ and\ \bibinfo {author} {\bibfnamefont {Y.-Z.}\ \bibnamefont {Fan}},\ }\href@noop {} {\bibfield  {journal} {\bibinfo  {journal} {arXiv eprints}\ } (\bibinfo {year} {2025})},\ \Eprint {https://arxiv.org/abs/2509.02047} {arXiv:2509.02047 [gr-qc]} \BibitemShut {NoStop}%
\bibitem [{\citenamefont {McGowan}\ \emph {et~al.}(2024)\citenamefont {McGowan} \emph {et~al.}}]{GW231123_dataquality}%
  \BibitemOpen
  \bibfield  {author} {\bibinfo {author} {\bibfnamefont {K.}~\bibnamefont {McGowan}} \emph {et~al.},\ }\href {https://dcc.ligo.org/LIGO-G2402555} {\bibinfo {title} {Gw231123 data quality report}} (\bibinfo {year} {2024})\BibitemShut {NoStop}%
\bibitem [{\citenamefont {Siegel}()}]{datarelease_thispaper}%
  \BibitemOpen
  \bibfield  {author} {\bibinfo {author} {\bibfnamefont {H.}~\bibnamefont {Siegel}},\ }\href {https://zenodo.org/records/17518871} {\bibinfo {title} {Data release}}\BibitemShut {NoStop}%
\bibitem [{\citenamefont {Baibhav}\ \emph {et~al.}(2023)\citenamefont {Baibhav}, \citenamefont {Cheung}, \citenamefont {Berti}, \citenamefont {Cardoso}, \citenamefont {Carullo}, \citenamefont {Cotesta}, \citenamefont {Del~Pozzo},\ and\ \citenamefont {Duque}}]{Baibhav:2023clw}%
  \BibitemOpen
  \bibfield  {author} {\bibinfo {author} {\bibfnamefont {V.}~\bibnamefont {Baibhav}}, \bibinfo {author} {\bibfnamefont {M.~H.-Y.}\ \bibnamefont {Cheung}}, \bibinfo {author} {\bibfnamefont {E.}~\bibnamefont {Berti}}, \bibinfo {author} {\bibfnamefont {V.}~\bibnamefont {Cardoso}}, \bibinfo {author} {\bibfnamefont {G.}~\bibnamefont {Carullo}}, \bibinfo {author} {\bibfnamefont {R.}~\bibnamefont {Cotesta}}, \bibinfo {author} {\bibfnamefont {W.}~\bibnamefont {Del~Pozzo}},\ and\ \bibinfo {author} {\bibfnamefont {F.}~\bibnamefont {Duque}},\ }\href {https://doi.org/10.1103/PhysRevD.108.104020} {\bibfield  {journal} {\bibinfo  {journal} {Phys. Rev. D}\ }\textbf {\bibinfo {volume} {108}},\ \bibinfo {pages} {104020} (\bibinfo {year} {2023})},\ \Eprint {https://arxiv.org/abs/2302.03050} {arXiv:2302.03050 [gr-qc]} \BibitemShut {NoStop}%
\bibitem [{\citenamefont {Giesler}\ \emph {et~al.}(2025)\citenamefont {Giesler} \emph {et~al.}}]{Giesler:2024hcr}%
  \BibitemOpen
  \bibfield  {author} {\bibinfo {author} {\bibfnamefont {M.}~\bibnamefont {Giesler}} \emph {et~al.},\ }\href {https://doi.org/10.1103/PhysRevD.111.084041} {\bibfield  {journal} {\bibinfo  {journal} {Phys. Rev. D}\ }\textbf {\bibinfo {volume} {111}},\ \bibinfo {pages} {084041} (\bibinfo {year} {2025})},\ \Eprint {https://arxiv.org/abs/2411.11269} {arXiv:2411.11269 [gr-qc]} \BibitemShut {NoStop}%
\bibitem [{\citenamefont {Mitman}\ \emph {et~al.}(2025)\citenamefont {Mitman} \emph {et~al.}}]{Mitman:2025hgy}%
  \BibitemOpen
  \bibfield  {author} {\bibinfo {author} {\bibfnamefont {K.}~\bibnamefont {Mitman}} \emph {et~al.},\ }\href {https://doi.org/10.1103/qq1g-jlnw} {\bibfield  {journal} {\bibinfo  {journal} {Phys. Rev. D}\ }\textbf {\bibinfo {volume} {112}},\ \bibinfo {pages} {064016} (\bibinfo {year} {2025})},\ \Eprint {https://arxiv.org/abs/2503.09678} {arXiv:2503.09678 [gr-qc]} \BibitemShut {NoStop}%
\bibitem [{\citenamefont {{Vehtari}}\ \emph {et~al.}(2015)\citenamefont {{Vehtari}}, \citenamefont {{Gelman}},\ and\ \citenamefont {{Gabry}}}]{LOO_Paper}%
  \BibitemOpen
  \bibfield  {author} {\bibinfo {author} {\bibfnamefont {A.}~\bibnamefont {{Vehtari}}}, \bibinfo {author} {\bibfnamefont {A.}~\bibnamefont {{Gelman}}},\ and\ \bibinfo {author} {\bibfnamefont {J.}~\bibnamefont {{Gabry}}},\ }\href@noop {} {\bibfield  {journal} {\bibinfo  {journal} {arXiv e-prints}\ ,\ \bibinfo {eid} {arXiv:1507.04544}} (\bibinfo {year} {2015})},\ \Eprint {https://arxiv.org/abs/1507.04544} {arXiv:1507.04544 [stat.CO]} \BibitemShut {NoStop}%
\bibitem [{\citenamefont {Vehtari}()}]{LOO_FAQ}%
  \BibitemOpen
  \bibfield  {author} {\bibinfo {author} {\bibfnamefont {A.}~\bibnamefont {Vehtari}},\ }\href {https://avehtari.github.io/modelselection/CV-FAQ.html} {\bibinfo {title} {Cross-validation faq}}\BibitemShut {NoStop}%
\bibitem [{\citenamefont {Gelman}\ \emph {et~al.}(2013)\citenamefont {Gelman}, \citenamefont {Carlin}, \citenamefont {Stern}, \citenamefont {Dunson}, \citenamefont {Vehtari},\ and\ \citenamefont {Rubin}}]{gelman2013bayesian}%
  \BibitemOpen
  \bibfield  {author} {\bibinfo {author} {\bibfnamefont {A.}~\bibnamefont {Gelman}}, \bibinfo {author} {\bibfnamefont {J.~B.}\ \bibnamefont {Carlin}}, \bibinfo {author} {\bibfnamefont {H.~S.}\ \bibnamefont {Stern}}, \bibinfo {author} {\bibfnamefont {D.~B.}\ \bibnamefont {Dunson}}, \bibinfo {author} {\bibfnamefont {A.}~\bibnamefont {Vehtari}},\ and\ \bibinfo {author} {\bibfnamefont {D.~B.}\ \bibnamefont {Rubin}},\ }\href {https://stat.columbia.edu/~gelman/book/} {\emph {\bibinfo {title} {Bayesian Data Analysis}}},\ \bibinfo {edition} {3rd}\ ed.,\ Chapman \& Hall/CRC Texts in Statistical Science Series\ (\bibinfo  {publisher} {CRC},\ \bibinfo {address} {Boca Raton, Florida},\ \bibinfo {year} {2013})\BibitemShut {NoStop}%
\bibitem [{\citenamefont {Hastie}\ \emph {et~al.}(2009)\citenamefont {Hastie}, \citenamefont {Tibshirani},\ and\ \citenamefont {Friedman}}]{hastie2009elements}%
  \BibitemOpen
  \bibfield  {author} {\bibinfo {author} {\bibfnamefont {T.}~\bibnamefont {Hastie}}, \bibinfo {author} {\bibfnamefont {R.}~\bibnamefont {Tibshirani}},\ and\ \bibinfo {author} {\bibfnamefont {J.}~\bibnamefont {Friedman}},\ }\href {https://books.google.com/books?id=eBSgoAEACAAJ} {\emph {\bibinfo {title} {The Elements of Statistical Learning: Data Mining, Inference, and Prediction}}},\ Springer series in statistics\ (\bibinfo  {publisher} {Springer},\ \bibinfo {year} {2009})\BibitemShut {NoStop}%
\bibitem [{\citenamefont {McLatchie}\ and\ \citenamefont {Vehtari}(2024)}]{mclatchie2024efficient}%
  \BibitemOpen
  \bibfield  {author} {\bibinfo {author} {\bibfnamefont {Y.}~\bibnamefont {McLatchie}}\ and\ \bibinfo {author} {\bibfnamefont {A.}~\bibnamefont {Vehtari}},\ }\href@noop {} {\bibfield  {journal} {\bibinfo  {journal} {Statistics and Computing}\ }\textbf {\bibinfo {volume} {34}},\ \bibinfo {pages} {132} (\bibinfo {year} {2024})}\BibitemShut {NoStop}%
\bibitem [{\citenamefont {Navarro}(2018)}]{navarro_2018_devildeepbluesea}%
  \BibitemOpen
  \bibfield  {author} {\bibinfo {author} {\bibfnamefont {D.}~\bibnamefont {Navarro}},\ }\href {https://doi.org/10.31234/osf.io/39q8y} {\bibinfo {title} {Between the devil and the deep blue sea: Tensions between scientific judgement and statistical model selection}} (\bibinfo {year} {2018})\BibitemShut {NoStop}%
\bibitem [{\citenamefont {Cutler}\ and\ \citenamefont {Vallisneri}(2007)}]{Cutler:2007mi_PEbias}%
  \BibitemOpen
  \bibfield  {author} {\bibinfo {author} {\bibfnamefont {C.}~\bibnamefont {Cutler}}\ and\ \bibinfo {author} {\bibfnamefont {M.}~\bibnamefont {Vallisneri}},\ }\href {https://doi.org/10.1103/PhysRevD.76.104018} {\bibfield  {journal} {\bibinfo  {journal} {Phys. Rev. D}\ }\textbf {\bibinfo {volume} {76}},\ \bibinfo {pages} {104018} (\bibinfo {year} {2007})},\ \Eprint {https://arxiv.org/abs/0707.2982} {arXiv:0707.2982 [gr-qc]} \BibitemShut {NoStop}%
\bibitem [{\citenamefont {Cornish}\ \emph {et~al.}(2011)\citenamefont {Cornish}, \citenamefont {Sampson}, \citenamefont {Yunes},\ and\ \citenamefont {Pretorius}}]{Cornish:2011ys_stealthbias}%
  \BibitemOpen
  \bibfield  {author} {\bibinfo {author} {\bibfnamefont {N.}~\bibnamefont {Cornish}}, \bibinfo {author} {\bibfnamefont {L.}~\bibnamefont {Sampson}}, \bibinfo {author} {\bibfnamefont {N.}~\bibnamefont {Yunes}},\ and\ \bibinfo {author} {\bibfnamefont {F.}~\bibnamefont {Pretorius}},\ }\href {https://doi.org/10.1103/PhysRevD.84.062003} {\bibfield  {journal} {\bibinfo  {journal} {Phys. Rev. D}\ }\textbf {\bibinfo {volume} {84}},\ \bibinfo {pages} {062003} (\bibinfo {year} {2011})},\ \Eprint {https://arxiv.org/abs/1105.2088} {arXiv:1105.2088 [gr-qc]} \BibitemShut {NoStop}%
\bibitem [{\citenamefont {Buscicchio}\ \emph {et~al.}(2019)\citenamefont {Buscicchio}, \citenamefont {Roebber}, \citenamefont {Goldstein},\ and\ \citenamefont {Moore}}]{Buscicchio:2019rir}%
  \BibitemOpen
  \bibfield  {author} {\bibinfo {author} {\bibfnamefont {R.}~\bibnamefont {Buscicchio}}, \bibinfo {author} {\bibfnamefont {E.}~\bibnamefont {Roebber}}, \bibinfo {author} {\bibfnamefont {J.~M.}\ \bibnamefont {Goldstein}},\ and\ \bibinfo {author} {\bibfnamefont {C.~J.}\ \bibnamefont {Moore}},\ }\href {https://doi.org/10.1103/PhysRevD.100.084041} {\bibfield  {journal} {\bibinfo  {journal} {Phys. Rev. D}\ }\textbf {\bibinfo {volume} {100}},\ \bibinfo {pages} {084041} (\bibinfo {year} {2019})},\ \Eprint {https://arxiv.org/abs/1907.11631} {arXiv:1907.11631 [astro-ph.IM]} \BibitemShut {NoStop}%
\bibitem [{\citenamefont {Gossan}\ \emph {et~al.}(2012{\natexlab{b}})\citenamefont {Gossan}, \citenamefont {Veitch},\ and\ \citenamefont {Sathyaprakash}}]{Gossan:2011ha}%
  \BibitemOpen
  \bibfield  {author} {\bibinfo {author} {\bibfnamefont {S.}~\bibnamefont {Gossan}}, \bibinfo {author} {\bibfnamefont {J.}~\bibnamefont {Veitch}},\ and\ \bibinfo {author} {\bibfnamefont {B.~S.}\ \bibnamefont {Sathyaprakash}},\ }\href {https://doi.org/10.1103/PhysRevD.85.124056} {\bibfield  {journal} {\bibinfo  {journal} {Phys. Rev. D}\ }\textbf {\bibinfo {volume} {85}},\ \bibinfo {pages} {124056} (\bibinfo {year} {2012}{\natexlab{b}})},\ \Eprint {https://arxiv.org/abs/1111.5819} {arXiv:1111.5819 [gr-qc]} \BibitemShut {NoStop}%
\bibitem [{\citenamefont {Farr}\ \emph {et~al.}()\citenamefont {Farr} \emph {et~al.}}]{amplitude_prior_note}%
  \BibitemOpen
  \bibfield  {author} {\bibinfo {author} {\bibfnamefont {W.~M.}\ \bibnamefont {Farr}} \emph {et~al.},\ }\href {https://raw.githubusercontent.com/farr/marginalization-prior/main-pdf/ms.pdf} {\bibinfo {title} {Marginalization prior technical note}}\BibitemShut {NoStop}%
\bibitem [{\citenamefont {Isi}(2023)}]{Isi:2022mbx}%
  \BibitemOpen
  \bibfield  {author} {\bibinfo {author} {\bibfnamefont {M.}~\bibnamefont {Isi}},\ }\href {https://doi.org/10.1088/1361-6382/acf28c} {\bibfield  {journal} {\bibinfo  {journal} {Class. Quant. Grav.}\ }\textbf {\bibinfo {volume} {40}},\ \bibinfo {pages} {203001} (\bibinfo {year} {2023})},\ \Eprint {https://arxiv.org/abs/2208.03372} {arXiv:2208.03372 [gr-qc]} \BibitemShut {NoStop}%
\bibitem [{\citenamefont {Kumar}\ \emph {et~al.}(2019)\citenamefont {Kumar}, \citenamefont {Carroll}, \citenamefont {Hartikainen},\ and\ \citenamefont {Martin}}]{arviz_2019}%
  \BibitemOpen
  \bibfield  {author} {\bibinfo {author} {\bibfnamefont {R.}~\bibnamefont {Kumar}}, \bibinfo {author} {\bibfnamefont {C.}~\bibnamefont {Carroll}}, \bibinfo {author} {\bibfnamefont {A.}~\bibnamefont {Hartikainen}},\ and\ \bibinfo {author} {\bibfnamefont {O.}~\bibnamefont {Martin}},\ }\href {https://doi.org/10.21105/joss.01143} {\bibfield  {journal} {\bibinfo  {journal} {Journal of Open Source Software}\ }\textbf {\bibinfo {volume} {4}},\ \bibinfo {pages} {1143} (\bibinfo {year} {2019})}\BibitemShut {NoStop}%
\bibitem [{\citenamefont {Stein}(2019)}]{qnmpackage_Stein}%
  \BibitemOpen
  \bibfield  {author} {\bibinfo {author} {\bibfnamefont {L.~C.}\ \bibnamefont {Stein}},\ }\href {https://doi.org/10.21105/joss.01683} {\bibfield  {journal} {\bibinfo  {journal} {J. Open Source Softw.}\ }\textbf {\bibinfo {volume} {4}},\ \bibinfo {pages} {1683} (\bibinfo {year} {2019})},\ \Eprint {https://arxiv.org/abs/1908.10377} {arXiv:1908.10377 [gr-qc]} \BibitemShut {NoStop}%
\bibitem [{\citenamefont {Carullo}\ \emph {et~al.}()\citenamefont {Carullo}, \citenamefont {Pozzo}, \citenamefont {Laghi}, \citenamefont {Isi},\ and\ \citenamefont {Veitch}}]{Pyring}%
  \BibitemOpen
  \bibfield  {author} {\bibinfo {author} {\bibfnamefont {G.}~\bibnamefont {Carullo}}, \bibinfo {author} {\bibfnamefont {W.~D.}\ \bibnamefont {Pozzo}}, \bibinfo {author} {\bibfnamefont {D.}~\bibnamefont {Laghi}}, \bibinfo {author} {\bibfnamefont {M.}~\bibnamefont {Isi}},\ and\ \bibinfo {author} {\bibfnamefont {J.}~\bibnamefont {Veitch}},\ }\href@noop {} {\bibinfo {title} {pyring}},\ \bibinfo {howpublished} {\url{https://git.ligo.org/lscsoft/pyring}}\BibitemShut {NoStop}%
\bibitem [{\citenamefont {{Abbott}}\ \emph {et~al.}(2021)\citenamefont {{Abbott}}, \citenamefont {{Abbott}}, \citenamefont {{Abraham}}, \citenamefont {{Acernese}}, \citenamefont {{Ackley}}, \citenamefont {{Adams}}, \citenamefont {{Adams}}, \citenamefont {{Adhikari}}, \citenamefont {{Adya}}, \citenamefont {{Affeldt}} \emph {et~al.}}]{TestingGR_LIGO_2ndCatalog}%
  \BibitemOpen
  \bibfield  {author} {\bibinfo {author} {\bibfnamefont {R.}~\bibnamefont {{Abbott}}}, \bibinfo {author} {\bibfnamefont {T.~D.}\ \bibnamefont {{Abbott}}}, \bibinfo {author} {\bibfnamefont {S.}~\bibnamefont {{Abraham}}}, \bibinfo {author} {\bibfnamefont {F.}~\bibnamefont {{Acernese}}}, \bibinfo {author} {\bibfnamefont {K.}~\bibnamefont {{Ackley}}}, \bibinfo {author} {\bibfnamefont {A.}~\bibnamefont {{Adams}}}, \bibinfo {author} {\bibfnamefont {C.}~\bibnamefont {{Adams}}}, \bibinfo {author} {\bibfnamefont {R.~X.}\ \bibnamefont {{Adhikari}}}, \bibinfo {author} {\bibfnamefont {V.~B.}\ \bibnamefont {{Adya}}}, \bibinfo {author} {\bibfnamefont {C.}~\bibnamefont {{Affeldt}}}, \emph {et~al.},\ }\href {https://doi.org/10.1103/PhysRevD.103.122002} {\bibfield  {journal} {\bibinfo  {journal} {\prd}\ }\textbf {\bibinfo {volume} {103}},\ \bibinfo {eid} {122002} (\bibinfo {year} {2021})},\ \Eprint {https://arxiv.org/abs/2010.14529} {arXiv:2010.14529 [gr-qc]} \BibitemShut {NoStop}%
\bibitem [{\citenamefont {Maga\~na Zertuche}\ \emph {et~al.}(2025)\citenamefont {Maga\~na Zertuche}, \citenamefont {Stein}, \citenamefont {Mitman}, \citenamefont {Field}, \citenamefont {Varma}, \citenamefont {Boyle}, \citenamefont {Deppe}, \citenamefont {Kidder}, \citenamefont {Moxon}, \citenamefont {Pfeiffer}, \citenamefont {Scheel}, \citenamefont {Nelli}, \citenamefont {Throwe},\ and\ \citenamefont {Vu}}]{QNMsurrogate_Zertuche}%
  \BibitemOpen
  \bibfield  {author} {\bibinfo {author} {\bibfnamefont {L.}~\bibnamefont {Maga\~na Zertuche}}, \bibinfo {author} {\bibfnamefont {L.~C.}\ \bibnamefont {Stein}}, \bibinfo {author} {\bibfnamefont {K.}~\bibnamefont {Mitman}}, \bibinfo {author} {\bibfnamefont {S.~E.}\ \bibnamefont {Field}}, \bibinfo {author} {\bibfnamefont {V.}~\bibnamefont {Varma}}, \bibinfo {author} {\bibfnamefont {M.}~\bibnamefont {Boyle}}, \bibinfo {author} {\bibfnamefont {N.}~\bibnamefont {Deppe}}, \bibinfo {author} {\bibfnamefont {L.~E.}\ \bibnamefont {Kidder}}, \bibinfo {author} {\bibfnamefont {J.}~\bibnamefont {Moxon}}, \bibinfo {author} {\bibfnamefont {H.~P.}\ \bibnamefont {Pfeiffer}}, \bibinfo {author} {\bibfnamefont {M.~A.}\ \bibnamefont {Scheel}}, \bibinfo {author} {\bibfnamefont {K.~C.}\ \bibnamefont {Nelli}}, \bibinfo {author} {\bibfnamefont {W.}~\bibnamefont {Throwe}},\ and\ \bibinfo {author} {\bibfnamefont {N.~L.}\ \bibnamefont {Vu}},\ }\href {https://doi.org/10.1103/q7sy-g3kl} {\bibfield  {journal} {\bibinfo  {journal} {Phys.
  Rev. D}\ }\textbf {\bibinfo {volume} {112}},\ \bibinfo {pages} {024077} (\bibinfo {year} {2025})}\BibitemShut {NoStop}%
\bibitem [{\citenamefont {Ghosh}\ \emph {et~al.}(2021)\citenamefont {Ghosh}, \citenamefont {Brito},\ and\ \citenamefont {Buonanno}}]{Ghosh:2021mrv}%
  \BibitemOpen
  \bibfield  {author} {\bibinfo {author} {\bibfnamefont {A.}~\bibnamefont {Ghosh}}, \bibinfo {author} {\bibfnamefont {R.}~\bibnamefont {Brito}},\ and\ \bibinfo {author} {\bibfnamefont {A.}~\bibnamefont {Buonanno}},\ }\href {https://doi.org/10.1103/PhysRevD.103.124041} {\bibfield  {journal} {\bibinfo  {journal} {Phys. Rev. D}\ }\textbf {\bibinfo {volume} {103}},\ \bibinfo {pages} {124041} (\bibinfo {year} {2021})},\ \Eprint {https://arxiv.org/abs/2104.01906} {arXiv:2104.01906 [gr-qc]} \BibitemShut {NoStop}%
\bibitem [{\citenamefont {Gennari}\ \emph {et~al.}(2024)\citenamefont {Gennari}, \citenamefont {Carullo},\ and\ \citenamefont {Del~Pozzo}}]{Gennari:2023gmx}%
  \BibitemOpen
  \bibfield  {author} {\bibinfo {author} {\bibfnamefont {V.}~\bibnamefont {Gennari}}, \bibinfo {author} {\bibfnamefont {G.}~\bibnamefont {Carullo}},\ and\ \bibinfo {author} {\bibfnamefont {W.}~\bibnamefont {Del~Pozzo}},\ }\href {https://doi.org/10.1140/epjc/s10052-024-12550-x} {\bibfield  {journal} {\bibinfo  {journal} {Eur. Phys. J. C}\ }\textbf {\bibinfo {volume} {84}},\ \bibinfo {pages} {233} (\bibinfo {year} {2024})},\ \Eprint {https://arxiv.org/abs/2312.12515} {arXiv:2312.12515 [gr-qc]} \BibitemShut {NoStop}%
\bibitem [{\citenamefont {Cheung}\ \emph {et~al.}(2024)\citenamefont {Cheung}, \citenamefont {Berti}, \citenamefont {Baibhav},\ and\ \citenamefont {Cotesta}}]{Cheung:2023vki}%
  \BibitemOpen
  \bibfield  {author} {\bibinfo {author} {\bibfnamefont {M.~H.-Y.}\ \bibnamefont {Cheung}}, \bibinfo {author} {\bibfnamefont {E.}~\bibnamefont {Berti}}, \bibinfo {author} {\bibfnamefont {V.}~\bibnamefont {Baibhav}},\ and\ \bibinfo {author} {\bibfnamefont {R.}~\bibnamefont {Cotesta}},\ }\href {https://doi.org/10.1103/PhysRevD.109.044069} {\bibfield  {journal} {\bibinfo  {journal} {Phys. Rev. D}\ }\textbf {\bibinfo {volume} {109}},\ \bibinfo {pages} {044069} (\bibinfo {year} {2024})},\ \bibinfo {note} {[Erratum: Phys.Rev.D 110, 049902 (2024)]},\ \Eprint {https://arxiv.org/abs/2310.04489} {arXiv:2310.04489 [gr-qc]} \BibitemShut {NoStop}%
\bibitem [{\citenamefont {Clarke}\ \emph {et~al.}(2024)\citenamefont {Clarke} \emph {et~al.}}]{Clarke:2024lwi}%
  \BibitemOpen
  \bibfield  {author} {\bibinfo {author} {\bibfnamefont {T.~A.}\ \bibnamefont {Clarke}} \emph {et~al.},\ }\href {https://doi.org/10.1103/PhysRevD.109.124030} {\bibfield  {journal} {\bibinfo  {journal} {Phys. Rev. D}\ }\textbf {\bibinfo {volume} {109}},\ \bibinfo {pages} {124030} (\bibinfo {year} {2024})},\ \Eprint {https://arxiv.org/abs/2402.02819} {arXiv:2402.02819 [gr-qc]} \BibitemShut {NoStop}%
\bibitem [{\citenamefont {Romero-Shaw}\ \emph {et~al.}(2020)\citenamefont {Romero-Shaw}, \citenamefont {Talbot}, \citenamefont {Biscoveanu}, \citenamefont {D’Emilio}, \citenamefont {Ashton}, \citenamefont {Berry}, \citenamefont {Coughlin}, \citenamefont {Galaudage}, \citenamefont {Hoy}, \citenamefont {Hübner}, \citenamefont {Phukon}, \citenamefont {Pitkin}, \citenamefont {Rizzo}, \citenamefont {Sarin}, \citenamefont {Smith}, \citenamefont {Stevenson}, \citenamefont {Vajpeyi}, \citenamefont {Arène}, \citenamefont {Athar}, \citenamefont {Banagiri}, \citenamefont {Bose}, \citenamefont {Carney}, \citenamefont {Chatziioannou}, \citenamefont {Clark}, \citenamefont {Colleoni}, \citenamefont {Cotesta}, \citenamefont {Edelman}, \citenamefont {Estellés}, \citenamefont {García-Quirós}, \citenamefont {Ghosh}, \citenamefont {Green}, \citenamefont {Haster}, \citenamefont {Husa}, \citenamefont {Keitel}, \citenamefont {Kim}, \citenamefont {Hernandez-Vivanco}, \citenamefont {Magaña Hernandez}, \citenamefont
  {Karathanasis}, \citenamefont {Lasky}, \citenamefont {De Lillo}, \citenamefont {Lower}, \citenamefont {Macleod}, \citenamefont {Mateu-Lucena}, \citenamefont {Miller}, \citenamefont {Millhouse}, \citenamefont {Morisaki}, \citenamefont {Oh}, \citenamefont {Ossokine}, \citenamefont {Payne}, \citenamefont {Powell}, \citenamefont {Pratten}, \citenamefont {Pürrer}, \citenamefont {Ramos-Buades}, \citenamefont {Raymond}, \citenamefont {Thrane}, \citenamefont {Veitch}, \citenamefont {Williams}, \citenamefont {Williams},\ and\ \citenamefont {Xiao}}]{10.1093/mnras/staa2850_bilbybayesianinference}%
  \BibitemOpen
  \bibfield  {author} {\bibinfo {author} {\bibfnamefont {I.~M.}\ \bibnamefont {Romero-Shaw}}, \bibinfo {author} {\bibfnamefont {C.}~\bibnamefont {Talbot}}, \bibinfo {author} {\bibfnamefont {S.}~\bibnamefont {Biscoveanu}}, \bibinfo {author} {\bibfnamefont {V.}~\bibnamefont {D’Emilio}}, \bibinfo {author} {\bibfnamefont {G.}~\bibnamefont {Ashton}}, \bibinfo {author} {\bibfnamefont {C.~P.~L.}\ \bibnamefont {Berry}}, \bibinfo {author} {\bibfnamefont {S.}~\bibnamefont {Coughlin}}, \bibinfo {author} {\bibfnamefont {S.}~\bibnamefont {Galaudage}}, \bibinfo {author} {\bibfnamefont {C.}~\bibnamefont {Hoy}}, \bibinfo {author} {\bibfnamefont {M.}~\bibnamefont {Hübner}}, \bibinfo {author} {\bibfnamefont {K.~S.}\ \bibnamefont {Phukon}}, \bibinfo {author} {\bibfnamefont {M.}~\bibnamefont {Pitkin}}, \bibinfo {author} {\bibfnamefont {M.}~\bibnamefont {Rizzo}}, \bibinfo {author} {\bibfnamefont {N.}~\bibnamefont {Sarin}}, \bibinfo {author} {\bibfnamefont {R.}~\bibnamefont {Smith}}, \bibinfo {author} {\bibfnamefont
  {S.}~\bibnamefont {Stevenson}}, \bibinfo {author} {\bibfnamefont {A.}~\bibnamefont {Vajpeyi}}, \bibinfo {author} {\bibfnamefont {M.}~\bibnamefont {Arène}}, \bibinfo {author} {\bibfnamefont {K.}~\bibnamefont {Athar}}, \bibinfo {author} {\bibfnamefont {S.}~\bibnamefont {Banagiri}}, \bibinfo {author} {\bibfnamefont {N.}~\bibnamefont {Bose}}, \bibinfo {author} {\bibfnamefont {M.}~\bibnamefont {Carney}}, \bibinfo {author} {\bibfnamefont {K.}~\bibnamefont {Chatziioannou}}, \bibinfo {author} {\bibfnamefont {J.~A.}\ \bibnamefont {Clark}}, \bibinfo {author} {\bibfnamefont {M.}~\bibnamefont {Colleoni}}, \bibinfo {author} {\bibfnamefont {R.}~\bibnamefont {Cotesta}}, \bibinfo {author} {\bibfnamefont {B.}~\bibnamefont {Edelman}}, \bibinfo {author} {\bibfnamefont {H.}~\bibnamefont {Estellés}}, \bibinfo {author} {\bibfnamefont {C.}~\bibnamefont {García-Quirós}}, \bibinfo {author} {\bibfnamefont {A.}~\bibnamefont {Ghosh}}, \bibinfo {author} {\bibfnamefont {R.}~\bibnamefont {Green}}, \bibinfo {author} {\bibfnamefont
  {C.-J.}\ \bibnamefont {Haster}}, \bibinfo {author} {\bibfnamefont {S.}~\bibnamefont {Husa}}, \bibinfo {author} {\bibfnamefont {D.}~\bibnamefont {Keitel}}, \bibinfo {author} {\bibfnamefont {A.~X.}\ \bibnamefont {Kim}}, \bibinfo {author} {\bibfnamefont {F.}~\bibnamefont {Hernandez-Vivanco}}, \bibinfo {author} {\bibfnamefont {I.}~\bibnamefont {Magaña Hernandez}}, \bibinfo {author} {\bibfnamefont {C.}~\bibnamefont {Karathanasis}}, \bibinfo {author} {\bibfnamefont {P.~D.}\ \bibnamefont {Lasky}}, \bibinfo {author} {\bibfnamefont {N.}~\bibnamefont {De Lillo}}, \bibinfo {author} {\bibfnamefont {M.~E.}\ \bibnamefont {Lower}}, \bibinfo {author} {\bibfnamefont {D.}~\bibnamefont {Macleod}}, \bibinfo {author} {\bibfnamefont {M.}~\bibnamefont {Mateu-Lucena}}, \bibinfo {author} {\bibfnamefont {A.}~\bibnamefont {Miller}}, \bibinfo {author} {\bibfnamefont {M.}~\bibnamefont {Millhouse}}, \bibinfo {author} {\bibfnamefont {S.}~\bibnamefont {Morisaki}}, \bibinfo {author} {\bibfnamefont {S.~H.}\ \bibnamefont {Oh}}, \bibinfo
  {author} {\bibfnamefont {S.}~\bibnamefont {Ossokine}}, \bibinfo {author} {\bibfnamefont {E.}~\bibnamefont {Payne}}, \bibinfo {author} {\bibfnamefont {J.}~\bibnamefont {Powell}}, \bibinfo {author} {\bibfnamefont {G.}~\bibnamefont {Pratten}}, \bibinfo {author} {\bibfnamefont {M.}~\bibnamefont {Pürrer}}, \bibinfo {author} {\bibfnamefont {A.}~\bibnamefont {Ramos-Buades}}, \bibinfo {author} {\bibfnamefont {V.}~\bibnamefont {Raymond}}, \bibinfo {author} {\bibfnamefont {E.}~\bibnamefont {Thrane}}, \bibinfo {author} {\bibfnamefont {J.}~\bibnamefont {Veitch}}, \bibinfo {author} {\bibfnamefont {D.}~\bibnamefont {Williams}}, \bibinfo {author} {\bibfnamefont {M.~J.}\ \bibnamefont {Williams}},\ and\ \bibinfo {author} {\bibfnamefont {L.}~\bibnamefont {Xiao}},\ }\href {https://doi.org/10.1093/mnras/staa2850} {\bibfield  {journal} {\bibinfo  {journal} {Monthly Notices of the Royal Astronomical Society}\ }\textbf {\bibinfo {volume} {499}},\ \bibinfo {pages} {3295} (\bibinfo {year} {2020})},\ \Eprint
  {https://arxiv.org/abs/https://academic.oup.com/mnras/article-pdf/499/3/3295/34052625/staa2850.pdf} {https://academic.oup.com/mnras/article-pdf/499/3/3295/34052625/staa2850.pdf} \BibitemShut {NoStop}%
\bibitem [{\citenamefont {Chavda}\ \emph {et~al.}(2025)\citenamefont {Chavda}, \citenamefont {Lagos},\ and\ \citenamefont {Hui}}]{Chavda:2024awq}%
  \BibitemOpen
  \bibfield  {author} {\bibinfo {author} {\bibfnamefont {A.}~\bibnamefont {Chavda}}, \bibinfo {author} {\bibfnamefont {M.}~\bibnamefont {Lagos}},\ and\ \bibinfo {author} {\bibfnamefont {L.}~\bibnamefont {Hui}},\ }\href {https://doi.org/10.1088/1475-7516/2025/07/084} {\bibfield  {journal} {\bibinfo  {journal} {JCAP}\ }\textbf {\bibinfo {volume} {07}},\ \bibinfo {pages} {084}},\ \Eprint {https://arxiv.org/abs/2412.03435} {arXiv:2412.03435 [gr-qc]} \BibitemShut {NoStop}%
\bibitem [{\citenamefont {Pacilio}\ \emph {et~al.}(2024)\citenamefont {Pacilio}, \citenamefont {Bhagwat}, \citenamefont {Nobili},\ and\ \citenamefont {Gerosa}}]{Pacilio:2024tdl}%
  \BibitemOpen
  \bibfield  {author} {\bibinfo {author} {\bibfnamefont {C.}~\bibnamefont {Pacilio}}, \bibinfo {author} {\bibfnamefont {S.}~\bibnamefont {Bhagwat}}, \bibinfo {author} {\bibfnamefont {F.}~\bibnamefont {Nobili}},\ and\ \bibinfo {author} {\bibfnamefont {D.}~\bibnamefont {Gerosa}},\ }\href {https://doi.org/10.1103/PhysRevD.110.103037} {\bibfield  {journal} {\bibinfo  {journal} {Phys. Rev. D}\ }\textbf {\bibinfo {volume} {110}},\ \bibinfo {pages} {103037} (\bibinfo {year} {2024})},\ \Eprint {https://arxiv.org/abs/2408.05276} {arXiv:2408.05276 [gr-qc]} \BibitemShut {NoStop}%
\bibitem [{\citenamefont {Forteza}\ \emph {et~al.}(2023)\citenamefont {Forteza}, \citenamefont {Bhagwat}, \citenamefont {Kumar},\ and\ \citenamefont {Pani}}]{Forteza:2022tgq}%
  \BibitemOpen
  \bibfield  {author} {\bibinfo {author} {\bibfnamefont {X.~J.}\ \bibnamefont {Forteza}}, \bibinfo {author} {\bibfnamefont {S.}~\bibnamefont {Bhagwat}}, \bibinfo {author} {\bibfnamefont {S.}~\bibnamefont {Kumar}},\ and\ \bibinfo {author} {\bibfnamefont {P.}~\bibnamefont {Pani}},\ }\href {https://doi.org/10.1103/PhysRevLett.130.021001} {\bibfield  {journal} {\bibinfo  {journal} {Phys. Rev. Lett.}\ }\textbf {\bibinfo {volume} {130}},\ \bibinfo {pages} {021001} (\bibinfo {year} {2023})},\ \Eprint {https://arxiv.org/abs/2205.14910} {arXiv:2205.14910 [gr-qc]} \BibitemShut {NoStop}%
\bibitem [{\citenamefont {Juli{\'e}}\ \emph {et~al.}(2025)\citenamefont {Juli{\'e}}, \citenamefont {Pompili},\ and\ \citenamefont {Buonanno}}]{Julie:2024fwy}%
  \BibitemOpen
  \bibfield  {author} {\bibinfo {author} {\bibfnamefont {F.-L.}\ \bibnamefont {Juli{\'e}}}, \bibinfo {author} {\bibfnamefont {L.}~\bibnamefont {Pompili}},\ and\ \bibinfo {author} {\bibfnamefont {A.}~\bibnamefont {Buonanno}},\ }\href {https://doi.org/10.1103/PhysRevD.111.024016} {\bibfield  {journal} {\bibinfo  {journal} {Phys. Rev. D}\ }\textbf {\bibinfo {volume} {111}},\ \bibinfo {pages} {024016} (\bibinfo {year} {2025})},\ \Eprint {https://arxiv.org/abs/2406.13654} {arXiv:2406.13654 [gr-qc]} \BibitemShut {NoStop}%
\bibitem [{\citenamefont {Doneva}\ \emph {et~al.}(2023)\citenamefont {Doneva}, \citenamefont {Arest{\'e}~Sal{\'o}}, \citenamefont {Clough}, \citenamefont {Figueras},\ and\ \citenamefont {Yazadjiev}}]{Doneva:2023oww}%
  \BibitemOpen
  \bibfield  {author} {\bibinfo {author} {\bibfnamefont {D.~D.}\ \bibnamefont {Doneva}}, \bibinfo {author} {\bibfnamefont {L.}~\bibnamefont {Arest{\'e}~Sal{\'o}}}, \bibinfo {author} {\bibfnamefont {K.}~\bibnamefont {Clough}}, \bibinfo {author} {\bibfnamefont {P.}~\bibnamefont {Figueras}},\ and\ \bibinfo {author} {\bibfnamefont {S.~S.}\ \bibnamefont {Yazadjiev}},\ }\href {https://doi.org/10.1103/PhysRevD.108.084017} {\bibfield  {journal} {\bibinfo  {journal} {Phys. Rev. D}\ }\textbf {\bibinfo {volume} {108}},\ \bibinfo {pages} {084017} (\bibinfo {year} {2023})},\ \Eprint {https://arxiv.org/abs/2307.06474} {arXiv:2307.06474 [gr-qc]} \BibitemShut {NoStop}%
\bibitem [{\citenamefont {Corman}\ \emph {et~al.}(2023)\citenamefont {Corman}, \citenamefont {Ripley},\ and\ \citenamefont {East}}]{Corman:2022xqg}%
  \BibitemOpen
  \bibfield  {author} {\bibinfo {author} {\bibfnamefont {M.}~\bibnamefont {Corman}}, \bibinfo {author} {\bibfnamefont {J.~L.}\ \bibnamefont {Ripley}},\ and\ \bibinfo {author} {\bibfnamefont {W.~E.}\ \bibnamefont {East}},\ }\href {https://doi.org/10.1103/PhysRevD.107.024014} {\bibfield  {journal} {\bibinfo  {journal} {Phys. Rev. D}\ }\textbf {\bibinfo {volume} {107}},\ \bibinfo {pages} {024014} (\bibinfo {year} {2023})},\ \Eprint {https://arxiv.org/abs/2210.09235} {arXiv:2210.09235 [gr-qc]} \BibitemShut {NoStop}%
\bibitem [{\citenamefont {Corman}\ \emph {et~al.}(2024)\citenamefont {Corman}, \citenamefont {Lehner}, \citenamefont {East},\ and\ \citenamefont {Dideron}}]{Corman:2024cdr}%
  \BibitemOpen
  \bibfield  {author} {\bibinfo {author} {\bibfnamefont {M.}~\bibnamefont {Corman}}, \bibinfo {author} {\bibfnamefont {L.}~\bibnamefont {Lehner}}, \bibinfo {author} {\bibfnamefont {W.~E.}\ \bibnamefont {East}},\ and\ \bibinfo {author} {\bibfnamefont {G.}~\bibnamefont {Dideron}},\ }\href {https://doi.org/10.1103/PhysRevD.110.084048} {\bibfield  {journal} {\bibinfo  {journal} {Phys. Rev. D}\ }\textbf {\bibinfo {volume} {110}},\ \bibinfo {pages} {084048} (\bibinfo {year} {2024})},\ \Eprint {https://arxiv.org/abs/2405.15581} {arXiv:2405.15581 [gr-qc]} \BibitemShut {NoStop}%
\bibitem [{\citenamefont {Cayuso}\ \emph {et~al.}(2017)\citenamefont {Cayuso}, \citenamefont {Ortiz},\ and\ \citenamefont {Lehner}}]{Cayuso:2017iqc}%
  \BibitemOpen
  \bibfield  {author} {\bibinfo {author} {\bibfnamefont {J.}~\bibnamefont {Cayuso}}, \bibinfo {author} {\bibfnamefont {N.}~\bibnamefont {Ortiz}},\ and\ \bibinfo {author} {\bibfnamefont {L.}~\bibnamefont {Lehner}},\ }\href {https://doi.org/10.1103/PhysRevD.96.084043} {\bibfield  {journal} {\bibinfo  {journal} {Phys. Rev. D}\ }\textbf {\bibinfo {volume} {96}},\ \bibinfo {pages} {084043} (\bibinfo {year} {2017})},\ \Eprint {https://arxiv.org/abs/1706.07421} {arXiv:1706.07421 [gr-qc]} \BibitemShut {NoStop}%
\bibitem [{\citenamefont {Bhagwat}\ \emph {et~al.}(2020)\citenamefont {Bhagwat}, \citenamefont {Forteza}, \citenamefont {Pani},\ and\ \citenamefont {Ferrari}}]{PhysRevD.101.044033}%
  \BibitemOpen
  \bibfield  {author} {\bibinfo {author} {\bibfnamefont {S.}~\bibnamefont {Bhagwat}}, \bibinfo {author} {\bibfnamefont {X.~J.}\ \bibnamefont {Forteza}}, \bibinfo {author} {\bibfnamefont {P.}~\bibnamefont {Pani}},\ and\ \bibinfo {author} {\bibfnamefont {V.}~\bibnamefont {Ferrari}},\ }\href {https://doi.org/10.1103/PhysRevD.101.044033} {\bibfield  {journal} {\bibinfo  {journal} {Phys. Rev. D}\ }\textbf {\bibinfo {volume} {101}},\ \bibinfo {pages} {044033} (\bibinfo {year} {2020})}\BibitemShut {NoStop}%
\bibitem [{\citenamefont {Chandra}\ and\ \citenamefont {Calder{\'o}n~Bustillo}(2025)}]{Chandra:2025ipu}%
  \BibitemOpen
  \bibfield  {author} {\bibinfo {author} {\bibfnamefont {K.}~\bibnamefont {Chandra}}\ and\ \bibinfo {author} {\bibfnamefont {J.}~\bibnamefont {Calder{\'o}n~Bustillo}},\ }\href@noop {} {\bibfield  {journal} {\bibinfo  {journal} {arXiv eprints}\ } (\bibinfo {year} {2025})},\ \Eprint {https://arxiv.org/abs/2509.17315} {arXiv:2509.17315 [gr-qc]} \BibitemShut {NoStop}%
\bibitem [{\citenamefont {Calder{\'o}n~Bustillo}\ \emph {et~al.}(2021{\natexlab{a}})\citenamefont {Calder{\'o}n~Bustillo}, \citenamefont {Lasky},\ and\ \citenamefont {Thrane}}]{CalderonBustillo:2020rmh}%
  \BibitemOpen
  \bibfield  {author} {\bibinfo {author} {\bibfnamefont {J.}~\bibnamefont {Calder{\'o}n~Bustillo}}, \bibinfo {author} {\bibfnamefont {P.~D.}\ \bibnamefont {Lasky}},\ and\ \bibinfo {author} {\bibfnamefont {E.}~\bibnamefont {Thrane}},\ }\href {https://doi.org/10.1103/PhysRevD.103.024041} {\bibfield  {journal} {\bibinfo  {journal} {Phys. Rev. D}\ }\textbf {\bibinfo {volume} {103}},\ \bibinfo {pages} {024041} (\bibinfo {year} {2021}{\natexlab{a}})},\ \Eprint {https://arxiv.org/abs/2010.01857} {arXiv:2010.01857 [gr-qc]} \BibitemShut {NoStop}%
\bibitem [{\citenamefont {Efron}\ and\ \citenamefont {Hastie}(2021)}]{Efron_Hastie_2021}%
  \BibitemOpen
  \bibfield  {author} {\bibinfo {author} {\bibfnamefont {B.}~\bibnamefont {Efron}}\ and\ \bibinfo {author} {\bibfnamefont {T.}~\bibnamefont {Hastie}},\ }\href@noop {} {\emph {\bibinfo {title} {Computer Age Statistical Inference, Student Edition: Algorithms, Evidence, and Data Science}}},\ Institute of Mathematical Statistics Monographs\ (\bibinfo  {publisher} {Cambridge University Press},\ \bibinfo {year} {2021})\BibitemShut {NoStop}%
\bibitem [{\citenamefont {Price}\ \emph {et~al.}(2016)\citenamefont {Price}, \citenamefont {Nampalliwar},\ and\ \citenamefont {Khanna}}]{Price:2015gia}%
  \BibitemOpen
  \bibfield  {author} {\bibinfo {author} {\bibfnamefont {R.~H.}\ \bibnamefont {Price}}, \bibinfo {author} {\bibfnamefont {S.}~\bibnamefont {Nampalliwar}},\ and\ \bibinfo {author} {\bibfnamefont {G.}~\bibnamefont {Khanna}},\ }\href {https://doi.org/10.1103/PhysRevD.93.044060} {\bibfield  {journal} {\bibinfo  {journal} {Phys. Rev. D}\ }\textbf {\bibinfo {volume} {93}},\ \bibinfo {pages} {044060} (\bibinfo {year} {2016})},\ \Eprint {https://arxiv.org/abs/1508.04797} {arXiv:1508.04797 [gr-qc]} \BibitemShut {NoStop}%
\bibitem [{\citenamefont {Oshita}\ \emph {et~al.}(2025)\citenamefont {Oshita}, \citenamefont {Ma}, \citenamefont {Chen},\ and\ \citenamefont {Yang}}]{Oshita:2025qmn}%
  \BibitemOpen
  \bibfield  {author} {\bibinfo {author} {\bibfnamefont {N.}~\bibnamefont {Oshita}}, \bibinfo {author} {\bibfnamefont {S.}~\bibnamefont {Ma}}, \bibinfo {author} {\bibfnamefont {Y.}~\bibnamefont {Chen}},\ and\ \bibinfo {author} {\bibfnamefont {H.}~\bibnamefont {Yang}},\ }\href@noop {} {\bibfield  {journal} {\bibinfo  {journal} {arXiv eprints}\ } (\bibinfo {year} {2025})},\ \Eprint {https://arxiv.org/abs/2509.09165} {arXiv:2509.09165 [gr-qc]} \BibitemShut {NoStop}%
\bibitem [{\citenamefont {Thompson}\ \emph {et~al.}(2025)\citenamefont {Thompson}, \citenamefont {Hoy}, \citenamefont {Fauchon-Jones},\ and\ \citenamefont {Hannam}}]{Thompson:2025hhc}%
  \BibitemOpen
  \bibfield  {author} {\bibinfo {author} {\bibfnamefont {J.~E.}\ \bibnamefont {Thompson}}, \bibinfo {author} {\bibfnamefont {C.}~\bibnamefont {Hoy}}, \bibinfo {author} {\bibfnamefont {E.}~\bibnamefont {Fauchon-Jones}},\ and\ \bibinfo {author} {\bibfnamefont {M.}~\bibnamefont {Hannam}},\ }\href {https://doi.org/10.1103/ddz7-x9zz} {\bibfield  {journal} {\bibinfo  {journal} {Phys. Rev. D}\ }\textbf {\bibinfo {volume} {112}},\ \bibinfo {pages} {064011} (\bibinfo {year} {2025})},\ \Eprint {https://arxiv.org/abs/2506.10530} {arXiv:2506.10530 [gr-qc]} \BibitemShut {NoStop}%
\bibitem [{\citenamefont {Boyle}\ \emph {et~al.}(2014)\citenamefont {Boyle}, \citenamefont {Kidder}, \citenamefont {Ossokine},\ and\ \citenamefont {Pfeiffer}}]{Boyle:2014ioa}%
  \BibitemOpen
  \bibfield  {author} {\bibinfo {author} {\bibfnamefont {M.}~\bibnamefont {Boyle}}, \bibinfo {author} {\bibfnamefont {L.~E.}\ \bibnamefont {Kidder}}, \bibinfo {author} {\bibfnamefont {S.}~\bibnamefont {Ossokine}},\ and\ \bibinfo {author} {\bibfnamefont {H.~P.}\ \bibnamefont {Pfeiffer}},\ }\href@noop {} {\bibinfo {title} {{Gravitational-wave modes from precessing black-hole binaries}}} (\bibinfo {year} {2014}),\ \Eprint {https://arxiv.org/abs/1409.4431} {arXiv:1409.4431 [gr-qc]} \BibitemShut {NoStop}%
\bibitem [{\citenamefont {Siegel}\ \emph {et~al.}(prep)\citenamefont {Siegel}, \citenamefont {Mitman}, \citenamefont {Isi} \emph {et~al.}}]{Siegel_eccentricity_unpublished}%
  \BibitemOpen
  \bibfield  {author} {\bibinfo {author} {\bibfnamefont {H.}~\bibnamefont {Siegel}}, \bibinfo {author} {\bibfnamefont {K.}~\bibnamefont {Mitman}}, \bibinfo {author} {\bibfnamefont {M.}~\bibnamefont {Isi}}, \emph {et~al.}} (\bibinfo {year} {in prep})\BibitemShut {NoStop}%
\bibitem [{\citenamefont {Vijaykumar}\ \emph {et~al.}(2024)\citenamefont {Vijaykumar}, \citenamefont {Hanselman},\ and\ \citenamefont {Zevin}}]{Vijaykumar:2024piy}%
  \BibitemOpen
  \bibfield  {author} {\bibinfo {author} {\bibfnamefont {A.}~\bibnamefont {Vijaykumar}}, \bibinfo {author} {\bibfnamefont {A.~G.}\ \bibnamefont {Hanselman}},\ and\ \bibinfo {author} {\bibfnamefont {M.}~\bibnamefont {Zevin}},\ }\href {https://doi.org/10.3847/1538-4357/ad4455} {\bibfield  {journal} {\bibinfo  {journal} {Astrophys. J.}\ }\textbf {\bibinfo {volume} {969}},\ \bibinfo {pages} {132} (\bibinfo {year} {2024})},\ \Eprint {https://arxiv.org/abs/2402.07892} {arXiv:2402.07892 [astro-ph.HE]} \BibitemShut {NoStop}%
\bibitem [{\citenamefont {Calder{\'o}n~Bustillo}\ \emph {et~al.}(2021{\natexlab{b}})\citenamefont {Calder{\'o}n~Bustillo}, \citenamefont {Sanchis-Gual}, \citenamefont {Torres-Forn{\'e}},\ and\ \citenamefont {Font}}]{CalderonBustillo:2020xms}%
  \BibitemOpen
  \bibfield  {author} {\bibinfo {author} {\bibfnamefont {J.}~\bibnamefont {Calder{\'o}n~Bustillo}}, \bibinfo {author} {\bibfnamefont {N.}~\bibnamefont {Sanchis-Gual}}, \bibinfo {author} {\bibfnamefont {A.}~\bibnamefont {Torres-Forn{\'e}}},\ and\ \bibinfo {author} {\bibfnamefont {J.~A.}\ \bibnamefont {Font}},\ }\href {https://doi.org/10.1103/PhysRevLett.126.201101} {\bibfield  {journal} {\bibinfo  {journal} {Phys. Rev. Lett.}\ }\textbf {\bibinfo {volume} {126}},\ \bibinfo {pages} {201101} (\bibinfo {year} {2021}{\natexlab{b}})},\ \Eprint {https://arxiv.org/abs/2009.01066} {arXiv:2009.01066 [gr-qc]} \BibitemShut {NoStop}%
\bibitem [{\citenamefont {Liu}\ \emph {et~al.}(2023)\citenamefont {Liu}, \citenamefont {Wong}, \citenamefont {Leong}, \citenamefont {More}, \citenamefont {Hannuksela},\ and\ \citenamefont {Li}}]{Liu:2023ikc}%
  \BibitemOpen
  \bibfield  {author} {\bibinfo {author} {\bibfnamefont {A.}~\bibnamefont {Liu}}, \bibinfo {author} {\bibfnamefont {I.~C.~F.}\ \bibnamefont {Wong}}, \bibinfo {author} {\bibfnamefont {S.~H.~W.}\ \bibnamefont {Leong}}, \bibinfo {author} {\bibfnamefont {A.}~\bibnamefont {More}}, \bibinfo {author} {\bibfnamefont {O.~A.}\ \bibnamefont {Hannuksela}},\ and\ \bibinfo {author} {\bibfnamefont {T.~G.~F.}\ \bibnamefont {Li}},\ }\href {https://doi.org/10.1093/mnras/stad1302} {\bibfield  {journal} {\bibinfo  {journal} {Mon. Not. Roy. Astron. Soc.}\ }\textbf {\bibinfo {volume} {525}},\ \bibinfo {pages} {4149} (\bibinfo {year} {2023})},\ \Eprint {https://arxiv.org/abs/2302.09870} {arXiv:2302.09870 [gr-qc]} \BibitemShut {NoStop}%
\bibitem [{\citenamefont {Hussain}\ and\ \citenamefont {Zimmerman}(2022)}]{Hussain:2022ins}%
  \BibitemOpen
  \bibfield  {author} {\bibinfo {author} {\bibfnamefont {A.}~\bibnamefont {Hussain}}\ and\ \bibinfo {author} {\bibfnamefont {A.}~\bibnamefont {Zimmerman}},\ }\href {https://doi.org/10.1103/PhysRevD.106.104018} {\bibfield  {journal} {\bibinfo  {journal} {Phys. Rev. D}\ }\textbf {\bibinfo {volume} {106}},\ \bibinfo {pages} {104018} (\bibinfo {year} {2022})},\ \Eprint {https://arxiv.org/abs/2206.10653} {arXiv:2206.10653 [gr-qc]} \BibitemShut {NoStop}%
\bibitem [{\citenamefont {Li}\ \emph {et~al.}(2023)\citenamefont {Li}, \citenamefont {Wagle}, \citenamefont {Chen},\ and\ \citenamefont {Yunes}}]{Li:2022pcy}%
  \BibitemOpen
  \bibfield  {author} {\bibinfo {author} {\bibfnamefont {D.}~\bibnamefont {Li}}, \bibinfo {author} {\bibfnamefont {P.}~\bibnamefont {Wagle}}, \bibinfo {author} {\bibfnamefont {Y.}~\bibnamefont {Chen}},\ and\ \bibinfo {author} {\bibfnamefont {N.}~\bibnamefont {Yunes}},\ }\href {https://doi.org/10.1103/PhysRevX.13.021029} {\bibfield  {journal} {\bibinfo  {journal} {Phys. Rev. X}\ }\textbf {\bibinfo {volume} {13}},\ \bibinfo {pages} {021029} (\bibinfo {year} {2023})},\ \Eprint {https://arxiv.org/abs/2206.10652} {arXiv:2206.10652 [gr-qc]} \BibitemShut {NoStop}%
\bibitem [{\citenamefont {Crescimbeni}\ \emph {et~al.}(2024)\citenamefont {Crescimbeni}, \citenamefont {Forteza}, \citenamefont {Bhagwat}, \citenamefont {Westerweck},\ and\ \citenamefont {Pani}}]{Crescimbeni:2024sam}%
  \BibitemOpen
  \bibfield  {author} {\bibinfo {author} {\bibfnamefont {F.}~\bibnamefont {Crescimbeni}}, \bibinfo {author} {\bibfnamefont {X.~J.}\ \bibnamefont {Forteza}}, \bibinfo {author} {\bibfnamefont {S.}~\bibnamefont {Bhagwat}}, \bibinfo {author} {\bibfnamefont {J.}~\bibnamefont {Westerweck}},\ and\ \bibinfo {author} {\bibfnamefont {P.}~\bibnamefont {Pani}},\ }\href@noop {} {\bibinfo {title} {{Theory-agnostic searches for non-gravitational modes in black hole ringdown}}} (\bibinfo {year} {2024}),\ \Eprint {https://arxiv.org/abs/2408.08956} {arXiv:2408.08956 [gr-qc]} \BibitemShut {NoStop}%
\bibitem [{\citenamefont {Lestingi}\ \emph {et~al.}(2025)\citenamefont {Lestingi}, \citenamefont {D'Addario},\ and\ \citenamefont {Sotiriou}}]{Lestingi:2025jyb}%
  \BibitemOpen
  \bibfield  {author} {\bibinfo {author} {\bibfnamefont {J.}~\bibnamefont {Lestingi}}, \bibinfo {author} {\bibfnamefont {G.}~\bibnamefont {D'Addario}},\ and\ \bibinfo {author} {\bibfnamefont {T.~P.}\ \bibnamefont {Sotiriou}},\ }\href {https://doi.org/10.1103/vdk4-tg61} {\bibfield  {journal} {\bibinfo  {journal} {Phys. Rev. D}\ }\textbf {\bibinfo {volume} {112}},\ \bibinfo {pages} {064070} (\bibinfo {year} {2025})},\ \Eprint {https://arxiv.org/abs/2505.18261} {arXiv:2505.18261 [gr-qc]} \BibitemShut {NoStop}%
\bibitem [{\citenamefont {Li}\ \emph {et~al.}(2024)\citenamefont {Li}, \citenamefont {Hussain}, \citenamefont {Wagle}, \citenamefont {Chen}, \citenamefont {Yunes},\ and\ \citenamefont {Zimmerman}}]{Li:2023ulk}%
  \BibitemOpen
  \bibfield  {author} {\bibinfo {author} {\bibfnamefont {D.}~\bibnamefont {Li}}, \bibinfo {author} {\bibfnamefont {A.}~\bibnamefont {Hussain}}, \bibinfo {author} {\bibfnamefont {P.}~\bibnamefont {Wagle}}, \bibinfo {author} {\bibfnamefont {Y.}~\bibnamefont {Chen}}, \bibinfo {author} {\bibfnamefont {N.}~\bibnamefont {Yunes}},\ and\ \bibinfo {author} {\bibfnamefont {A.}~\bibnamefont {Zimmerman}},\ }\href {https://doi.org/10.1103/PhysRevD.109.104026} {\bibfield  {journal} {\bibinfo  {journal} {Phys. Rev. D}\ }\textbf {\bibinfo {volume} {109}},\ \bibinfo {pages} {104026} (\bibinfo {year} {2024})},\ \Eprint {https://arxiv.org/abs/2310.06033} {arXiv:2310.06033 [gr-qc]} \BibitemShut {NoStop}%
\bibitem [{\citenamefont {Maenaut}\ \emph {et~al.}(2024)\citenamefont {Maenaut}, \citenamefont {Carullo}, \citenamefont {Cano}, \citenamefont {Liu}, \citenamefont {Cardoso}, \citenamefont {Hertog},\ and\ \citenamefont {Li}}]{Maenaut:2024oci}%
  \BibitemOpen
  \bibfield  {author} {\bibinfo {author} {\bibfnamefont {S.}~\bibnamefont {Maenaut}}, \bibinfo {author} {\bibfnamefont {G.}~\bibnamefont {Carullo}}, \bibinfo {author} {\bibfnamefont {P.~A.}\ \bibnamefont {Cano}}, \bibinfo {author} {\bibfnamefont {A.}~\bibnamefont {Liu}}, \bibinfo {author} {\bibfnamefont {V.}~\bibnamefont {Cardoso}}, \bibinfo {author} {\bibfnamefont {T.}~\bibnamefont {Hertog}},\ and\ \bibinfo {author} {\bibfnamefont {T.~G.~F.}\ \bibnamefont {Li}},\ }\href@noop {} {\bibfield  {journal} {\bibinfo  {journal} {arXiv eprints}\ } (\bibinfo {year} {2024})},\ \Eprint {https://arxiv.org/abs/2411.17893} {arXiv:2411.17893 [gr-qc]} \BibitemShut {NoStop}%
\bibitem [{\citenamefont {Salvatier}\ \emph {et~al.}(2016)\citenamefont {Salvatier}, \citenamefont {Wiecki},\ and\ \citenamefont {Fonnesbeck}}]{Pymc}%
  \BibitemOpen
  \bibfield  {author} {\bibinfo {author} {\bibfnamefont {J.}~\bibnamefont {Salvatier}}, \bibinfo {author} {\bibfnamefont {T.~V.}\ \bibnamefont {Wiecki}},\ and\ \bibinfo {author} {\bibfnamefont {C.}~\bibnamefont {Fonnesbeck}},\ }\href {https://doi.org/10.7717/peerj-cs.55} {\bibfield  {journal} {\bibinfo  {journal} {{PeerJ} Computer Science}\ }\textbf {\bibinfo {volume} {2}},\ \bibinfo {pages} {e55} (\bibinfo {year} {2016})}\BibitemShut {NoStop}%
\bibitem [{\citenamefont {Waskom}(2021)}]{Seaborn}%
  \BibitemOpen
  \bibfield  {author} {\bibinfo {author} {\bibfnamefont {M.~L.}\ \bibnamefont {Waskom}},\ }\href {https://doi.org/10.21105/joss.03021} {\bibfield  {journal} {\bibinfo  {journal} {Journal of Open Source Software}\ }\textbf {\bibinfo {volume} {6}},\ \bibinfo {pages} {3021} (\bibinfo {year} {2021})}\BibitemShut {NoStop}%
\bibitem [{\citenamefont {Hunter}(2007)}]{matplotlib}%
  \BibitemOpen
  \bibfield  {author} {\bibinfo {author} {\bibfnamefont {J.~D.}\ \bibnamefont {Hunter}},\ }\href {https://doi.org/10.1109/MCSE.2007.55} {\bibfield  {journal} {\bibinfo  {journal} {Computing in Science \& Engineering}\ }\textbf {\bibinfo {volume} {9}},\ \bibinfo {pages} {90} (\bibinfo {year} {2007})}\BibitemShut {NoStop}%
\bibitem [{\citenamefont {Kluyver}\ \emph {et~al.}(2016)\citenamefont {Kluyver} \emph {et~al.}}]{jupyter}%
  \BibitemOpen
  \bibfield  {author} {\bibinfo {author} {\bibfnamefont {T.}~\bibnamefont {Kluyver}} \emph {et~al.},\ }\href@noop {} {\bibinfo {title} {Jupyter notebooks -- a publishing format for reproducible computational workflows}} (\bibinfo {year} {2016})\BibitemShut {NoStop}%
\bibitem [{\citenamefont {Harris}\ \emph {et~al.}(2020)\citenamefont {Harris} \emph {et~al.}}]{numpy}%
  \BibitemOpen
  \bibfield  {author} {\bibinfo {author} {\bibfnamefont {C.~R.}\ \bibnamefont {Harris}} \emph {et~al.},\ }\href {https://doi.org/10.1038/s41586-020-2649-2} {\bibfield  {journal} {\bibinfo  {journal} {Nature}\ }\textbf {\bibinfo {volume} {585}},\ \bibinfo {pages} {357} (\bibinfo {year} {2020})}\BibitemShut {NoStop}%
\bibitem [{\citenamefont {Virtanen}\ \emph {et~al.}(2020)\citenamefont {Virtanen} \emph {et~al.}}]{scipy}%
  \BibitemOpen
  \bibfield  {author} {\bibinfo {author} {\bibfnamefont {P.}~\bibnamefont {Virtanen}} \emph {et~al.},\ }\href {https://doi.org/10.1038/s41592-019-0686-2} {\bibfield  {journal} {\bibinfo  {journal} {Nature Methods}\ }\textbf {\bibinfo {volume} {17}},\ \bibinfo {pages} {261} (\bibinfo {year} {2020})}\BibitemShut {NoStop}%
\bibitem [{\citenamefont {Institute}()}]{disbatch}%
  \BibitemOpen
  \bibinfo {author} {\bibfnamefont {F.}~\bibnamefont {Institute}}\BibitemShut {NoStop}%
\bibitem [{\citenamefont {pandas~development team}(2020)}]{pandas}%
  \BibitemOpen
\bibfield  {author} {  }\bibfield  {author} {\bibinfo {author} {\bibfnamefont {T.}~\bibnamefont {pandas~development team}},\ }\href {https://doi.org/10.5281/zenodo.3509134} {\bibinfo {title} {pandas-dev/pandas: Pandas}} (\bibinfo {year} {2020})\BibitemShut {NoStop}%
\bibitem [{\citenamefont {Van~Rossum}\ and\ \citenamefont {Drake}(2009)}]{Python3}%
  \BibitemOpen
  \bibfield  {author} {\bibinfo {author} {\bibfnamefont {G.}~\bibnamefont {Van~Rossum}}\ and\ \bibinfo {author} {\bibfnamefont {F.~L.}\ \bibnamefont {Drake}},\ }\href@noop {} {\emph {\bibinfo {title} {Python 3 Reference Manual}}}\ (\bibinfo  {publisher} {CreateSpace},\ \bibinfo {address} {Scotts Valley, CA},\ \bibinfo {year} {2009})\BibitemShut {NoStop}%
\bibitem [{\citenamefont {OpenAI}(2024)}]{openai2024chatgpt}%
  \BibitemOpen
  \bibfield  {author} {\bibinfo {author} {\bibnamefont {OpenAI}},\ }\href@noop {} {\bibinfo {title} {Chatgpt}},\ \bibinfo {howpublished} {\url{https://chat.openai.com}} (\bibinfo {year} {2024}),\ \bibinfo {note} {large language model}\BibitemShut {NoStop}%
\end{thebibliography}%

\end{document}